\documentclass[article,12pt]{article} 
\usepackage{url}
\usepackage{bm}
\usepackage{rotating}
\usepackage{bbm}
\usepackage{float}
\usepackage{flushend}
\usepackage{verbatim}
\usepackage{tabularx}
\usepackage{multirow} 
\usepackage{arydshln}
\usepackage{hyperref}
\usepackage{subfigure}
\usepackage{pifont}
\usepackage[toc,page]{appendix}
\usepackage{bigints}
\usepackage{diagbox}
\usepackage{mathtools}
\usepackage{color}

\graphicspath{{figures/}}

\RequirePackage{amsthm,amsmath,amsfonts,amssymb}
\RequirePackage[numbers]{natbib}
\theoremstyle{plain}

\theoremstyle{definition}

\newtheorem{rem}{{\bf Remark}}

\theoremstyle{remark}

\newcommand{\E}{\mathbb{E}}

\newcommand{\x}{{\bm \theta}}

\newcommand{\z}{{\bf z}}
\newcommand{\post}{\bar{\pi}}

\newcommand{\norm}[1]{\left\lVert#1\right\rVert}

\UseRawInputEncoding

\title{Optimality in importance sampling: a gentle survey}
\author{F. LLorente$^\Diamond$, L. Martino$^\star$ \\
{\small$^\Diamond$   Stony Brook University, Long Island (NY), USA. }\\
{\small $^\star$ Universit{\'a} degli Studi di Catania, Catania, Italy.}
}

\begin{document}

\maketitle

\thispagestyle{empty}

%%%%%%%%%%%%%%%%%%%%%%%%%%%%%%%%%%%%%%%%%%%%%%%
%% Only one address is permitted per author. %%
%% Only division, organization and e-mail is %%
%% included in the address.                  %%
%% Additional information can be included in %%
%% the Acknowledgments section if necessary. %%
%% ORCID can be inserted by command:         %%
%% \orcid{0000-0000-0000-0000}               %%
%%%%%%%%%%%%%%%%%%%%%%%%%%%%%%%%%%%%%%%%%%%%%%%

%%%%%%%%%%%%%%%%%%%%%%%%%%%%%%%%%%%%%%%%%%%%%%
%% Addresses                                %%
%%%%%%%%%%%%%%%%%%%%%%%%%%%%%%%%%%%%%%%%%%%
\begin{abstract}
The performance of the Monte Carlo sampling methods relies on the crucial choice of a proposal density.
The notion of optimality is fundamental to design suitable adaptive procedures of the proposal density within Monte Carlo schemes. This work is an exhaustive review around the notion of optimality in importance sampling. Several frameworks are described and analyzed, including the approximation of the marginal likelihood for model selection, the use of multiple proposal densities {(including  also the computer graphics context)}, sequences of tempered posteriors {with an optimal tempering schedule}, and noisy scenarios encompassing applications to {energy-based models}, approximate Bayesian computation (ABC), and reinforcement learning, among others. {The behavior of different divergences used to adapt the proposal density so as to approximate the optimal proposal is described.} Some theoretical and empirical comparisons are also provided.
\end{abstract}

%%%%%%%%%%%%%%%%%%%%%%%%%%%%%%%%%%%%%%%%%%%%%%
%% Please use \tableofcontents for articles %%
%% with 50 pages and more                   %%
%%%%%%%%%%%%%%%%%%%%%%%%%%%%%%%%%%%%%%%%%%%%%%
%\tableofcontents

\section{Introduction}
Monte Carlo (MC) methods are  powerful tools for numerical inference and optimization widely employed in statistics, signal processing and machine learning \cite{Liu04b,Robert04}. They are mainly used for computing approximately   the solution of definite integrals, and by extension, of differential equations (for this reason, MC schemes can be considered stochastic quadrature rules). Although exact analytical solutions to integrals are always desirable, such ``unicorns'' are rarely available, specially in real-world systems. Many applications inevitably require the approximation of intractable integrals. Specifically, Bayesian methods need the computation of expectations with respect to posterior  probability density function (pdf) which, generally, are analytically intractable \cite{gelman2013bayesian}. The MC methods can be divided in four main families: direct methods (based on transformations or random variables), accept-reject techniques,  Markov chain Monte Carlo (MCMC) algorithms, and importance sampling (IS)  schemes \cite{LuengoMartino2020,martino2018independent}. The last two families are the most popular for the facility and universality of their possible application \cite{Liang10,Liu04b,Robert04}.
\newline
\newline
All the MC sampling methods require the choice of a suitable proposal density that is crucial for their performance (except a direct transformation) \cite{LuengoMartino2020,Robert04}. For this reason, adaptive strategies that update the proposal density are often employed \cite{Bugallo15,bugallo2017adaptive,Cappe04,Liang10}. In order to design a suitable adaptation procedure, the notion of optimal proposal density (at  least associated to a specific task)  is required. For instance, let us consider a parametric proposal family of densities $q_{{\bm \xi}}(\x)$ (where ${\bm \xi}$ is a parameter vector), that is used in a MC scheme for approximating a specific integral. 
 One idea could be minimizing a divergence $D(q_\text{opt},q_{{\bm \xi}})$ between the optimal proposal for the specific MC scheme and integral to compute, $q_\text{opt}$, and the parametric proposal $q_{{\bm \xi}}$ \cite{akyildiz2024global,Akyildiz2021,Dieng_NIPS2017,perello2023adaptively}. Hence, the optimal parameter vector would be
$$
\widehat{{\bm \xi}}=\arg\min_{{\bm \xi}} J({\bm \xi})= \arg\min_{{\bm \xi}}D(q_\text{opt}, q_{{\bm \xi}}).
$$
However, in order to minimize $J({\bm \xi})$, it is essential the knowledge of $q_\text{opt}(\x)$ for the specific that task we desire to solve.  Note that the parametric family of $q_{{\bm \xi}}(\x)$ must be chosen such that: {\bf (a)} we are  able to draw samples  from $q_{{\bm \xi}}(\x)$ and {\bf (b)} we are able to evaluate point-wise  $q_{{\bm \xi}}(\x)$, for each possible value of $\x$ and the parameter vector ${\bm \xi}$ \cite{akyildiz2024global,perello2023adaptively}.  A  divergence, which naturally arises by the IS theory, is the chi-squared $\chi^2$ divergence, a.k.a., Pearson divergence \cite{Agapiou17,Akyildiz2021,CHEN2005,Dieng_NIPS2017}. For other variational approaches see for instance \cite{Su2021}. { Sections \ref{Unique_var_form_Sect}, \ref{VarInference_Sect}, \ref{Divergence_Num_Sect} and Appedix \ref{FVsect} in this work are specifically devoted to analyze this topic.}
%{ In Section~\ref{Divergence_Num_Sect}, we illustrate the landscapes of three different divergences, highlighting their distinct behaviors during a parametric proposal adaptation.}
% Importance sampling is
% For this reason, several adaptive importance sampling (AIS) schemes have been proposed in the literature.}
\newline
\newline
In this work, we focus on the IS class of methods. It is important to remark that an IS scheme employing a proposal density close to the optimal one (with respect to the specific framework of application) is able to outperform the ideal Monte Carlo technique. This is the reason why the IS approaches are also known as variance reduction methods \cite{Arouna2004,Lapeyre2011,owen2013monte}.
 {After introducing the required background in Sections \ref{ProbStatSect} and \ref{Sect2},} we address several frameworks of practical interest and provide the corresponding optimal proposal density $q_\text{opt}$ \cite{gelman1998simulating,meng2002warp,Owen00,rainforth2020target}.  For this purpose, we have the opportunity to review numerous IS schemes proposed in the literature during the last years, describing also several related properties and results.
 \newline
{In a first part of this survey}, i.e.,  in Sections \ref{FirstPart} and \ref{sec_ISconvariasprop}, we exhaustively analyze the use of a unique or multiple proposal densities for approximating one integral or several integrals \cite{CORNUET12,ElviraMIS15,Owen00,Veach95}. { In Section \ref{sec_ISconvariasprop}, we discuss the optimality (in terms of optimal weights and optimal sample allocation) in different multiple proposal settings, including estimators widely used in computer graphics,  in particular for global illumination and physically based rendering \cite{he2014optimal,Veach95}.}
  Moreover, we consider the joint approximation 
of several integrals in different contexts \cite{Llorente2023_TBI,rainforth2020target}, including a sequence of tempered posteriors \cite{Locatelli00,neal1996sampling,Neal01}.  
\newline
{The second part of the survey begins in Section~\ref{NoisyIS_schemes}, where more specialized and less conventional application frameworks are considered, together with specific strategies for proposal adaptation and construction}. The noisy framework i.e., when the evaluation of the posterior is a random variable itself is addressed in Section \ref{NoisyIS_schemes}: it includes the reinforcement learning and approximate Bayesian computation (ABC) as special cases \cite{DenizNoisy,LLORENTEnoisyIS,LlorenteABC_RF,newton1994approximate}.
{ A suitable design of an optimal proposal density for IS schemes applied to inference in non-normalized models, such as energy-based models, is described in Section~\ref{EBMSect} \cite{dawid2024introduction,Geyer1994Convergence}.}
In Section \ref{sec_marglike}, the specific scenario of the approximation of the marginal likelihood for model selection purpose is also discussed \cite{llorenteREV_ML,gelman1998simulating,meng2002warp}. { An optimal tempering schedule for thermodynamic integration \cite{CalderheadGirolami2009} is described in detail in  Section \ref{OptimalTDI_sect}.}
 In Section \ref{SpecPropSect}, we present the optimal construction of piecewise constant proposal densities, e.g., employed in the {\it Vegas algorithm} \cite{lepage1978new}. { In Section~\ref{VarInference_Sect}, we analyze the behavior of different divergences for adapting the proposal density in order to approximate the optimal proposal.} {  Finally, Section~\ref{NumSect} presents numerical experiments, together with theoretical results, that support and clarify the concepts introduced throughout the previous sections.} 
\newline
{Thus, this work provides an exhaustive survey centered on the notion of optimality in importance sampling, presenting a unified treatment of diverse scenarios through consistent notation and detailed derivations, including full proofs, with the aim of facilitating a clear and rigorous understanding for the reader. The presentation is then intentionally self-contained, making the material accessible to both newcomers and experienced researchers interested in the design and analysis of optimal importance sampling schemes.} The range of applications is wider than only the Monte Carlo world: recently, related notions of optimality have also acquired a relevant place in other related fields such as {\it contrastive learning} \cite{chehab2023,PaperScaffidiMangano}, that has been proved to have a close theoretical development to importance sampling    
 \cite{chehab2023,gutmann12a,Gutman_2019}.
\newline 
\newline 
 { {\bf Clarifying the notion of optimality.} In this survey, the term optimality refers {\it mainly} to minimum-variance optimality, that is, to proposal densities that minimize the variance (or equivalently, the mean-squared error for unbiased estimators) of the IS estimator under consideration. This definition is mathematically tractable and unifies many results across IS theory. Other notions of optimality, such as minimizing Kullback-Leibler (KL) or Renyi divergences, maximizing effective sample size, or achieving robustness to model misspecification, are also meaningful in practice but lie beyond the scope of the present review. We highlight these alternative perspectives where relevant, but our derivations adopt the minimum-variance criterion as the central measure of efficiency.}

\setcounter{tocdepth}{3}
\tableofcontents

%%%%%%%%%%%%%%%%%%%%%%%%%%%%%%%%%%%%%
\section{Problem statement and main notation}\label{ProbStatSect}
%%%%%%%%%%%%%%%%%%%%%%%%%%%%%%%%%%%%%

%%%%%%%%%%%%%%%%%%
\subsection{Bayesian inference}
%%%%%%%%%%%%%%%%%%

In Bayesian inference, the goal is to extract information from the posterior density $\post(\x)=p(\x|{\bf y})$ of a parameter vector $\x = [\theta_1,\dots,\theta_{D}]^\top\subset  \Theta$ given the data ${\bf y}\in \mathbb{R}^{D_y}$, i.e.,   
\begin{equation}
\post(\x)=p(\x|{\bf y})=\frac{\ell({\bf y}|\x) g(\x)}{p({\bf y})},
\end{equation}
where $\ell({\bf y}|\x)$ is the likelihood function, $g(\x)$ is the prior density, and 
\begin{equation}\label{MargLike}
Z=p({\bf y})= \int_\Theta \ell({\bf y}|\x) g(\x) d\x= \int_\Theta \pi(\x) d\x,
\end{equation}
represents the marginal likelihood (a.k.a., Bayesian evidence) \cite{Liu04b,owen2013monte,Robert04}. Moreover, above we have defined the unnormalized posterior 
$$
\pi(\x)= \ell({\bf y}|\x) g(\x),
$$
 i.e., $\pi(\x)\propto \post(\x)$ and $ \post(\x)=\frac{1}{Z}\pi(\x)$. 
The marginal likelihood $Z=p({\bf y})$, which plays the role of a normalizing constant, is particularly important for model selection purposes \cite{llorenteREV_ML}.

%%%%%%%%%%%%%%%%%%%%%%%%%%%%%%%%%%
\subsection{Integrals of interest}
%%%%%%%%%%%%%%%%%%%%%%%%%%%%%%%%%%%

In order to extract  information about the posterior $\post(\x)$, often we are interested in computing integrals which generally involve the product of a generic function $f$ and the posterior $\post$. We can distinguish four cases depending on the possible vectorial nature of $f$ and/or $\post$. We also highlight the corresponding application frameworks.
\newline
\newline
{\bf Case 1: both scalar functions.} In this scenario we have an integral of form
\begin{align}\label{eq_I_of_interest0}
I = \int_\Theta  f(\x)\post(\x)d\x=\frac{1}{Z}\int_\Theta f(\x)\pi(\x)d\x,
\end{align}
where $\post(\x)=\frac{1}{Z} \pi(\x)$ is a pdf with support $\Theta$, and $f(\x): \Theta \to \mathbb{R}$  is a generic integrable function. The  function $f(\x)$ defines the specific expectation with respect to the posterior, that we are interested in computing. We also desire to calculate the normalizing constant of $\pi(\x)$, i.e., 
\begin{align}\label{eq_Z_of_int}
	Z = \int_\Theta\pi(\x)d\x.
\end{align}
{\bf Case 2: vectorial function ${\bf f}(\x)$.} Moreover, since $\x$ is a multidimensional vector, in many other cases we need to consider a vectorial function ${\bf f}(\x)=[f_1(\x),...,f_P(\x)]^{\top}: \Theta \to \mathbb{R}^{P}$ with $P\geq 1$. For instance,  in order to express a moment of order $\alpha$ of a random variable with density $\post(\x)$, for instance, we could set ${\bf f}(\x)=\x^\alpha$ (where $P=D$). In these scenarios, we have a multidimensional integral of interest,
\begin{align}\label{eq_I_of_interest}
{\bf I} = \int_\Theta {\bf f}(\x)\post(\x)d\x=\frac{1}{Z}\int_\Theta {\bf f}(\x)\pi(\x)d\x,
\end{align}
 %moreover ${\bf f}(\x)=[f_1(\x),...,f_P(\x)]^{\top}: \Theta \to \mathbb{R}^{P}$ is a generic integrable function on $\Theta$ with $P\geq 1$ (scalar or vector-valued), and 
 with
$$
{\bf I}=[I_1,...,I_p,...,I_P]^{\top}, \quad \mbox{ where } \quad  I_p=\int_\Theta f_p(\x)\post(\x)d\x.
$$
 For instance, setting ${\bf f}(\x)=\x$, the integral ${\bf I}$  represents the expected value of the r.v. $\x \sim \post(\x)$, that is also known as  minimum mean square error (MMSE) estimator. An interesting special case is addressed in Section \ref{FantSectionLuca},
 where $P=2$ and the estimation of the vector ${\bf I} = [I_1,I_2=Z]^\top=[I,Z]^\top$, i.e.,  $f_1(\x)=f(\x)$ and $f_2(\x)=Z$.\footnote{Note that if $f(\x)=Z$ then $I=\frac{1}{Z}\int_\Theta Z\pi(\x)d\x=\int_\Theta \pi(\x)d\x=Z$.} \footnote{This special case is also related to the approach in Section \ref{Twoor3prop}.} Whereas the general case for a generic ${\bf f}(\x)$ (and $P$) is addressed in Section \ref{SectVariasF}.
 %that also includes the evidence $Z$ as a component. 
 \newline
\newline
{\bf Case 3: $f$ scalar but several posteriors.}   Above, we have to compute a set of $P$ different integrals. We can have other scenarios of interest with several integrals due to the use of $M$ different target pdfs, i.e.,
\begin{align}\label{eq_I_of_interest2}
{\bf I} = \int_\Theta  f(\x){\bm \post}(\x)d\x
\end{align}
 %moreover ${\bf f}(\x)=[f_1(\x),...,f_P(\x)]^{\top}: \Theta \to \mathbb{R}^{P}$ is a generic integrable function on $\Theta$ with $P\geq 1$ (scalar or vector-valued), and 
 with
\begin{align}\label{eq_I_of_interest3}
{\bf I}=[I_1,...,I_m,...,I_M], \quad \mbox{ where } \quad  I_m=\int_\Theta f(\x)\post_m(\x)d\x,
\end{align}
and ${\bm \post}(\x)=[\post_1(\x),...,\post_M(\x)]$. Note that the different target pdfs can be produced by a sequence of tempered posterior distributions \cite{Locatelli00,neal1996sampling,Neal01}.  This case is discussed in Section \ref{VariasTargets}.
\newline
\newline
{\bf Case 4: both vectorial functions.}  The most general scenario is when ${\bf I}$ and ${\bm \pi}$ are both vectorial functions, i.e.,  
\begin{align}\label{eq_I_of_interest4_0}
{\bf I}=[I_1,...,I_p,...,I_P]^{\top} \quad \mbox{ where } \quad  I_{p}=\int_\Theta f_p(\x)\post_p(\x)d\x
\end{align} 
that is addressed in Section \ref{MoreGenCaseSect}.
Generally, the integrals ${\bf I}$ and/or $Z$ cannot be computed analytically, and their computations require approximations by quadrature rules, variational algorithms and/or Monte Carlo methods. Here, we focus on the importance sampling (IS) family of techniques. 

\subsection{The baseline ideal  Monte Carlo (MC) estimator}
For simplicity, let us consider the integral $I =\mathbb{E}_{\bar{\pi}}\left[f(\x)\right]= \int_\Theta  f(\x)\post(\x)d\x$, in Eq. \eqref{eq_I_of_interest0}. The MC estimator of $I$ is given by
\begin{align}\label{MCest}
\widehat{I}_{\texttt{MC}} = \frac{1}{N}\sum_{n=1}^N  f(\x_n), \qquad \x_n \sim \post(\x). 
\end{align}
The estimator above is an unbiased estimation of $I$, i.e., 
\begin{align}\label{MCest2}
\mathbb{E}_{\bar{\pi}}\left[\widehat{I}_{\texttt{MC}}\right] = \frac{1}{N}\sum_{n=1}^N  \mathbb{E}_{\bar{\pi}}\left[f(\x_n)\right]= \frac{1}{N} \left(N  I\right)=I, 
\end{align}
 since $\x_n \sim \post(\x)$ and $\mathbb{E}_{\bar{\pi}}\left[f(\x_n)\right]=I$. Hence, $\mbox{Bias}_{\bar{\pi}}\big[\widehat{I}_{\texttt{MC}}\big]=0$, so that the mean squared error (MSE) is 
 $\mbox{MSE}_{\bar{\pi}}\big[\widehat{I}_{\texttt{MC}}\big]= \mbox{Var}_{\bar{\pi}}\big[\widehat{I}_{\texttt{MC}}\big] +\mbox{Bias}_{\bar{\pi}}\big[\widehat{I}_{\texttt{MC}}\big]^2=\mbox{Var}_{\bar{\pi}}\big[\widehat{I}_{\texttt{MC}}\big]$,
where
\begin{align*}
\mbox{Var}_{\bar{\pi}}\big[\widehat{I}_{\texttt{MC}}\big]=\frac{1}{N} \mbox{Var}_{\bar{\pi}}\left[f(\boldsymbol{\theta})\right]&=\frac{1}{N} \int_{\Theta}(f(\boldsymbol{\theta})-I)^2 \bar{\pi}(\boldsymbol{\theta}) d \boldsymbol{\theta}=\frac{1}{N} \left(\mathbb{E}_{\bar{\pi}}\left[f(\boldsymbol{\theta})^2\right]-I^2 \right).
\end{align*}
This variance and, as a consequence, the MSE converge to zero as $N\rightarrow 0$. It also depends on the variance of the random variable $F=f(\boldsymbol{\theta})$ with $\x \sim \post(\x)$. Unfortunately,  the estimator $\widehat{I}_{\texttt{MC}} $  cannot be applied in many practical problems, since we cannot draw samples directly from $\bar{\pi}(\boldsymbol{\theta})$. However, other types of MC sampling algorithms, such as rejection sampling schemes, Markov chain Monte Carlo (MCMC) techniques, importance sampling (IS) methods can be applied.
Generally, these alternative estimators have bigger variances than the baseline MC estimator. However, we will show the IS schemes using the optimal proposal densities can beat the baseline MC estimator, in terms of efficiency.
%Note that (5) is equivalent to stating that $\widehat{I}_M \xrightarrow{d} \mathcal{N}\left(I, V_M\right)$ as $M \rightarrow \infty$.
%Unfortunately, Algorithm 1 cannot be applied in many practical problems, because we cannot draw samples directly from $\bar{\pi}(\boldsymbol{\theta} | \mathbf{y})$. In these cases, if we can perform point-wise evaluations of the target function, $\pi(\boldsymbol{\theta} | \mathbf{y})=\ell(\mathbf{y} | \boldsymbol{\theta}) p_0(\boldsymbol{\theta})$, we can apply other types of Monte Carlo algorithms: Rejection Sampling (RS) schemes, Markov chain Monte Carlo (MCMC) techniques and importance sampling (IS) methods. These two large classes of algorithms, MCMC and IS, are the core of this paper and will be described in detail in the rest of this work. Before, we briefly recall the basis of the RS approach, which is one of the key ingredients of MCMC methods, in the following section.

%%%%%%%%%%%%%%%%%%%%%%%%%%%%%%%%%%%%%%%%
%\section{Part 1: minimizing the MSE with an unique proposal density}\label{FirstPart}
%%%%%%%%%%%%%%%%%%%%%%%%%%%%%%%%%%%%%%%%%

%%%%%%%%%%%%%%%%%%%%%%%%%%%%%%%%%%
%%%%%%%%%%%%%%%%%%%%%%%%%%%%%%%%%%
{
\subsection{{ Why  importance sampling? } }\label{ISvsMCsec}

Among the four main families of MC techniques - direct methods, accept-reject algorithms, importance sampling (IS), and MCMC methods -  IS and MCMC are the two approaches that can be  applied universally across a wide range of problems. For this reason, they are the most diffused techniques in the literature.
IS schemes are more widely used in signal processing, largely due to the success of particle filtering methods (i.e., sequential IS algorithms with resampling steps) in tracking problems \cite{Akyildiz2020,Djuric03}. In contrast, MCMC algorithms have primarily gained prominence within the statistics community. The number of weighted samples $N$ in IS, and the length $T$ of the Markov chain in MCMC play analogous roles in ensuring the consistency of the resulting estimators, as $N$ and/or $T$ tend to infinity.  In our view, the key advantages of IS with respect to MCMC (or rejecting sampling) are the following:
\begin{itemize}
\item {\bf Deterministic weighting.} Other MC families can also be interpreted as performing an implicit {\it stochastic weighting} of the generated samples. For instance, in rejection sampling, rejected samples effectively receive a weight of  $0$,  while accepted samples receive a weight of $1$. Similarly, in Metropolis-Hastings-type MCMC algorithms, the possible repetition of states can be viewed as assigning random multiplicities to those samples. In contrast, IS assigns deterministic weights to each generated sample, {\it removing a source of variability}. Consequently, when considering the same (static) proposal distribution and the same application setting, IS is expected to yield more efficient estimators, i.e., having lower variance.
\item {\bf Variance reduction.} Closely related to the previous point, an IS scheme can even outperform ideal Monte Carlo when an adequate proposal density (sufficiently close to the optimal one) is employed, as illustrated in Section \ref{FirstSectNum}. To achieve this does not require additional variance-reduction techniques, such as control variates or other more sophisticated schemes.
\item {\bf Normalizing constants.} IS schemes allow for the straightforward estimation of normalizing constants, such as the marginal likelihood, $Z=p({\bf y})$. In fact, virtually any estimator of the marginal likelihood (even when combined with MCMC algorithms) relies either explicitly or implicitly on some IS step (see \cite{llorenteREV_ML}).
\item {\bf Range of applications.} More sophisticated IS schemes can be readily extended to more complex frameworks, such as sequential settings with factorized posteriors, as in particle filtering \cite{Akyildiz2020,Djuric03}.
\end{itemize}
On the other hand, one of the main advantages of MCMC algorithms lies in the use of so-called {\it random-walk proposal densities}, which can be easily implemented even in the simplest MCMC variants. The use of a random-walk proposal promotes thorough exploration of the state space, ensuring an inherently explorative behavior. Moreover, in the random-walk scenario, the range of potential allowed proposals is typically broader than in a standard IS scheme.  This is likely the primary reason why MCMC algorithms have gained widespread adoption within the statistics community.
Several attempts to combine the explorative capabilities of MCMC with the efficiency benefits of importance sampling have been explored in the literature \cite{llorente2022mcmc,LAIS17}.

}
%%%%%%%%%%%%%%%%%%%%%%%%%%%%%%%%%%
%%%%%%%%%%%%%%%%%%%%%%%%%%%%%%%%%%

%%%%%%%%%%%%%%%%%%%%%%%%%%%%%%%%%%
%%%%%%%%%%%%%%%%%%%%%%%%%%%%%%%%%%
\section{The two fundamental families of  IS estimators }\label{Sect2}
%%%%%%%%%%%%%%%%%%%%%%%%%%%%%%%%%%
%%%%%%%%%%%%%%%%%%%%%%%%%%%%%%%%%%

In this section, we describe the two  basic IS schemes in the simplest scenario: considering one proposal density $q$ and one integral $I$. Note that, throughout the manuscript, the proposal density $q(\x)$ is always  normalized, i.e., $\int_{\Theta} q(\x) d\x=1$.

%%%%%%%%%%%%%%%%%%%%%%%%%%%%%%%%%%
\subsection{$Z$ known: standard importance sampling estimator $\widehat{I}_\text{IS}$}
%%%%%%%%%%%%%%%%%%%%%%%%%%%%%%%%%
 Let $q(\x)$ be a pdf with support $\Theta$, and denote as $\x_i$ a sample drawn from it, i.e., $\x_i\sim q(\x)$. The IS estimator of $I$ in Eq. \eqref{eq_I_of_interest} based on ${q}$ is given by
\begin{align}\label{eq_std_IS_est}
	\widehat{I}_\text{IS} = \frac{1}{N}\sum_{i=1}^N \frac{f(\x_i)\post(\x_i)}{q(\x_i)} 
	&=\frac{1}{NZ}\sum_{i=1}^N \frac{\pi(\x_i)}{q(\x_i)} f(\x_i),  \nonumber\\
	&=\frac{1}{NZ}\sum_{i=1}^N w_i f(\x_i),   \qquad \x_i\sim q(\x),
\end{align}
where we have set $w_i=\frac{\pi(\x_i)}{q(\x_i)}\geq 0$ for the non-negative weight assigned to sample $\x_i$.
The estimator $\widehat{I}_\text{IS}$ is an unbiased estimator of $I$  in Eq. \eqref{eq_I_of_interest0}, obtained with the sufficient condition that $q(\x) >0$ whenever $\post(\x)>0$. 

%%%%%%%%%%%%%%
\subsection{ $Z$ unknown: the self-normalized IS estimator $\widehat{I}_\text{SNIS} $}
%%%%%%%%%%%%%%%%%%
 When $\post(\x) = \frac{1}{Z}\pi(\x)$ and $Z$ is unknown, we need to resort to the so-called {\it self-normalized} IS (SNIS) estimator. 
 By using a standard IS estimator, we can estimate $Z$ ({\it reusing} the set of samples drawn from $q(\x)$) as 
 \begin{align}\label{eq_Z_est}
	\widehat{Z}_{\text{IS}}=\widehat{Z} =\frac{1}{N}\sum_{k=1}^{N}w_k= \frac{1}{N}\sum_{k=1}^N\frac{\pi(\x_k)}{q(\x_k)}, \qquad \x_k\sim q(\x).
\end{align}
 Then,  we can replace $Z$ with $\widehat{Z}$ into the  standard IS estimator $\widehat{I}_\text{IS}$ in Eq. \eqref{eq_std_IS_est}, obtaining 
\begin{align}\label{eq_SNIS_est}
	\widehat{I}_\text{SNIS} =\frac{1}{N\widehat{Z} }\sum_{i=1}^N w_i f(\x_i)&= \frac{1}{\sum_{k=1}^{N}w_k}\sum_{i= 1}^{N}w_i f(\x_i), \\ 
	  &=\sum_{i= 1}^{N}\bar{w}_i f(\x_i),  \qquad \x_i\sim q(\x), \nonumber 
\end{align}
where $w_i=\frac{\pi(\x_i)}{q(\x_i)}$ and $\bar{w}_i=\frac{w_i}{\sum_{k=1}^{N}w_k}$, so that $\sum_{i=1}^{N}\bar{w}_i=1$. Unlike  $\widehat{I}_\text{IS}$,  the SNIS estimator is biased, but is still a consistent estimator of Eq. \eqref{eq_I_of_interest}. However, $\widehat{I}_\text{SNIS}$ can be generally more efficient than $\widehat{I}_\text{IS}$ and, in many real-world applications, it is the {\it only} applicable IS estimator \cite{Robert04,Liu04b}. Since $\widehat{I}_\text{SNIS}$ is a {\it convex}  combination of $ f(\x_i)$, $\widehat{I}_\text{SNIS}$ is always bounded (unlike $\widehat{I}_\text{IS}$). Indeed,  we can write 
\begin{align}
	\min_i f(\x_i) \leq \widehat{I}_\text{SNIS} \leq \max_i f(\x_i).
\end{align} 
This is a good property that allows, in some cases,  $\widehat{I}_\text{SNIS}$ to have better performance than $\widehat{I}_\text{IS}$. For instance, see Remark \ref{Rem4}. Note also that $\widehat{I}_\text{SNIS}$ can be seen as a quotient of two standard IS estimators, using the same set of samples drawn from $q(\x)$, i.e., 
\begin{align}
	\widehat{I}_\text{SNIS}= \frac{\widehat{E}}{\widehat{Z}} =\frac{\frac{1}{N} \sum_{i= 1}^{N}w_i f(\x_i)}{\frac{1}{N}\sum_{k=1}^{N}w_k},  \qquad \x_i\sim q(\x),
	% = \frac{\frac{1}{N}\sum_{i=1}^N\frac{f(\x_i)\pi(\x_i)}{q(\x_i)}}{\frac{1}{N}\sum_{i=1}^N\frac{\pi(\x_i)}{q(\x_i)}},
\end{align}
where the numerator $\widehat{E}$ is the estimator of $E=\int_{\Theta} f(\x)\pi(\x)d\x$, whereas the denominator $\widehat{Z}$ in Eq. \eqref{eq_Z_est} is the estimator of $Z=\int_{\Theta} \pi(\x)d\x$. Observe that both estimators are using the same proposal $q(\x)$ and also the same set of samples $\x_i$'s, so that they are correlated. {Alternative procedures will be presented in Section \ref{sec_ISconvariasprop}.} 

%%%%%%%%%%%%%%%%%%%%%%%%%%%%%%%%%%%%%%%%
\section{Optimal IS schemes with a unique proposal density}\label{FirstPart}
%%%%%%%%%%%%%%%%%%%%%%%%%%%%%%%%%%%%%%%%%

%%%%%%%%%%%%%%%%%%%%%%%%%%%%%%%%%%%%%
\subsection{One proposal pdf, one function $f$ and one target $\post$}\label{SectAquiIStand}
%%%%%%%%%%%%%%%%%%%%%%%%%%%%%%%%%%%%%
In this section, we recall the optimal proposals of classical IS estimators: standard IS where $Z$ is assumed known, and the self-normalized IS where  $Z$ is unknown. Hence, here we consider the simplest case, i.e., ${\bf f}(\x)=f(\x)$ and ${\bf I}=I$ as in Eq. \eqref{eq_I_of_interest0}.

\subsubsection{Optimal proposal density for the standard IS estimator}\label{Op_in_StandIS}

Recall that the equality $\mbox{Var}_q[A]=\mbox{E}_q[A^2]-\left(\E_q[A]\right)^2$, and since the samples $\x_i$ are independent and identically distributed from $q$, the variance of the estimator $\widehat{I}_\text{IS}=\frac{1}{N}\sum_{i=1}^N \frac{f(\x_i)\post(\x_i)}{q(\x_i)}$ above is 
\begin{align}
	\mbox{Var}_q[\widehat{I}_\text{IS}]&=\frac{1}{N^2} \cdot N \ \mbox{Var}_{q}\left[\frac{f(\x)\post(\x)}{q(\x)}\right] 
	= \frac{1}{N}\left(\E_{q}\left[\left(\frac{f(\x)\post(\x)}{q(\x)}\right)^2\right] - I^2 \right), \nonumber \\
	&= \frac{1}{N}\sigma_\text{IS}^2, \label{eq_var_std_IS}
%	&= \int_{\Theta_{q}}\frac{(f(\x)\post(\x) - Iq(\x))^2}{q(\x)}d\x.
\end{align}
where we have used $\E_{q}\left[\frac{f(\x)\post(\x)}{q(\x)}\right]=I$, and we have set 
$$
\sigma_\text{IS}^2=\E_{q}\left[\left(\frac{f(\x)\post(\x)}{q(\x)}\right)^2\right] - I^2.
$$
Thus, by applying Jensen's inequality in Eq. \eqref{eq_var_std_IS},  we have that
\begin{align}\label{eq_Jensen1}
\mathbb{E}_{q}\left[\left(\frac{f(\x)\post(\x)}{q(\x)}\right)^2\right] \geq \left(\mathbb{E}_{q}\left[\frac{f(\x)\post(\x)}{q(\x)}\right] \right)^2,
\end{align}
and the equality holds if and only if $\frac{f(\x)\post(\x)}{q(\x)}$ is constant, i.e., we should have $q(\x) \propto f(\x)\post(\x)$. However, for $f(\x)$ taking both negative and positive values, the product $f(\x)\post(\x)$ does not define a pdf. In this case, the only possibility is to take
%\framebox{dsds}
\begin{align}\label{OptimalProposalStandIS}
\fbox{$q_\text{opt}(\x)= \frac{|f(\x)|\post(\x)}{\int_{\Theta}   |f(\x)|\post(\x) d\x} \propto |f(\x)|\post(\x).$}
\end{align}
{\rem Note that the normalization $\int_{\Theta}   |f(\x)|\post(\x) d\x$ of $q_\text{opt}(\x)$ is unknown, hence we can evaluate $q_\text{opt}(\x)$ only approximately or up to the normalizing constant.  Moreover, it is difficult to sample from it.
}
\newline
\newline
The minimum possible value of the variance is
%However, contrary to the previous case, this choice does not yield a zero variance estimator. Indeed we have that
\begin{align}\label{aquifer}
	\mbox{Var}_{q_\text{opt}}[\widehat{I}_\text{IS}] = \frac{1}{N} \left[\left(\int_\Theta |f(\x)|\post(\x)d\x\right)^2 - I^2\right].
\end{align}
Hence, when $f(\x)$ assumes both negative and positive values, the  lowest possible variance the standard IS can achieve is given by the expression above. If  $f(\x)$ is non-negative or non-positive, i.e., $f(\x)>0$ or $f(\x)<0$ for all $\x$, the minimum  variance is zero  since the terms within parenthesis in Eq. \eqref{aquifer} cancel out.

{\rem $\mbox{Var}_{q_\text{opt}}[\widehat{I}_\text{IS}]=0$ when  $f(\x)>0$ or $f(\x)<0$ for all $\x\in \Theta$. }
\newline
\newline
Hence, if  $f(\x)>0$ for all $\x$, then
\begin{align}\label{OptimalProposalStandIS}
\fbox{$q_\text{opt}(\x)= \frac{f(\x)\post(\x)}{\int_{\Theta}   f(\x)\post(\x) d\x} =\frac{f(\x)\post(\x)}{I}$,}
\end{align}
and the minimum possible variance is zero, reached using $q_\text{opt}(\x)$.  Note that the normalizing constant of the optimal proposal coincides with the integral we are trying to approximate. That gives the idea that the optimal proposal is in most cases intractable and cannot be achieved. However, the numerator can be evaluated and this can be exploited to design adaptive mechanisms as shown, for instance, in Section \ref{VarInference_Sect}. 

{\rem Note that $q_\text{opt}(\x)$ in Eq. \eqref{OptimalProposalStandIS} is the optimal proposal for a specific integral, i.e., considering a specific function $f(\x)$ \cite{Liu04b,Robert04}. %For every different expectation with respect to $\post(\x)$, we have a different optimal proposal $q_\text{opt}(\x)$. See Section \ref{AquiGenCase} for a further details.
}

%We consider two scenarios: (i) $f(\x)$ is non-negative, (ii) $f(\x)$ is arbitrary.
%%\newline
%For $f(\x)\geq 0$, and using the result from Eq. \eqref{eq_Jensen1}, the variance of $\widehat{I}_\text{IS}$ is minimized when
%\begin{align}\label{eq_q_opt_f_pos}
%	q_\text{opt}(\x) \propto f(\x)\post(\x).
%\end{align} 
%Indeed, this choice yields $\mbox{Var}[\widehat{I}_\text{IS}] = 0$. This implies that $\widehat{I}_\text{IS}$ is always exact (i.e. $\widehat{I}_\text{IS}=I$) regardless the specific values of $\x_i$'s or the sample size $N$, since
%\begin{align}
%	\widehat{I}_\text{IS}^{\text{opt}} = \frac{1}{N}\sum_{i=1}^N\frac{f(\x_i)\post(\x_i)}{f(\x_i)\post(\x_i)/I} = \frac{1}{N}\sum_{i=1}^N I = I.
%\end{align}
%Note that this choice is possible since $f(\x)\geq 0$ for all $\x$. 

\subsubsection{Optimal proposal density for the SNIS estimator}\label{Sect_optSNIS} 
%Let use denote $\widehat{E} = \frac{1}{N}\sum_{i=1}^N\frac{f(\x_i)\pi(\x_i)}{q(\x_i)}$ and $\widehat{Z} = \frac{1}{N}\sum_{i=1}^N\frac{\pi(\x_i)}{q(\x_i)}$. Hence, $\widehat{I}_\text{SNIS}= \frac{\widehat{E}}{\widehat{Z}}$.
We have seen that $\widehat{I}_\text{SNIS}= \frac{\widehat{E}}{\widehat{Z}}$. If $N$ is large enough, the variance of the ratio $\widehat{I}_\text{SNIS}= \frac{\widehat{E}}{\widehat{Z}}$ can be approximated  as \cite{Robert04},
\begin{align*}
\text{Var}_{q}[\widehat{I}_{\text{SNIS}}] = \text{Var}_{q}\left[\frac{\widehat{E}}{\widehat{Z}}\right] 
\approx \frac{1}{Z^2}\text{Var}_{q}\big[\widehat{E}\big] - 2\frac{E}{Z}\text{Cov}_{q}\big[\widehat{E},\widehat{Z}\big]+ \frac{E^2}{Z^4}\text{Var}_{q}\big[\widehat{Z}\big].
\end{align*}
After some algebra over $\text{Cov}_{q}\big[\widehat{E},\widehat{Z}\big]$,  it is possible to show that 
\begin{align}\label{eq_var_SNIS}
	\mbox{Var}_{q}[\widehat{I}_\text{SNIS}] \approx
%	\widetilde{\mbox{var}}_{q}(\widehat{I}_\text{SNIS}) = 
	\frac{\sigma^2_{\text{SNIS}}}{N} =\frac{1}{N} \mathbb{E}_{q}\left[\left(\frac{\post(\x)}{q(\x)}\left(f(\x) - I\right)\right)^2\right].
\end{align}
 The optimal choice of $q(\x)$, for a specific $f(\x)$ (i.e., for a specific integral), is thus
\begin{align}\label{eq_q_opt_SNIS}
\fbox{$q_\text{opt}(\x)=\frac{|f(\x) - I|\post(\x)}{C_q}   \propto |f(\x) - I|\post(\x)$}.
\end{align} 
The normalizing constant 
$$
C_q=\int_{\Theta} |f(\x) - I|\post(\x) d\x=\mathbb{E}_{\post}[|f(\x)-I|],
$$
 is again unknown. The minimum reachable variance is
\begin{align}
 \mbox{Var}_{q_\text{opt}} [\widehat{I}_\text{SNIS}] & \approx
 \frac{1}{N} \int_{\Theta}\left(\frac{\post(\x)(f(\x) - I)}{q_\text{opt}(\x)} \right)^2 q_\text{opt}(\x) d\x, \nonumber\\
 & \approx
 \frac{1}{N} \int_{\Theta}C_q^2 \frac{1}{C_q} |f(\x) - I|\post(\x)d\x, \nonumber\\
  & \approx
 \frac{1}{N} C_q \int_{\Theta} |f(\x) - I|\post(\x)d\x, \nonumber\\
 & \approx
 \frac{1}{N} C_q^2=
  \frac{1}{N} \Big[\mathbb{E}_{\post}[|f(\x)-I|]\Big]^2.
\end{align}
%{(deberiamos distinguir la expresion general $\sigma_\text{SNIS}^2$ de cuando se ha cogido la $q_\text{opt}$...)}
The above expression defines a fundamental lower bound for any SNIS estimator. 

{\rem\label{Rem4} In this case, unlike for $\widehat{I}_\text{IS}$, there does not exist a proposal density $q(\x)$ such that $\sigma^2_{\text{SNIS}}=0$   (even if $f(\x)$ is non-negative or non-positive). However, an interesting special case is when $f(\x)= c$, i.e., $f(\x)$ is a constant value. Indeed, if $f(\x)= c$, we will always have $\widehat{I}_\text{SNIS}=c$, i.e., we have zero bias and zero variance, while generally $\widehat{I}_\text{IS} \neq c$.}

{\rem Note that $q_\text{opt}(\x)$  in Eq. \eqref{eq_q_opt_SNIS} depends on the unknown integral  $I$. However, this expression has a theoretical value. Furthermore, the value of the integral $I$ could be also replaced  with an estimator $\widehat{I}$ (using also iterative procedures that we discuss in Section \ref{BridgeSect}).}
%\newline
%\newline
%{
\subsubsection{Optimal proposal for estimating $Z$}\label{Sect_optZ}

The estimator $\widehat{Z}$ in Eq. \eqref{eq_Z_est} is a standard IS estimator of the integral in Eq. \eqref{eq_Z_of_int} and its variance is given by
\begin{align}\label{eq_var_Z}
\text{Var}[\widehat{Z}] = \frac{1}{N}\E_q\left[\frac{\pi(\x)^2}{q(\x)^2}\right] - \frac{1}{N}Z^2.
\end{align}
The optimal proposal is thus
\begin{align}
\fbox{$q_\text{opt}(\x) = \post(\x) \propto \pi(\x)$.}
\end{align}
 In this scenario, the optimal (minimum) variance is zero, i.e., $\text{Var}_{q_\text{opt}}[\widehat{Z}]=0$. Table  \ref{TablaIS_menos_1} summarizes all the considerations so far.

%}
\begin{table}[!h]	
	\caption{Summary of optimal proposal pdfs in Section \ref{SectAquiIStand}.}\label{TablaIS_menos_1}
	\vspace{-0.3cm}
	\begin{center}
		\begin{tabular}{|c|c|c| } 
			\hline
		 \diagbox{{\bf Scheme}}{{\bf Target}}
%		 {\bf Integral of interest}	
		 &  $I=\int_\Theta f(\x)\post(\x)d\x$  &  $Z=\int_\Theta \pi(\x)d\x$      \\
%			&&& \\
			\hline 
			\hline		 
			&& \\
	Stand IS	&	$q_\text{opt}(\x)\propto |f(\x)|\post(\x)$ &
			  ---------------    \\
			  && \\
		  SNIS &  $q_\text{opt}(\x)\propto |f(\x) - I|\post(\x)$ 
			  &    $q_\text{opt}(\x)=\post(\x)$ \\
			  && \\
%		     \hline
%		      \hline
%		     {\bf Section}  & \ref{SectAquiIStand} & \ref{SectVariasF}& \ref{AquiGenCase} \\
			\hline
		\end{tabular}
	\end{center}
\end{table}

%%%%%%%%%%%%%%%%%%%%%%%%%%%%%%%%%%%%%%%%%%%%%%%%%
%\subsubsection{Optimal proposal density for a generic particle approximation} \label{AquiGenCase}
%%%%%%%%%%%%%%%%%%%%%%%%%%%%%%%%%%%%%%%%%%%%%%%%%
%
%
%We have seen that the IS techniques provide estimators for specific expectations with respect to $\post(\x)$. 
% More generally, the IS scheme provides the following particle approximation of the measure of $\post(\x)$, i.e.,
%\begin{align}\label{eq_part_approx}
%	\post(\x)\approx \ \widehat{\pi}(\x) = \sum_{i=1}^N \bar{w}_i \delta(\x-\x_i), \qquad \x_i \sim q(\x),
%\end{align}
%where $\bar{w}_i =\frac{1}{\widehat{Z}} w(\x_i)$ and $w(\x_i) = \pi(\x_i)/q(\x_i)$ .
%%{\bf Optimal proposal for SNIS ``in general''. {(lo de Deniz y Joaquin...)}} In the general setting when one is not interested in a specific $f(\x)$, it is usually stated that the ``optimal'' $q(\x)$ in SNIS corresponds to the target density $\post(\x)$. 

\subsubsection{Related theoretical results} \label{AquiGenCase}

In the literature, MSE bounds for the SNIS estimator can be found \cite{Akyildiz2020,Akyildiz2021}, for instance,
\begin{align}
	\mathbb{E}\left[\left(I - \widehat{I}_\text{SNIS}\right)^2\right] \leq \frac{c_f\rho}{N},
\end{align}
where $c_f = 4\norm{f}_\infty$ and $\rho = \mathbb{E}_{q}\left[\frac{\post(\x)^2}{q(\x)^2}\right]= \mathbb{E}_{q}\left[\frac{\pi(\x)^2}{Z^2q(\x)^2}\right]=\mathbb{E}_{q}\left[\frac{1}{Z^2}w(\x)^2\right]$ is the second moment of  $\post(\x)/q(\x)$ \cite{Akyildiz2020}.
The variance of the unnormalized weight $\mbox{Var}_{q}[w(\x)]$  can be related to a measure of divergence between the posterior and proposal \cite{Agapiou17,CHEN2005}, \cite[App. A.2]{llorente2021deep},
\begin{align}
\mbox{Var}_{q}[w(\x)]&= 
\int_\Theta (w(\x) - Z)^2q(\x)d\x, \nonumber \\ 
&= Z^2 \int_\Theta \left(\frac{\pi(\x)}{Zq(\x)} - 1\right)^2q(\x)d\x,  \nonumber  \\
		&= Z^2\int\frac{(\post(\x)- q(\x))^2}{q(\x)}d\x = Z^2 D_{\chi^2}(\post,q), \label{Eqchi}
\end{align} 
where we have used $\post(\x)=\frac{1}{Z}\pi(\x)$ and $D_{\chi^2}(\post,q)$ denotes the Pearson divergence between the posterior $\post$ and proposal $q$. Since $\E_q[w(\x)]=Z$, the relative MSE is 
\begin{align}
\mbox{rel-MSE}=\frac{\E_q[(w(\x)-Z)^2]}{Z^2}=\frac{\mbox{Var}_{q}[w(\x)]}{Z^2}&\propto D_{\chi^2}(\post,q). \label{Eqchi0}
\end{align} 
 Note that $D_{\chi^2}(\post,q) = \norm{(\post-q)\left(\frac{\post-q}{q}\right)}_{L_1}$ and, by Holder's inequality, we have 
\begin{align}\label{Eqchi2}
	D_{\chi^2}(\post,q) = \norm{(\post-q)\left(\frac{\post-q}{q}\right)}_{L_1}
	\leq \norm{\post- q}_{L_2}\norm{\frac{\post-q}{q}}_{L_2}.
\end{align}
Hence, by reducing the $L_2$ distance between $\post$ and $q$, we are diminishing the chi-squared divergence and equivalently the variance of the weight function $w(\x)=\pi(\x)/q(\x)$ \cite[App. A.3]{llorente2021deep}. Moreover, the MSE of $\widehat{I}_\text{SNIS}$ is shown to be bounded also in terms of  $D_{\chi^2}(\post,q)$ \cite{Agapiou17,Akyildiz2021,CHEN2005}.
In this sense, we can assert that the optimal particle approximation can be obtained with $q(\x)=\post(\x)$, as we have seen for $Z$ \cite{Dieng_NIPS2017}. { For a related discussion regarding the minimization of different type of divergences for adapting the proposal density, see Section \ref{VarInference_Sect}.}

%%%%%%%%%%%%%%%%%%%%%%%%%%%%%%%%%%%
\subsection{Unique optimal proposal pdf for multiple related integrals}
%%%%%%%%%%%%%%%%%%%%%%%%%%%%%%%%%%%

In this section, we address the problem of estimating multiple related integrals using a single proposal pdf, and derive the optimal choice for the different cases. Namely, here we search for a {\it unique} optimal proposal density for the simultaneous estimation of several quantities.

% ---------------------------------------------------- %

\subsubsection{Optimal proposal for the simultaneous estimation of $I$ and $Z$}\label{FantSectionLuca} 
%esto me lo enviaste en el whatapp

The choice $q(\x)=\post(\x)$ is very common when one is considering the SNIS estimator.  %as we have discussed in Section \ref{AquiGenCase}. 
With this choice, the variance of the estimator $\widehat{Z}$ is zero, but it is not optimal for estimating $I$ using SNIS.  Conversely, if one uses the optimal proposal for SNIS in Eq. \eqref{eq_q_opt_SNIS}, this choice is not optimal for estimating $Z$. 
\newline
Let us consider the case, where we seek an optimal proposal density for {\it simultaneously} estimating both $Z$ and $I$, using respectively standard IS and SNIS.
We think of the problem of estimating the vector of multiple integrals, where $P=2$ and the estimation of the vector
$$
{\bf I} =[I,Z]^\top,
$$ 
(i.e.,  $f_1(\x)=f(\x)$ and $f_2(\x)=Z$),  using a single proposal $q(\x)$, resulting in the vector estimator $\widehat{{\bf I}} = [\widehat{I}_\text{SNIS},\widehat{Z}]^\top$, where $\widehat{Z}$ from Eq. \eqref{eq_Z_est} and $\widehat{I}_\text{SNS}$ from Eq. \eqref{eq_SNIS_est}.
\newline
\newline
Since we are considering a vector-valued estimator, the variance $\text{Var}[\widehat{{\bf I}}]$ corresponds to a $2 \times 2$ covariance matrix.
We aim to find the proposal which minimizes the sum of variances in the diagonal of the covariance matrix.
For a scalar $f(\x)$, we have in \eqref{eq_var_SNIS} that
$$
\text{Var}[\widehat{I}_\text{SNIS}] \approx \frac{1}{NZ^2}\E\left[\frac{\pi(\x)^2(f(\x)-I)^2}{q(\x)^2}\right].
$$
Recall  also that $\text{Var}[\widehat{Z}] = \frac{1}{N}\left\{\E\left[\frac{\pi(\x)^2}{q(\x)^2}\right]-Z^2\right\}$. 
Thus, 
considering  the following definition of optimal density
\begin{align}
q_\text{opt}(\x) &= \arg\min_{q} \left( \text{Var}[\widehat{I}_\text{SNIS}] + \text{Var}[\widehat{Z}] \right) \\
&=  \arg\min_{q} \E \left[
\frac{Z^2\pi(\x)^2 + \pi(\x)^2(f(\x) - I)^2}{q(\x)^2}.
\right]
\end{align}
%{ using the  Jensen's inequality as we did above (a lo mejor necesitamos un paso intermedio o hacer un appendix...)} { - bien appendix,} 
Using the  Jensen's inequality as in the previous sections, we have that
\begin{align}
\fbox{$q_\text{opt}(\x) \propto \pi(\x)\sqrt{(f(\x)- I)^2 + Z^2}$.}
\end{align}
Note again that  the density $q_\text{opt}(\x)$  depends on two unknowns $I$ and $Z$. However, this expression has a theoretical value and  iterative procedures could be employed, as shown in Section \ref{BridgeSect}.

%%{ add proof in appendix}

% ---------------------------------------------------- %

%%%%%%%%%%%%%%%%%%%%%%%%%%%%%%%%%%%%%
\subsubsection{Optimal proposal for vector-valued functions}\label{SectVariasF}
%%%%%%%%%%%%%%%%%%%%%%%%%%%%%%%%%%%%%

Let us consider a vector-valued function ${\bf f}(\x) = [f_1(\x),\dots,f_P(\x)]^\top$
%$${\bf f}(\x) = \begin{bmatrix}
%f_1(\x) \\
%f_2(\x) \\
%\vdots \\
%f_P(\x)
%\end{bmatrix}$$
with $P$ components (and $M=1$).
%where the $p$-th component
% {$f_p(\x) \geq 0$ (creo que podemos quitar esta condicion...)}. 
We are interested in a vector of integrals
$${\bf I} = \begin{bmatrix}
I_1 \\
I_2 \\
\vdots \\
I_P
\end{bmatrix}
= \begin{bmatrix}
\int_\Theta  f_1(\x)\post(\x)d\x \\
\int_\Theta  f_2(\x)\post(\x)d\x \\
\vdots \\
\int_\Theta  f_P(\x)\post(\x)d\x
\end{bmatrix}.$$
Note that all the integrals share the presence of the posterior $\post(\x)$ (hence, they are in some sense connected). If one function $f_p(\x) =Z$ for some $p$, then we have $I_p=Z$. 
We aim to obtain the optimal proposal for estimating the whole vector ${\bf I}$ in standard IS and in SNIS. Since the posterior $\post(\x)=\frac{1}{Z} \pi(\x)$ is the same in each component of ${\bf I}$, so that we have also a unique normalizing constant $Z$.
\newline
\newline
{\bf Knowing $Z$.} Using the standard IS scheme,
we already know that the $p$-th integral can be estimated through IS with optimal proposal  $q_{p,\text{opt}}(\x) \propto |f_p(\x)|\post(\x)$  for generic $f_p$, for which the variance is minimum.  On the contrary, if we consider a unique proposal for estimating all $I_p$'s, we need to study the variance of the vector-valued estimator ${\bf \widehat{I}}$ whose $p$-th component is
\begin{align}
	\left({\bf \widehat{I}}\right)_p =  \widehat{I}_p = \frac{1}{N}\sum_{i=1}^N\frac{\post(\x_i)f_p(\x_i)}{q(\x_i)},
\end{align}
for $p=1,\dots, P$. Note that all $\widehat{I}_p$'s use the same set $\{\x_i\}_{i=1}^N \sim q$.
%The variance of ${\bf \widehat{I}}$ is a covariance matrix
%$
%\mbox{Var}({\bf \widehat{I}}) = \left(\mbox{Cov}(\widehat{I}_{n}, \widehat{I}_{m})\right)_{n,m}$,  $1\leq n,m \leq P,$
%where
%\begin{align}
%	\mbox{Cov}(\widehat{I}_n,\widehat{I}_m)
%	 =
%	 \frac{1}{N}\E_{q}\left[\frac{\post(\x)^2}{q(\x)^2}f_n(\x)f_m(\x)\right] - \frac{1}{N}I_nI_m.
%\end{align}
%If we use $q(\x) = q_{p,\text{opt}}(\x) \propto f_p(\x)\post(\x)$, we have $\mbox{cov}(\widehat{I}_p, \widehat{I}_j)=0$ for $j= 1, \dots, P$.
%{toda la discusion anterior sobre los terminos de covarianza puede confundir... en la sect anterior simplemente dijimos de minimizar la suma de las varianzas y aqui parece que tenemos que justificarlo...} 
It is natural to look for the proposal that minimizes the sum of the variance of each component, i.e., 
\begin{align}
	q_\text{opt}(\x) = \arg \min_{q} \sum_{p=1}^{P}\mbox{Var}_{q}[\widehat{I}_p].
\end{align}
This is justified from the MSE of ${\bf \widehat{I}}$
\begin{align}\label{eq_MSE_vec}
	\mbox{MSE}({\bf \widehat{I}}) =	\mathbb{E}[({\bf \widehat{I}} - {\bf I})^\top ({\bf \widehat{I}} - {\bf I})]
		=\sum_{p=1}^{P}\E\left[(\widehat{I}_p - I_p)^2\right]
	=\sum_{p=1}^{P}\mbox{Var}_{q}[\widehat{I}_p],
\end{align}
where in the last equality we use $\mathbb{E}[\widehat{I}_p] = I_p$ for $p=1,\dots,P$, i.e., we have unbiased estimators. Hence the $q_\text{opt}$ above is the choice for which ${\bf \widehat{I}}$ is the MMSE estimator of ${\bf I}$. 
Let us rewrite the sum of variances as follows
\begin{align*}
	\sum_{p=1}^{P}\mbox{Var}_{q}(\widehat{I}_p) &= \frac{1}{N}\sum_{p=1}^{P}\mbox{Var}\left[\frac{\post(\x) f_p(\x)}{q(\x)}\right] \\
	&=\frac{1}{N}\sum_{p=1}^{P}\left(
	\mathbb{E}_{q}\left[\frac{\post(\x)^2 f_p(\x)^2}{q(\x)^2}\right] - I_p^2 
	\right) \\
	&=\frac{1}{N}\sum_{p=1}^{P}
	\E_{q}\left[\frac{\post(\x)^2 f_p(\x)^2}{q(\x)^2}\right] - \frac{1}{N}\sum_{p=1}^{P}I_p^2 \\
	&=\frac{1}{N}	\E_{q}\left[\frac{\post(\x)^2 \sum_{p=1}^{P}
	f_p(\x)^2}{q(\x)^2}\right] - \frac{1}{N}\sum_{p=1}^{P}I_p^2. 
\end{align*}
Thus, by Jensen's inequality, we have that
\begin{align}
	\E_{q}\left[\frac{\post(\x)^2 \sum_{p=1}^{P}
		f_p(\x)^2}{q(\x)^2}\right] 
	\geq \left(\E_{q}\left[\frac{\post(\x) \sqrt{\sum_{p=1}^{P}
			f_p(\x)^2
		}}{q(\x)}\right]\right)^2 = \left(\E_{q}\left[\frac{\post(\x) \norm{{\bf 
	f}(\x)}_2}{q(\x)}\right]\right)^2,
\end{align}
where the equality holds if and only if 
$\frac{\post(\x) \norm{{\bf f}(\x)}_2}{q(\x)}$ is constant.
Thus, we have that 
\begin{align}
\fbox{$q_\text{opt}(\x) \propto \post(\x)\norm{{\bf f}(\x)}_2.$}
\end{align}
%and
%\begin{align}
% \sum_{p=1}^{P}\mbox{var}_{q_\text{opt}}(\widehat{I}_p) = \frac{1}{N}\left( \int_\Theta  \post(\x)\norm{{\bf f}(\x)}_2d\x\right)^2 - \frac{1}{N}\sum_{p=1}^{P}I_p^2. 
%\end{align}
%\newline
\newline
{\bf With an unknown $Z$.} Let us consider now estimating the vector ${\bf I}$ using the SNIS approach, i.e., the $p$-th estimator is
\begin{align}
\left({\bf \widehat{I}}\right)_p =  \widehat{I}_p = \frac{1}{\sum_{j=1}^N\frac{\pi(\x_j)}{q(\x_j)}}\sum_{i=1}^N\frac{\pi(\x_i)f_p(\x_i)}{q(\x_i)}.
\end{align}
In this case, the MSE in Eq. \eqref{eq_MSE_vec} does not correspond exactly to the sum of variances, since the SNIS estimators are biased, i.e.,
\begin{align}
	\text{MSE}\left(\widehat{{\bf I}}_\text{SNIS}\right) = \sum_{p=1}^{P}\left\{\text{Var}\left(\widehat{I}^\text{SNIS}_p\right) + \text{Bias}\left(\widehat{I}^\text{SNIS}_p\right)^2\right\} \approx \sum_{p=1}^{P}\text{Var}\left(\widehat{I}^\text{SNIS}_p\right),
\end{align}
where the last approximation fulfills if the sample size $N$ is big enough (since the bias terms are dominated by the variances in the limit $N \to \infty$).
%{la ecuacion de arriba me gusta pero, de nuevo, cuando derivamos la opt en SNIS por primera vez no escribimos lo del sesgo y aqui si....}
Its variance is given (approximately) by
	$$ 
	\mbox{Var}\left[\widehat{I}_p\right] \approx \frac{1}{N}\E_{q}\left[\left(\frac{\post(\x)}{q(\x)}(f_p(\x) - I_p)\right)^2\right].
	$$
The sum of diagonal variances is thus
\begin{align}
\sum_{p=1}^P\mbox{Var}\left[\widehat{I}_p\right] &\approx	\sum_{p= 1}^P\E_{q}\left[\left(\frac{\post(\x)}{q(\x)}(f_p(\x) - I_p)\right)^2\right] \\
&= \E_{q}\left[\frac{\post(\x)^2}{q(\x)^2}\sum_{p= 1}^P(f_p(\x) - I_p)^2\right].
\end{align}
Hence, the optimal proposal is given by
\begin{align}
\fbox{	$q_\text{opt}(\x) 
%	&= \arg \min_{q}	\sum_{p= 1}^P\E_{q}\left[\left(\frac{\post(\x)}{q(\x)}(f_p(\x) - I_p)\right)^2\right] \nonumber \\
	\propto \post(\x)\norm{{\bf f}(\x) - {\bf I}}_2.$}
\end{align}

% ------------------------------------------------------------- %

%%%%%%%%%%%%%%%%%%%%%%%%%%%%%%%%%%%%%%%%%%%%
\subsubsection{Optimal proposal for integrals involving several target pdfs}\label{VariasTargets}
%%%%%%%%%%%%%%%%%%%%%%%%%%%%%%%%%%%%%%%%%%%%

 Now, instead of a set of functions as in the previous section, we are interested in a vector of integrals induced by having a set of target pdfs and a fixed scalar function. This setting corresponds to, e.g., robust Bayesian analysis, where one is interested in computing a lower bound on expectations of a specific function with respect to a family of posterior distributions \cite{cruz2022iterative}.
Let us denote with $\post_m(\x)$ for $m=1,\dots,M$ a set of target pdfs.
We are interested in the following vector of integrals
$${\bf I} = \begin{bmatrix}
	I_1 \\
	I_2 \\
	\vdots \\
	I_M
\end{bmatrix}
= \begin{bmatrix}
	\int_\Theta  f(\x)\post_1(\x)d\x \\
	\int_\Theta  f(\x)\post_2(\x)d\x \\
	\vdots \\
	\int_\Theta  f(\x)\post_M(\x)d\x
\end{bmatrix}.$$
Note that we consider the same $f(\x)$ for all the integrals, each w.r.t. $\post_m(\x)=\frac{1}{Z_m}\pi_m(\x)$. Note that, in this scenario, we also have $P$ different normalizing constants $Z_m$.
\newline
\newline
{\bf All the $Z_m$ are known.} In the case we can evaluate $\post_m(\x)$ for all $m$ (i.e. we have available $Z_m=\int_\Theta  \pi_m(\x)d\x$ for all $p$), the variance of each
$$
\widehat{I}_m = \frac{1}{N}\sum_{i=1}^{N}\frac{\post_m(\x_i)f(\x_i)}{q(\x_i)}, \quad  \x_i\sim q(\x_i),
$$
is given by
$$
\mbox{Var}(\widehat{I}_m) = \frac{1}{N}\mbox{Var}\left[\frac{\post_m(\x)f(\x)}{q(\x)}\right].
$$
Applying the Jensen's inequality, we can see that the sum of these variances $\sum_{m=1}^M \mbox{Var}(\widehat{I}_m)$ is minimized when we take the proposal as
\begin{align}
&\fbox{$q_\text{opt}(\x) \propto |f(\x)|\sqrt{\post_1(\x)^2+\dots+\post_M(\x)^2}$,}  \nonumber\\
&\fbox{$ q_\text{opt}(\x)\propto f(\x)\|\bm{\bar{\pi}}(\x)\|_2$,}
\end{align}
where we have defined 
$$
\bm{\bar{\pi}}(\x)=[\post_1(\x),...,\post_M(\x)].
$$
%\newline
%\newline
{\bf The $Z_m$ are unknown.} In the case we can only evaluate $\pi_m(\x)$ for all $m$ (i.e. $Z_m=\int_\Theta  \pi_m(\x)d\x$  are not availanle for all $m$), we need to consider the self-normalized estimators
$$
\widehat{I}_m = \frac{1}{\sum_{j=1}^N\frac{\pi_m(\x_j)}{q(\x_j)}}\sum_{i=1}^{N}\frac{\pi_m(\x_i)f(\x_i)}{q(\x_i)}, \quad  \x_i\sim q(\x_i)
$$
whose asymptotic variance is
$$
\mbox{var}(\widehat{I}_m) \approx
\frac{1}{N}\E_{q}\left[\left(\frac{\post_m(\x)}{q(\x)}(f(\x) - I_m)\right)^2\right], \qquad \mbox{ as } \qquad N\rightarrow \infty.
$$
 Again using the Jensen's inequality, we can see that the sum of asymptotic variances is minimized when we take the following proposal density:
\begin{align}
&\fbox{$q_\text{opt}(\x) \propto \sqrt{\post_1(\x)^2(f(\x)-I_1)^2+\dots+\post_M(\x)^2(f(\x)-I_M)^2}$,} \nonumber \\
&\fbox{$q_\text{opt}(\x)\propto \|\bm{\bar{\pi}}(\x)\odot(f(\x)\bm{1}_M - {\bf I})\|_2$,}
\end{align}
where $\bm{1}_M=[1,....1]$ is a $1\times M$ unit vector and $\odot$ denotes the element-wise product. Table \ref{TablaIS_0} summarizes these results.

\begin{table}[!h]	
	\caption{Summary for scalar and vector integrals. The related sections are also provided.}\label{TablaIS_0}
	\vspace{-0.5cm}
	\begin{center}
		\begin{tabular}{|c|c|c|c| } 
			\hline
%			\multicolumn{6}{|c|}{ } \\
%			\multicolumn{6}{|c|}{ $\widehat{Z}_{IS1} = \frac{1}{N} \sum_{i=1}^N\frac{g(\x_i)}{q(\x_i)} \ell(\y|\x_i)=\frac{1}{N}\sum_{i=1}^N \rho_i \ell(\y|\x_i)$, \quad  $\rho_i=\frac{g(\x_i)}{q(\x_i)}$}\\
%			\multicolumn{6}{|c|}{ } \\ 
%%			\hline
			 &&& \\
			{\bf Integrals} &  $I=\int_\Theta f(\x)\post(\x)d\x$  &  ${\bf I}=\int_\Theta{\bf f}(\x)\post(\x)d\x$   &  ${\bf I}=\int_\Theta f(\x)\bm{\bar{\pi}}(\x)d\x$   \\
			 {\bf of interest} &&& \\
			\hline 
			\hline		 
			&&& \\
	Stand IS	&	$q_\text{opt}(\x)\propto |f(\x)|\post(\x)$ &
			  $q_\text{opt}(\x)\propto \|{\bf f}(\x)\|_2\post(\x)$ & 
			  $q_\text{opt}(\x)\propto f(\x)\|\bm{\bar{\pi}}(\x)\|_2$  \\
			  &&& \\
		  SNIS &  $q_\text{opt}(\x)\propto |f(\x) - I|\post(\x)$ 
			  & $q_\text{opt}(\x)\propto \|{\bf f}(\x) - {\bf I}\|_2\post(\x)$ &
			  $q_\text{opt}(\x)\propto \|\bm{\bar{\pi}}(\x)\odot(f(\x)\bm{1}_P - {\bf I})\|_2$ \\
			  &&& \\
		     \hline
		      \hline
		     			 {\bf Section}  &  \ref{Op_in_StandIS}-\ref{Sect_optSNIS} & \ref{SectVariasF}& \ref{VariasTargets} \\
			\hline
		\end{tabular}
	\end{center}
	\begin{center}
		\begin{tabular}{|c|c|c|} 
		\hline
		    &  &  \\ 
		{\bf Integrals}  & ${\bf I} = \int_\Theta {\bf f}(\x) \odot {\bm \post}(\x)d\x$   &    $ \left[I,Z\right]$\\
		   {\bf of interest}      & &  \\   
	       \hline	
	         \hline	
	       & & \\ 
	  %      & & \\
	Stand IS	&  $q_\text{opt}(\x)\propto  \|{\bf f}(\x)\odot\post(\x)\|_2$  &  --------------- \\
			 & & \\
		  SNIS & $q_\text{opt}(\x)\propto \|\bm{\bar{\pi}}(\x)\odot({\bf f}(\x) - {\bf I})\|_2$  &  $q_\text{opt}(\x) \propto \pi(\x)\sqrt{(f(\x)- I)^2 + Z^2}$  \\
			&  & \\
	         \hline
	          \hline
	           {\bf Section}  & \ref{MoreGenCaseSect} & \ref{FantSectionLuca} \\
	           \hline	
       	\end{tabular}
	\end{center}
\end{table}

%%%%%%%%%%%%%%%%%%%%%%%%%%%
\subsubsection{Optimal proposal for vector-valued functions and several target densities}\label{MoreGenCaseSect}
%%%%%%%%%%%%%%%%%%%%%%%%%%%

%{ {\bf esto es solo el caso `diagonal' ...no con la matriz}}

Let us consider now the following vector of integrals
$${\bf I} = \begin{bmatrix}
	I_1 \\
	I_2 \\
	\vdots \\
	I_P
\end{bmatrix}
= \begin{bmatrix}
	\int_\Theta  f_1(\x)\post_1(\x)d\x \\
	\int_\Theta  f_2(\x)\post_2(\x)d\x \\
	\vdots \\
	\int_\Theta  f_P(\x)\post_P(\x)d\x
\end{bmatrix},
$$
which can be summarized in the following vectorial form,
\begin{align}\label{eq_I_of_interest_gen}
{\bf I} = \int_\Theta {\bf f}(\x) \odot {\bm \post}(\x)d\x,
\end{align}
where both ${\bf f}(\x)$ and ${\bm \post}(\x)$ are vector-valued functions with $P$ components. This scenario can appear when using tempered posteriors or posteriors considering different mini-batches of data, for instance. 
\newline
\newline
%{Luca says: se puede hacer este caso? a–adir a tabla resumen si fuera posible}
%yo creo que va ser algo as'...
{\bf With known $Z_p$'s.}  We look for the proposal that optimizes the variance of the vector-valued estimator whose $p$-th component is $\widehat{I}_p = \frac{1}{N}\sum_{i=1}^{N}\frac{f_p(\x_i)\post_p(\x_i)}{q(\x_i)}$, since $\widehat{I}_p$ are all unbiased. Hence, we have
\begin{align}
	\text{Var}_q[\widehat{{\bf I}}_\text{IS}] &= \frac{1}{N} \sum_{p=1}^P \text{Var}_q\left[\frac{f_p(\x)\post_p(\x)}{q(\x)}\right],\nonumber \\
	&= \frac{1}{N} \sum_{p=1}^{P} \E\left[\frac{f_p(\x)^2\post_p(\x)^2}{q(\x)^2}\right] + \text{constant terms}, \nonumber \\
	&=  \frac{1}{N} \E\left[\frac{\sum_{p=1}^{P}f_p(\x)^2\post_p(\x)^2}{q(\x)^2}\right] + \text{constant terms}.
\end{align}
Then, applying the Jensen's inequality as in the previous sections, we obtain  
\begin{align}
\fbox{$ q_\text{opt}(\x)\propto \|{\bf f}(\x) \odot \bm{\bar{\pi}}(\x)\|_2$.}
\end{align}
\newline
{\bf With unknown $Z_p$'s.} The sum of the variances of the SNIS estimators is
\begin{align}
	\text{Var}_q[\widehat{{\bf I}}_\text{SNIS}] &\approx \sum_{p=1}^P \frac{1}{NZ_p^2} \E_q\left[\frac{\pi_p(\x)^2(f_p(\x)-I_p)^2}{q(\x)}\right] \\
	&=  \frac{1}{N} \E_q\left[\frac{\sum_{p=1}^P\post_p(\x)^2(f_p(\x)-I_p)^2}{q(\x)}\right].
\end{align}
Hence, following the same procedure, we finally get
\begin{align}
&\fbox{$q_\text{opt}(\x) \propto \| 
{\bf \post}(\x) \odot ({\bf f}(\x) - {\bf I})		
 \|_2$.} 
\end{align}

\noindent
{\rem All the optimal proposal densities $q_\text{opt}(\x)$ in this section are just known up to a normalizing constant, and their point-wise evaluation is also intractable in unnormalized form, since they depend on the unknown quantities we want to estimate, such as $I$ or $Z$. 
% Therefore, their use for computing the marginal likelihood in Eqs. \eqref{MargLike} would require the  additional approximation of the normalizing constant of the proposal pdf as well.
}

{
\subsection{{Unique variational formulation of the optimal proposal problem}}\label{Unique_var_form_Sect}

 All optimal proposal results derived above can be viewed as instances of a single variational optimization problem. Indeed, the proposal density $q(\x)$ that minimizes the variance of an IS estimator is equivalent to solving
\begin{align}
\min _q J[q]=\int_{\Theta} \frac{\|h(\x)\|_2^2}{q(\x)} d \x, \quad \text { subject to } \quad q(\x) \geq 0, \int_{\Theta} q(\x) d \x=1.
\end{align}
where $h(\x)$ represents the specific estimation goal, for instance, $h(\x)=f(\x) \pi(\x)$, or $h(\x)=(f(\x)-I) \pi(\x)$, etc. Introducing a constrained optimization problem, we can obtain:
$$
q_{\mathrm{opt}}(\x) \propto\|h(\x)\|_2.
$$
More specifically, let us consider the constrained functional equation
$$
\mathcal{L}[q]=\int_{\Theta} \frac{\|h(\x)\|_2^2}{q(\x)} d \x+\lambda\left(\int_{\Theta} q(\x) d \x-1\right),
$$
where $\lambda$ is a Lagrange multiplier enforcing normalization. Taking the first variation with respect to $q$ and setting the functional derivative to zero (see Appendix \ref{FVsect} for all the details), we obtain 
$$
\frac{\delta \mathcal{L}}{\delta q}=-\frac{\|h(\x)\|_2^2}{q(\x)^2}+\lambda=0,
$$
 yielding  $q(\x)^2=\frac{\|h(\x)\|_2^2}{\lambda}$ as solution, and hence
$$
q_{\mathrm{opt}}(\x)=\frac{\|h(\x)\|_2}{\int_{\Theta} \|h(\x')\|_2d \x'}.
$$
Indeed, note that the Lagrange multiplier is determined by enforcing the normalization constraint, i.e., $ \int_{\Theta} q(\x) d \x=1$, and substituting into the expression above, we obtain $
\frac{1}{\sqrt{\lambda}} \int_{\Theta} \|h(\x)\|_2 d \x=1$. Solving for $\lambda$, we have
$$
\sqrt{\lambda}=\int_{\Theta}\|h(\x)\|_2 d \x, \quad \Longrightarrow \quad \lambda=\left(\int_{\Theta} \|h(\x)\|_2 d \x\right)^2.
$$
Thus, the optimal proposal in different settings can be interpreted as a special case of this general variational problem, and depends on the form of the integrand $\|h(\x)\|_2$. This point of view highlights that the essence of IS optimality is a calculus-of-variations problem with a single, unifying functional structure. Many cases in this work fit into this general framework. Some example is given below:
\newline
\newline
\fbox{\parbox{0.98\textwidth}{
\begin{itemize} %{\bf Section}  & \ref{SectAquiIStand} & \ref{SectVariasF}& \ref{VariasTargets} \\ 
\item Standard IS, Section \ref{Op_in_StandIS}: $h(\x)=|f(\x)| \pi(\x)$.
\item Self-normalized IS, Section \ref{Sect_optSNIS}: $h(\x)=|f(\x)-I | \pi(\x)$
\item Evidence estimation, Section \ref{Sect_optZ}: $h(\x)=\pi(\x)$
\item Joint $I, Z$ estimation, Section \ref{FantSectionLuca}: $h(\x)=\pi(\x)\sqrt{(f(\x)- I)^2 + Z^2}$.
\item Vector or multiple targets, Sections \ref{SectVariasF}-\ref{VariasTargets}: $h(\x)=\pi(\x) ||\mathbf{f}(\x)||_2$ and $h(\x)=f(\x) ||{\bm \bar{\pi}}(\x)||_2$.
\item Noisy target, Section \ref{NoisyIS_schemes}: $h(\x)=\left[m(\x)^2+s^2(\x)\right]^{1 / 2}$.
\end{itemize} 
}}
\newline
\newline
See Section \ref{VarInference_Sect} for further details regarding the use of different divergences in order to obtain the optimal proposal.
}

%%%%%%%%%%%%%%%%%%%%%%%%%%%%%%%%%%%%%%%%%%%%%
%\section{Part 2: Optimal IS schemes for posterior expectations with multiple proposal pdfs}
\section{Optimality in IS with multiple proposal pdfs}
\label{sec_ISconvariasprop}
%%%%%%%%%%%%%%%%%%%%%%%%%%%%%%%%%%%%%%%%%%%%%%

%{- con el self-normalized los ingles y franceses dice que es mejor dos.... $\Longleftarrow$ esta en Sect. \ref{sec_gen_results_IS} 
%}
 It is interesting to note that we can beat optimal {\it one-proposal} IS estimators by introducing (new) additional proposals, each one tuned and/or optimized for a specific task. These optimal {\it two/three-proposal} IS estimators can overcome the performance limits of previous analyzed estimators \cite{rainforth2020target}.
\newline
In this section, we present some results regarding the optimality in IS schemes where the use of more than one proposal pdfs is jointly considered. In Sections \ref{Twoor3prop}- \ref{Twoor3prop2}, we describe an optimal use of two and three proposal pdfs in the standard IS and SNIS schemes, respectively.
In Sect. \ref{sec_MIS}, we present the general setting of multiple IS (MIS), and discuss the optimal MIS scheme.
Here we focus in the approximation of the posterior expectation $I$ in Eq. \eqref{eq_I_of_interest0}.

%In the next Section \ref{sec_marglike}, we consider the case of the marginal likelihood computation with  multiple proposals.

%{ {\bf faltan mas refs/citas, no olvidar  - tambien la nuestra Computational statistics}}

\subsection{Two proposals for the standard IS estimator $\widehat{I}_\text{IS}$}\label{Twoor3prop}

%The variance of the classical estimators (standard IS and SNIS), even with the optimal proposal choice, is greater than a lower bound. 
%In other words, there is a limit in the performance these estimators can have for a given problem and a given sample size.
%The use of more than one proposals allows to bypass this limitation and hence produce estimators with arbitrary small variance, for a given problem and sample size. 
%\newline
%{\bf Standard IS with two proposals.}
In a standard IS scheme (i.e., when $Z$ is known), with a generic function $f(\x)$, is possible to obtain an estimator with zero variance with the so-called `positivisation trick' of $f$ and the use of two proposals $q_1(\x)$ and $q_2(\x)$ \cite{Owen00,rainforth2020target}. 
The positivisation trick consists in dividing the integral of interest \cite{Llorente2023_TBI,rainforth2020target},
$$
I=\int_\Theta f(\x)\post(\x)d\x,
$$
 in two different integrals,
\begin{align}
I = I_+ - I_- = \int_\Theta f_+(\x)\post(\x)d\x - \int_\Theta f_-(\x)\post(\x)d\x,
\end{align}
where $f_+(\x) = \max\{0, f(\x)\}$ and $f_-(\x) = \max\{0, -f(\x)\}$ are non-negative functions. 
Thus, we can address the approximation of the two integrals, 
$$
I_+ = \E_{q_1}\left[\frac{f_+(\x)\post(\x)}{q_1(\x)}\right] \quad  \text{and} \quad  I_-= \E_{q_2}\left[\frac{f_-(\x)\post(\x)}{q_2(\x)}\right].
$$
Hence, we can design the following two-proposal IS estimator using {\it two different sets of samples}, $\x_i$ and $\widetilde{\x}_j$,
\begin{align}
	\widehat{I}_\text{IS-2q} = \widehat{I}_+ - \widehat{I}_- 
	=&\frac{1}{N_1}\sum_{i=1}^{N_1}\frac{f_+(\x_i)\post(\x_i)}{q_{1}(\x_i)} - 
	\frac{1}{N_2}\sum_{j=1}^{N_2}\frac{f_-(\widetilde{\x}_j)\post(\widetilde{\x}_j)}{q_{2}(\widetilde{\x}_j)}, \\
	& \quad\quad \x_i\sim q_{1}(\x), \quad\quad\quad\quad\quad  \widetilde{\x}_j \sim q_2(\x),  \nonumber
\end{align}
with $i=1,\dots,N_1$ and $j=1,\dots,N_2$.
A zero-variance estimator could be obtained by choosing, respectively, 
\begin{align}\label{eq_2_q_opt}
\fbox{$q_{1,\text{opt}}(\x) \propto f_+(\x)\post(\x),$} \quad \text{and} \quad 	\fbox{$q_{2,\text{opt}}(\x) \propto f_-(\x)\post(\x).$}
\end{align}

%Finally, the resulting estimator is the difference of two IS estimators
%\begin{align}
%\widehat{I}_\text{IS-2q} = \widehat{I}_+ - \widehat{I}_- 
%=\frac{1}{N_1}\sum_{i=1}^{N_!}\frac{f_+(\x_i)\post(\x_i)}{q_{1,\text{opt}}(\x_i)} - 
%\frac{1}{N_2}\sum_{i=1}^{N_2}\frac{f_-(\widetilde{\x}_i)\post(\widetilde{\x}_i)}{q_{2,\text{opt}}(\widetilde{\x}_i)},
%\end{align}
%{no se si poner la expression con la q optimas merece la pena (es trivial $\widehat{I}_+-\widehat{I}_- = I_+ - I_-$).... seria mejor ponerla con q genericas y luego decir las optimas...{ es decir, lo que he puesto en rojo arriba, y luego simplemente decir que se puede hacer Var$[\widehat{I}_{2q}]=0$}...}
%where $\x_i \sim q_{1,\text{opt}}(\x)$ ($i=1,\dots,N_1$) and $\widetilde{\x}_i \sim q_{2,\text{opt}}(\x)$ ($i=1,\dots,N_2$).
%\newline

%%%%%%%%%%%%%%%%%%%%%%%%%%%%%%%
\subsection{Two proposals for the SNIS estimator $\widehat{I}_\text{SNIS}$}\label{Twoor2prop2}
%%%%%%%%%%%%%%%%%%%%%%%%%%%%%%%
If  $Z$ is unknown, we should use a SNIS approach.
In this case, if we want to have zero variance, we need to consider more than two proposal pdfs \cite{rainforth2020target}. Indeed, the SNIS estimator is the ratio of two standard IS estimators, 
$$  I = \frac{E}{Z} = \frac{\E_q\left[\frac{f(\x)\pi(\x)}{q(\x)}\right]}{\E_q\left[\frac{\pi(\x)}{q(\x)}\right]}\approx \widehat{I}_\text{SNIS}=\frac{\widehat{E}}{\widehat{Z}}
=\dfrac{
	\dfrac{1}{N}\sum_{i=1}^{N}\frac{f(\x_i)\pi(\x_i)}{q(\x_i)} 
}{
\dfrac{1}{N}\sum_{i=1}^{N}\frac{\pi(\x_i)}{q(\x_i)}
}, \qquad \x_i \sim q(\x).
$$ 
So far, for both estimators $\widehat{E}$ and $\widehat{Z}$ we employ the same samples from the same unique proposal pdf $q(\x)$. Note that it is impossible to design a proposal density that works arbitrary well for both numerator and denominator \cite{Llorente2023_TBI,rainforth2020target}.
Hence, we could employ a different proposal density for each estimator,
$$ 
 I = \frac{E}{Z} = \frac{\E_{q_1}\left[\frac{f(\x)\pi(\x)}{q_1(\x)}\right]}{\E_{q_2}\left[\frac{\pi(\x)}{q_2(\x)}\right]},
$$ 
so that the final estimator is the ratio of two estimators using different samples from different proposal pdfs,
\begin{align}
 \widehat{I}_\text{SNIS-2q}=\frac{\widehat{E}_{q_1}}{\widehat{Z}_{q2}}=
 \dfrac{
	\dfrac{1}{N}\sum_{i=1}^{N_1}\frac{f(\x_i)\pi(\x_i)}{q_1(\x_i)} 
}{
\dfrac{1}{N_2}\sum_{k=1}^{N_2}\frac{\pi(\z_k)}{q_2(\z_k)}
}, \qquad  \x_i \sim q_1(\x), \quad \z_k \sim q_2(\z).
\end{align}
We can use the two proposals $q_{1,\text{opt}}(\x)$ as in Section \ref{SectAquiIStand} for estimating the numerator $E$, and take a second optimal proposal as $q_{2,\text{opt}}(\x) \propto \pi(\x)$ for estimating $Z$, i.e.,
\begin{align}
\fbox{$q_{1,\text{opt}}(\x) \propto |f(\x)|\post(\x),$}   \quad \text{and} \quad \fbox{$q_{2,\text{opt}}(\x) \propto \pi(\x)$.}
\end{align}
Note that $\widehat{I}_\text{SNS-2q}$ can provide better performance than $\widehat{I}_\text{SNS}$ (considering the same number of total samples and evaluation of $\pi(\x)$, i.e.,  $N=N_1+N_2$), since each optimal proposal is tailored to each specific estimator (instead of a unique proposal addressing the whole ratio of estimators).

%%%%%%%%%%%%%%%%%%%%%%%%%%%%%%%
\subsection{Three proposals for the SNIS estimator $\widehat{I}_\text{SNIS}$}\label{Twoor3prop2}
%%%%%%%%%%%%%%%%%%%%%%%%%%%%%%%

The previous estimator $ \widehat{I}_\text{SNS-2q}$ can be improved using an additional proposal density. Indeed, we can split  $f(\x)$ as we have done in Section \ref{Twoor3prop}. %.... can generally positive and/or negative, for different values of $\x$.
%Instead of using the same $q(\x)$, we can devise two optimal proposals, one for $E$, and another one for $Z$. However, if $f(\x)$ is not non-negative, the optimal standard IS estimator $\widehat{E}$ does not have zero variance (see positivisation trick above).
The idea is to divide the integral $I$ in three different parts, i.e.,
$$ 
 I = \frac{E^+ - E^-}{Z} = \frac{\E_{q_1}\left[\frac{f_+(\x)\pi(\x)}{q_1(\x)}\right] - \E_{q_2}\left[\frac{f_-(\x)\pi(\x)}{q_2(\x)}\right]}{\E_{q_3}\left[\frac{\pi(\x)}{q_3(\x)}\right]},
 $$ 
 where we have applied the positivisation trick of $f$ in the numerator, i.e., we denote as $f_+(\x) = \max\{0, f(\x)\}$ and $f_-(\x) = \max\{0, -f(\x)\}$ two non-negative functions. The resulting estimator is, in this case,
\begin{align}
\widehat{I}_\text{SNIS-3q}=  \dfrac{
	\dfrac{1}{N_1}\sum_{i=1}^{N_1}\frac{f_+(\x_i)\pi(\x_i)}{q_{1}(\x_i)} - 
	\dfrac{1}{N_2}\sum_{j=1}^{N_2}\frac{f_-(\widetilde{\x}_j)\pi(\widetilde{\x}_j)}{q_{2}(\widetilde{\x}_j)}
}{
	\dfrac{1}{N_3}\sum_{k=1}^{N_3}\frac{\pi(\z_k)}{q_{3}(\z_k)}},
\end{align}
where $\x_i \sim q_{1}(\x)$ ($i=1,\dots,N_1$), $\widetilde{\x}_j \sim q_{2}(\x)$ ($j=1,\dots,N_2$) and $\z_k \sim q_{3}(\x)$ ($k=1,\dots,N_3$).
The estimator above with three generic proposal pdfs $q_i$, with $i=1,2,3$, is generally more efficient than SNIS when there is significant mismatch between $\pi(\x)$ and $f(\x)\pi(\x)$, since in that scenario, it is difficult to find a proposal that produces low variance estimates of both numerator and denominator \cite{rainforth2020target}. 
Regarding the optimal choices of the three proposal pdfs, we can use the two proposals $q_{1,\text{opt}}(\x)$ and $q_{2,\text{opt}}(\x)$ as in Eq. \eqref{eq_2_q_opt} for building a zero variance estimators for $E^+$ and $E^-$, and take a third optimal proposal as $q_{3,\text{opt}}(\x) \propto \pi(\x)$ for estimating $Z$, i.e.,
\begin{align}
\fbox{$q_{1,\text{opt}}(\x) \propto f_+(\x)\post(\x),$} \quad	\fbox{$q_{2,\text{opt}}(\x) \propto f_-(\x)\post(\x).$}  \quad \text{and} \quad \fbox{$q_{3,\text{opt}}(\x) \propto \pi(\x).$}
\end{align}
Note that $\widehat{I}_{\text{IS}-2q}$ and $\widehat{I}_{\text{SNIS}-3q}$ can achieve zero variance with suitable choices of proposal pdfs, contrary to standard IS, where we could only have zero variance when $Z$ is known and $f(\x)$ is either non-positive or non-negative. Table \ref{TablaIS_superSUM} summarizes the different optimal IS schemes for the scalar integral $I=\int_\Theta f(\x)\post(\x)d\x$, considering the possible use of a different numbers of proposal densities.
%{
%\subsection{Summary: several strategies for estimating one integral $I$ by IS}\label{sec_obs}
%
%In this section, we can do a summary of the results in the previous sections.
%Recall the (scalar) integral of interest
%\begin{align}\label{eq_I_of_interest0_2}
%I = \int_\Theta  f(\x)\post(\x)d\x=\frac{1}{Z}\int_\Theta f(\x)\pi(\x)d\x,
%\end{align}
% Looking $I$ as unique/sole integral, if $Z$ is known,
%the optimal proposal densities are
% \begin{align}\label{OptimalProposalStandIS_2}
%\fbox{$q_\text{opt}(\x) \propto |f(\x)|\post(\x).$}
%\end{align}
%or, if $Z$ is unknown, 
%\begin{align}\label{eq_q_opt_SNIS_2}
%\fbox{$q_\text{opt}(\x) \propto |f(\x) - I|\post(\x)$,}
%\end{align} 
%
% Looking $I$ as formed by two integrals,  i.e.,
% $$
% I =
% $$
%
%Important observation:
%I can use one proposal, two proposals, or three proposals....

\begin{table}[!h]	
	\caption{Different optimal IS approximations for the scalar integral $I=\int_\Theta f(\x)\post(\x)d\x$. The column regarding the possible reachable zero variance takes into account a generic non-constant function $f(\x)$ (that takes both positive and negative values), and the use of the optimal proposal pdfs. }\label{TablaIS_superSUM}
	\vspace{-0.5cm}
	\begin{center}
		\begin{tabular}{|c|c|c|c|} 
			\hline
%			\multicolumn{6}{|c|}{ } \\
%			\multicolumn{6}{|c|}{ $\widehat{Z}_{IS1} = \frac{1}{N} \sum_{i=1}^N\frac{g(\x_i)}{q(\x_i)} \ell(\y|\x_i)=\frac{1}{N}\sum_{i=1}^N \rho_i \ell(\y|\x_i)$, \quad  $\rho_i=\frac{g(\x_i)}{q(\x_i)}$}\\
%			\multicolumn{6}{|c|}{ } \\ 
%%			\hline
{\bf $Z$ known}		& {\bf Identity}  & {\bf Optimal proposals} & {\bf Min. zero variance} \\
			\hline
		        \hline
		              && & only if $f$ is positive \\ 
\checkmark		& $I= \E_{q}\left[\frac{f(\x)\post(\x)}{q(\x)}\right]$ &  $q_\text{opt}(\x) \propto |f(\x)|\post(\x)$  & or negative \\
		       && & \\ 
\multirow{2}{*}{\checkmark}		& 
		\multirow{2}{*}{$I= \E_{q_1}\left[\frac{f_+(\x)\post(\x)}{q_1(\x)}\right] - \E_{q_2}\left[\frac{f_-(\x)\post(\x)}{q_2(\x)}\right]$}
		 & $q_{1,\text{opt}}(\x) \propto f_+(\x)\post(\x)$ & \multirow{2}{*}{\checkmark} \\
		 && $q_{2,\text{opt}}(\x) \propto f_-(\x)\post(\x)$  & \\ 
		 	 &&  &   \\ 
		 \hline
		 \hline
				 && & \\ 
\ding{55}	&	 $ 
 I =  \frac{\E_{q}\left[\frac{f(\x)\pi(\x)}{q(\x)}\right]}{\E_{q}\left[\frac{\pi(\x)}{q(\x)}\right]}
 $ &  $q_\text{opt}(\x) \propto |f(\x) - I|\post(\x)$ & \ding{55}\\	
				 && & \\ 
\multirow{2}{*}{\ding{55}}				 &
				\multirow{2}{*}{ $ I =  \frac{\E_{q_1}\left[\frac{f(\x)\pi(\x)}{q_1(\x)}\right]}{\E_{q_2}\left[\frac{\pi(\x)}{q_2(\x)}\right]}
 $		}		 
				 & $q_\text{1,opt}(\x) \propto |f(\x)|\post(\x)$ & only if $f$ is positive \\ 
	&	  &  $q_\text{2,opt}(\x) \propto \post(\x)$ & or negative \\		 
		 &&  &\\ 
\multirow{3}{*}{\ding{55}}		  &
		\multirow{3}{*}{    $ I =  \frac{\E_{q_1}\left[\frac{f_+(\x)\pi(\x)}{q_1(\x)}\right] - \E_{q_2}\left[\frac{f_-(\x)\pi(\x)}{q_2(\x)}\right]}{\E_{q_3}\left[\frac{\pi(\x)}{q_3(\x)}\right]}$ }		  
		  & $q_{1,\text{opt}}(\x) \propto f_+(\x)\post(\x)$ &\\ 
	&& $q_{2,\text{opt}}(\x) \propto f_-(\x)\post(\x)$  & \checkmark	  \\
  && $q_\text{3,opt}(\x) \propto \post(\x)$ &\\ 
   &&  & \\ 
			\hline
		\end{tabular}
	\end{center}
	\end{table}

\subsection{Multiple importance sampling (MIS): optimal weights and sampling scheme}\label{sec_MIS}

%{no se si esta sect pega mucho en este paper... aqui solo se habla de como muestrear y pesar.... quiza seria mejor hacer una sect pequen\~a donde hablemos de ``other optimality criteria in IS''....}

So far, we have found optimal proposal densities in order to minimize the MSE in approximation of an integral (or several integrals). Here, we show the optimal form of the importance weights and optimal sampling scheme to reduce the variance of the final IS estimators when multiple proposal pdfs are employed. Below, we describe {\it the optimal sampling and weighting} in this scenario. 
%We consider to use of $R$ pro
\newline
\newline
{\bf Optimal sampling.} Regarding the sampling, the best strategy is to employ the {\it deterministic mixture approach} \cite{CORNUET12,EfficientMIS,HereticalMIS,ElviraMIS15}.
This scheme can be used each time the number of samples $N$ is a multiple of the number $K$ of proposal densities, $N=KR$ where $R$ is an integer. 
Indeed, let us consider the joint use of $K$ different proposal pdfs $q_K(\x)$,  and  we can draw one sample from each one, i.e.,
$$
\x_{k,r} \sim q_K(\x), \qquad \mbox{for  $r=1,...,R$ and $k=1,...,K$}.
$$
Collecting all these samples $\{\x_{k,r}\}$ in the same ``urn''  and using  the samples $\{\x_{k,r}\}$  all together indiscriminately, they are distributed according to the mixture of $q_K$'s with equal weights. Clearly, we have avoided the random selection of the components so that this strategy has less variance with respect to the standard one. The deterministic mixture approach can be also applied when the weights  of the mixture are not equal {\it but} are rational numbers (i.e., they can still expressed as fractions). In that case, the number of samples from each proposal density should be different (according to the weigths).
\newline
\newline
{\bf Proper weighting schemes.}  In a multiple proposal scenario, different {\it proper} importance weights can be employed \cite{ElviraMIS15,HereticalMIS,EfficientMIS}, i.e.,
\begin{eqnarray}
w_{k,r}=\frac{\pi(\x_{k,r})}{\psi(\x_{k,r})},
\end{eqnarray}
which differ for the possible denominator $\psi(\x_{k,r})$. The easiest and cheaper possibility (but the worst in terms of performance) is the classical choice $\psi(\x)=q_K(\x)$. For other possible choices see \cite{ElviraMIS15,HereticalMIS,EfficientMIS}.
\newline
\newline
{\bf Optimal weighting.}  Considering the sampling scheme above (with the same number of samples $K$ per proposal), it is possible to show that the best choice for the denominator \cite{ElviraMIS15,he2014optimal,Veach95}, in terms of minimum variance of the resulting estimator, is 
\begin{eqnarray}
\psi_{\text{opt}}(\x_{m,r})=\frac{1}{K} \sum_{K=1}^{K} q_k(\x_{k,r}),
\end{eqnarray}
which is usually called {\it full-deterministic mixture} denominator ({\it f}-DM).
Hence, the optimal MIS estimators employ the weights 
\begin{eqnarray}
w_{k,r}^{\texttt{(opt)}}=\frac{\pi(\x_{k,r})}{\frac{1}{K} \sum_{k=1}^{K}  q_k(\x_{k,r})}.
\end{eqnarray}
It can be shown that $\text{Var}[\widehat{I}_\text{ful-DM}] \leq \text{Var}[\widehat{I}_\text{MIS}]$ (where $\widehat{I}_\text{ful-DM}$ uses the optimal weights $w_{k,r}^{\texttt{(opt)}}$ above) for any $\widehat{I}_\text{MIS}$ that is built using other valid sampling and weighting strategy \cite{ElviraMIS15}. 

{
%%%%%%%%%%%%%%%%%%%%%%%%%%%%%%%%%%%%
\subsection{{Combinations of standard IS estimators with different proposals}}
%%%%%%%%%%%%%%%%%%%%%%%%%%%%%%%%%%%%
In this section, for simplicity, let consider $Z$ known.
We could combine different standard IS estimators with a combination, i.e.,
\begin{align}\label{LinearComb}
\widehat{I} &= 
\sum_{k=1}^K {\bar \rho}_k \underbrace{\left(\frac{1}{N_k} \sum_{i=1}^{N_k}
\frac{f({\bm \theta}_{k,i}) \bar{\pi}({\bm \theta}_{k,i})}
     {q_k({\bm \theta}_{k,i})} \right)}_{\mbox{stand-IS}}, \qquad {\bm \theta}_{k,i} \sim q_k({\bm \theta}). \\
    &=  \sum_{k=1}^K {\bar \rho}_k \widehat{I}_\text{IS}^{(k)}.
\end{align}
where $N_k$ are the number of samples from the $k$-th proposal $q_k({\bm \theta})$. Since we are  considering $Z$ known, we can use standard IS estimators that are unbiased, i.e., $E[\widehat{I}_\text{IS}^{(k)}]=I$ for all $k$. Thus, we could force that $\widehat{I}$ be also unbiased. Indeed, since
 $$
 E[\widehat{I}]=\sum_{k=1}^K {\bar \rho}_k E[\widehat{I}_\text{IS}^{(k)}]=  \left(\sum_{k=1}^K {\bar \rho}_k\right) \cdot I,
 $$
  we can require the condition
\begin{align}\label{CondUnbias}
\sum_{k=1}^K {\bar \rho}_k=1, \quad \mbox{(note that ${\bar \rho}_k$ can be also negative, ${\bar \rho}_k \lessgtr 0$).}
\end{align} 
 Hence, we seek the coefficients ${\bm {\bar \rho}}=[{\bar \rho}_1,...,{\bar \rho}_k]^{\top}$ such that minimize the MSE that, requiring the condition \eqref{CondUnbias}, becomes the variance of $\widehat{I}$.   Assuming that the estimators are independent, $\widehat{I}_\text{IS}^{(k)}$, {(that is clearly true, if the samples are generated independently from the different proposals)}, then we have 
\begin{align}
{\bar \rho}_k \propto 1/ \mbox{var}\left[\widehat{I}_\text{IS}^{(k)}\right],
\end{align}
that is an inverse-variance weighting. Note that, for the condition \eqref{CondUnbias}, we have ${\bar \rho}_k=\frac{1/ \mbox{var}\left[\widehat{I}_\text{IS}^{(k)}\right]}{\sum_{j=1}^K 1/ \mbox{var}\left[\widehat{I}_\text{IS}^{(j)}\right]}$ and, in this case, ${\bar \rho}_k>0$ for all $k=0$.
If the samples, drawn from the different proposals, would be generated with correlation the solution should take into account this covariance matrix considered in the sampling scheme and some ${\bar \rho}_k$ could be negative \cite{aitken1936least,rao1973linear}.

}

{
%%%%%%%%%%%%%%%%%%%%%%%%%%%%%%%%%%%%
\subsection{{A similar estimator in computer graphics}}
%%%%%%%%%%%%%%%%%%%%%%%%%%%%%%%%%%%%

Computer graphics - and in particular global illumination and physically based rendering - is one of the main application domains of MIS. In this context, MIS allows different sampling techniques to be combined through properly designed weights, producing significantly lower variance renderings.
In computer graphics, the integrand function $f({\bm \theta})\post({\bm \theta})$ may represent the total contribution of photons reaching a point in an image. If there are multiple light sources and reflective objects in the scene, then mixture sampling is suitable with components $q_j$ representing different light paths reaching the image. Considering for simplicity $Z$ known, the {\it heuristic mixture estimator} \cite{he2014optimal,Veach95} is given by:
\begin{align}\label{CompVisionApp}
\widehat{I} = 
\sum_{k=1}^K \left(\frac{1}{N_k} \sum_{i=1}^{N_k}
 \bar{\rho}_k({\bm \theta}_{k,i})\frac{f({\bm \theta}_{k,i}) \bar{\pi}({\bm \theta}_{k,i})}
     {q_k({\bm \theta}_{k,i})} \right),
\qquad {\bm \theta}_{k,i} \sim q_k(\x), 
\end{align}
where 
$$
\sum_{k=1}^K \bar{\rho}_k({\bm \theta})=1,  \quad  \mbox{ and \quad $\bar{\rho}_k({\bm \theta})> 0$, \quad if \quad $f({\bm \theta}) \bar{\pi}({\bm \theta}) \neq 0$,}
$$
otherwise, $\bar{\rho}_k(\x)=0$ for all $k$ if $f(\x) \bar{\pi}(\x)= 0$.
{\rem Note that the estimator in Eq. \eqref{CompVisionApp} resembles the estimator in \eqref{LinearComb}, but they are quite different: here, the coefficients depend on the samples ${\bm \theta}_{k,i}$.}
\newline
\newline
 However, similarly as in \eqref{CompVisionApp}, here  we have another degree of freedom, due to the choices of the (normalized) {\it heuristic} weights $\bar{\rho}_k(\x)$ (following a nomenclature used in computer graphics). Note that the estimator $\widehat{I} $ is still consistent as all $N_k \rightarrow \infty$ for all $k$, and unbiased, indeed (using the independence of the samples),
\begin{align}
\mathbb{E}_{q_{1:K}}[\widehat{I}]&= 
\sum_{k=1}^K \frac{1}{N_k} \sum_{i=1}^{N_k} \mathbb{E}_{q_{k}}\left[
 \bar{\rho}_k({\bm \theta})\frac{f({\bm \theta}) \bar{\pi}({\bm \theta})}
     {q_k({\bm \theta})}\right], \\
      &=\sum_{k=1}^K \frac{1}{N_k} N_k \int_{\Theta}  \bar{\rho}_k({\bm \theta})\frac{f({\bm \theta}) \bar{\pi}({\bm \theta})}{q_k({\bm \theta})} q_k({\bm \theta}) d{\bm \theta}, \\
      &=\sum_{k=1}^K \int_{\Theta}  \bar{\rho}_k({\bm \theta})f({\bm \theta}) \bar{\pi}({\bm \theta}) d{\bm \theta}, \\  
      &= \int_{\Theta} \left[\sum_{k=1}^K  \bar{\rho}_k({\bm \theta})\right]f({\bm \theta}) \bar{\pi}({\bm \theta}) d{\bm \theta},  \\
      &= \int_{\Theta} f({\bm \theta}) \bar{\pi}({\bm \theta}) d{\bm \theta} = I.    
\end{align}
Moreover, with similar arguments, the variance is
\begin{align}
\mbox{var}_{q_{1:K}}[\widehat{I}] &=
\sum_{k=1}^K \frac{1}{N_k^2} N_k \left[\mathbb{E}_{q_{k}}\left[\left(
 \bar{\rho}_k({\bm \theta})\frac{f({\bm \theta}) \bar{\pi}({\bm \theta})}{q_k({\bm \theta})}
     \right)^2\right]-\mathbb{E}_{q_{k}}\left[
 \bar{\rho}_k({\bm \theta})\frac{f({\bm \theta}) \bar{\pi}({\bm \theta})}{q_k({\bm \theta})}\right]^2 \right], \nonumber \\
      &=\sum_{k=1}^K \frac{1}{N_k} \left[\int_{\Theta}  \frac{\left(\bar{\rho}_k({\bm \theta})f({\bm \theta}) \bar{\pi}({\bm \theta})\right)^2}{q_k({\bm \theta})}d{\bm \theta}-
      \left( \int_{\Theta}
 \bar{\rho}_k({\bm \theta})f({\bm \theta}) \bar{\pi}({\bm \theta}) d{\bm \theta}\right)^2\right]    \nonumber 
       \\
      &=\sum_{k=1}^K \frac{1}{N_k} \left[ \int_{\Theta}  \frac{\left(\bar{\rho}_k({\bm \theta})f({\bm \theta}) \bar{\pi}({\bm \theta})\right)^2}{q_k({\bm \theta})}d{\bm \theta}-
        \mu_k^2 \right],   \\
      &=\sum_{k=1}^K \frac{V_k}{N_k},     \label{VarDecEQ}              
\end{align}
where
\begin{align} \label{VarDecEQ2}
V_k&=  \int_{\Theta}  \frac{\bar{\rho}_k({\bm \theta})^2f({\bm \theta})^2 \bar{\pi}({\bm \theta})^2}{q_k({\bm \theta})}d{\bm \theta}- \mu_k^2,  \qquad \mu_k= \int_{\Theta}
 \bar{\rho}_k({\bm \theta})f({\bm \theta}) \bar{\pi}({\bm \theta}) d{\bm \theta}.               
\end{align}
Note that the independence assumption across sample blocks is crucial. If samples
        between proposals are correlated the cross-covariance terms must be included.
Below we provide some examples of  possible choices of $\bar{\rho}_k(\x)$:
\begin{itemize}
\item The simplest choice is $\bar{\rho}_k(\x)=\frac{1}{K}$, i.e., we have an average of $K$ standard IS estimators using different proposals $q_k$. Thus, in this scenario, this approach becomes more similar to Eq. \eqref{LinearComb}.
\item The expression above recall also as a stratified sampling scheme. If we define $\bar{\rho}_k(\x)=1$ for $x \in \mathcal{D}_k \subset \mathcal{X}$ and 0 otherwise, then equation above becomes independent importance sampling within strata. If we further take $q_k$ to be $f(\x)\post(\x)$ restricted to $\mathcal{D}_k$ then the estimator above becomes stratified sampling. Thus, the heuristic mixture estimator   can be seem as a generalization of stratified sampling in which strata are allowed to overlap.
\item Other possible choices have been considered in computer graphics: for instance,  the so called {\it power heuristic},  $\bar{\rho}_k(\x) \propto\left(n_k q_k(\x)\right)^\beta$ for $\beta \geqslant 1$ and the {\it cutoff heuristic}, 
$$
\bar{\rho}_k(\x) \propto \left\{ 1, \mbox{ if }  n_k q_k(\x) \geqslant \alpha \max n_{j} q_{j}(\x), \mbox{ or } 0 \mbox{ otherwise} \right\},
$$
 for $\alpha>0$. In all cases, the coefficient $\bar{\rho}_k(\x)$ must be normalized,  summing 1. The idea is to add extra weight on the component with the largest $n_j q_j(\x)$ values that are more locally (around $\x$) important. Finally, the {\it maximum heuristic} is the cutoff heuristic with $\alpha=1$, or the power heuristic with $\beta \rightarrow \infty$. It puts all weight on the component with the largest value of $n_k q_k(\x)$ or shares that weight equally when multiple components are tied for the maximum, in specific $\x$ (the component  with the largest value of $n_k q_k(\x)$ can changes with $\x$).
\end{itemize}
%%%%%%%%%%%%%%%%%%%%%%%%%%%%%%%%%%%%%%%%%%%
\subsubsection{{Nearly optimal choice of the heuristic weights}}\label{NOH_MIS} 
%%%%%%%%%%%%%%%%%%%%%%%%%%%%%%%%%%%%%%%%%
  Among the possible choices for $\rho_k({\bm \theta})$, the most studied  is the so-called {\it balance heuristic} where $\rho_k(\x) \propto N_k q_k(\x)$, i.e., that power heuristic with $\beta=1$; that is
\begin{align}
\rho_k(\x)= \frac{N_k q_k(\x)}{\sum_{j=1}^K N_j q_j(\x)}.
\end{align}
It is possible to show that the balance heuristic is ``nearly optimal'' in terms of variance. For the definition of nearly-optimal estimator and the corresponding proof,  see Theorem 1 in \cite{Veach95}.
\newline
Now, denoting $N=\sum_{j=1}^K N_j$, defining $\alpha_k=N_k / N$, and considering the balance heuristic weights, we can rewrite the estimator in Eq. \eqref{CompVisionApp}  as:
\begin{align}
\widehat{I} &= 
\sum_{k=1}^K \frac{1}{N_k} \sum_{i=1}^{N_k}
\frac{N_k q_k({\bm \theta}_{k,i})}{\sum_{j=1}^K N_j q_j({\bm \theta}_{k,i})}  \frac{f({\bm \theta}_{k,i}) \bar{\pi}({\bm \theta}_{k,i})}
     {q_k({\bm \theta}_{k,i})}, \nonumber \\
    &= \sum_{k=1}^K \frac{1}{N_k} \sum_{i=1}^{N_k}
\frac{N_k f({\bm \theta}_{k,i}) \bar{\pi}({\bm \theta}_{k,i})}{\sum_{j=1}^K N_j q_j({\bm \theta}_{k,i})},  \nonumber \\ 
&= \sum_{k=1}^K \sum_{i=1}^{N_k}
\frac{f({\bm \theta}_{k,i}) \bar{\pi}({\bm \theta}_{k,i})}{\sum_{j=1}^K N_j q_j({\bm \theta}_{k,i})}, \label{FormUtilforVar} \\ 
&=\frac{1}{N} \sum_{k=1}^K \sum_{i=1}^{N_k}
\frac{f({\bm \theta}_{k,i}) \bar{\pi}({\bm \theta}_{k,i})}{\sum_{j=1}^K \alpha_j q_j({\bm \theta}_{k,i})}, \qquad {\bm \theta}_{k,i} \sim q_k(\x),  \\
&=\frac{1}{N} \sum_{k=1}^K \sum_{i=1}^{N_k}
\frac{f({\bm \theta}_{k,i}) \bar{\pi}({\bm \theta}_{k,i})}{ q_{\texttt{MIS}}({\bm \theta}_{k,i})}, \qquad {\bm \theta}_{k,i} \sim q_{\texttt{MIS}}(\x),  \label{FinalMixDensity}
\end{align}
where we set $q_{\texttt{MIS}}(\x)=\sum_{j=1}^K \alpha_j q_j({\bm \theta})$ and we draw from this mixture {\it deterministically}, generating $N_k$ samples from the $k$-th component.

{\rem Hence, with the  balance heuristic weights, the heuristic mixture estimator reduces to the MIS estimator we would use a deterministic mixture strategy, drawing different samples per proposal density, $N_j$ and, as a consequence, different  mixture weights equal to $\alpha_j=N_j / N$. Once again, the weight associated with a given sampled value  $\x_{i j}$ does not depend {\it solely} on the mixture component from which it was drawn.}

%%%%%%%%%%%%%%%%%%%%%%%%%%%%%%%%%%%%%%%%
\subsubsection{{Optimal sample allocation in MIS schemes and summary}}
%%%%%%%%%%%%%%%%%%%%%%%%%%%%%%%%%%%%%%%%
Let us recall  the decomposition of the variance in Eq. \eqref{VarDecEQ},
\begin{align}
\mbox{var}_{q_{1:K}}[\widehat{I}] =\sum_{k=1}^K \frac{V_k}{N_k}.                  
\end{align}
Minimizing the total variance, under the constraint $\sum_{k=1}^K N_k = N$,
leads to the {\it Cochran-Neyman optimal allocation} \cite{cochran1977,neyman1934}:
\begin{align}
N_k^\star &\propto \sqrt{V_k},  \label{eq:Cochran0}\\
N_k^\star &= N \frac{\sqrt{V_k}}{\sum_{j=1}^K \sqrt{V_j}}, 
\label{eq:Cochran}
\end{align}
where $V_k$ is given in Eq. \eqref{VarDecEQ2}.
Namely, the optimal number of samples drawn from proposal $q_k$ is proportional
to the \emph{standard deviation} $\sqrt{V_k}$ of its contribution in the MIS
estimator. In other words, more samples must be allocated to proposals that induce larger variance in their MIS contribution, but only in proportion to the square root of $V_k$. This is a well-known result in sample allocation for stratified random sampling \cite{cochran1977,neyman1934}.
\newline
\newline
{\bf Summary.}  Let us define a total number of samples $N$ (total budget), and choose $K$ different proposal densities $q_k(\x)$. Considering jointly all the results in Sections \ref{sec_MIS}-\ref{NOH_MIS} and in this section, an optimal sampling strategy with multiple proposals is:
\begin{enumerate}
\item  Set $N_k^\star \propto \sqrt{V_k}$, with $k=1,...,K$, as in Eq \eqref{eq:Cochran}. 
\item Draw $N_k^\star$ samples from each $q_k(\x)$, in a deterministic mixture fashion.
\item Consider in the denominator of the IS weights, the complete mixture
$$
q_{\texttt{MIS}}(\x)=\sum_{j=1}^K \alpha_j q_j({\bm \theta}),
$$
 with different coefficients $\alpha_k=N_k/N$ for each component, as in Eq. \eqref{FinalMixDensity}.
\end{enumerate}
%%%%%%%%%%%%%%%%%%%%%%%%%%%%%%%%%%%%%%%%%%
%%%%%%%%%%%%%%%%%%%%%%%%%%%%%%%%%%%%%%%%%%

%---------------------------------------------------------

}

%%%%%%%%%%%%%%%%%%%%%%%%%%%%%%%%%%%%
\section{Optimal proposal with noisy evaluations of the target density}\label{NoisyIS_schemes}
%%%%%%%%%%%%%%%%%%%%%%%%%%%%%%%%%%%%

In many applications, the direct pointwise evaluation of  $\pi(\x)$ is not possible \cite{LLORENTEnoisyIS,LlorenteABC_RF}. In this section, we deal with noisy evaluations $\widetilde{\pi}(\x)$ that is related to $\pi(\x)$, instead of direct evaluations of $\pi(\x)$. More specifically,  $\widetilde{\pi}(\x)$ is a random variable and we have access to realizations of this random variable. Furthermore,  let denote with
$$
m(\x) = \E[\widetilde{\pi}(\x)|\x], \quad \text{and} \quad s(\x)^2 = \text{Var}[\widetilde{\pi}(\x)|\x],
$$
 the expectation and variance of $\widetilde{\pi}(\x)$ respectively, given a fixed value of $\x$.  Then, we can consider the following noisy IS estimators
\begin{align}
	\widetilde{Z}=\frac{1}{N}\sum_{n=1}^N \frac{\widetilde{\pi}(\x)}{q(\x)}, 
\end{align}
and
\begin{align}
	%\label{eq_std_IS}
	\widetilde{{\bf I}}_\text{IS}=\frac{1}{N \bar{Z}} \sum_{n=1}^N \frac{\widetilde{\pi}(\x)}{q(\x)} {\bf f}(\x_n),  \qquad 
	\widetilde{{\bf I}}_\text{SNIS}=\frac{1}{N \widetilde{Z}} \sum_{n=1}^N \frac{\widetilde{\pi}(\x)}{q(\x)} {\bf f}(\x_n). \label{AquiSelfNormISnoisy}
\end{align}
where $\bar{Z} =\int_{\Theta}  m(\x) d\x$. The above estimators converge, respectively, to \cite{LLORENTEnoisyIS,tran2013importance,fearnhead2010random}
\begin{equation}\label{Goal_in_Integral}
	\bar{Z}=\int_{\Theta}  m(\x) d\x, \quad \bar{{\bf I}}=\frac{1}{\bar{Z}}\int_{\Theta} {\bf f}(\x) m(\x) d\x.
\end{equation}
{\bf Unbiased scenario.} In the unbiased case, we would have $\E[\widetilde{\pi}(\x)|\x]=m(\x)=\pi(\x)$, we have $\bar{Z}=Z$ in Eq. \eqref{eq_Z_of_int} and $\bar{{\bf I}} = {\bf I}$ in Eq. \eqref{eq_I_of_interest}.
\newline
As in the non-noisy framework, the estimator $\widetilde{{\bf I}}_\text{IS}$ requires the knowledge of $\bar{Z}$, that is not needed in the so-called self-normalized estimator, $\widetilde{{\bf I}}_\text{SNIS}$.
In the following, we show the optimal proposals for $\widetilde{Z}$, $\widetilde{{\bf I}}_\text{IS}$ and $\widetilde{{\bf I}}_\text{SNIS}$.
%Below, we describe the optimality of the proposal for the estimators above.

%%%%%%%%%%%%%%%%%%%%%%%%%%%%%
\subsection{Optimal proposal pdf for estimating $\bar{Z}$}
%%%%%%%%%%%%%%%%%%%%%%%%%%%%%
The variance of $\widetilde{Z}$ (w.r.t. the samples and the noisy realizations) is given by \cite{LLORENTEnoisyIS},
\begin{align}
	\mbox{Var}[\widetilde{Z}] = \frac{1}{N}\mathbb{E}\left[\frac{m(\x)^2+s(\x)^2}{q(\x)^2}\right] - \frac{1}{N}\bar{Z}^2.
\end{align}
The minimum variance, denoted as $\mbox{V}_\text{opt}$, is attained at
\begin{align}
	\fbox{$q_\text{opt}(\x)= \frac{1}{\widetilde{C}_q}\sqrt{m(\x)^2+s(\x)^2}\propto \sqrt{m(\x)^2+s(\x)^2}$},
\end{align}
where $\widetilde{C}_q=\int_\Theta  \sqrt{m(\x)^2+s(\x)^2}d\x$.
Note that $\mbox{V}_\text{opt}=\min_q\mbox{Var}[\widetilde{Z}]$ is always greater than 0, specifically,
\begin{align}
\mbox{V}_\text{opt} &= \frac{1}{N}\mathbb{E}\left[\widetilde{C}_q^2\right] - \frac{1}{N}\bar{Z}^2, \nonumber\\
&=\frac{1}{N}\widetilde{C}_q^2 - \frac{1}{N}\bar{Z}^2,\nonumber \\
&=\frac{1}{N}\left[\int_\Theta  \sqrt{m(\x)^2+s(\x)^2}d\x\right]^2 - \frac{1}{N}\bar{Z}^2.
\end{align}
Hence, differently from the non-noisy setting in Section \ref{Sect_optZ}, in the noisy IS scenario the optimal estimator of $Z$ does not reach a null variance, i.e., is not equal to 0, as long as $s(\x)$ is not null everywhere.  Recall that with $s(\x)=0$ we recover the non-noisy scenario. 
%\newline
%\noindent{\bf Illustration of $q_\text{opt}$.} Assume a multiplicative noisy $\widetilde{\pi}(\x) = \epsilon \pi(\x)$ with Var$[\epsilon]=\sigma^2$, hence $s(\x)^2 = \pi(\x)^2\mbox{Var}[\epsilon] = \sigma^2 \pi(\x)^2$. In this case, the optimal proposal coincides with the optimal one in the non-noisy setting, since 
%\begin{align}
%	q_\text{opt} &\propto \sqrt{\sigma^2\pi^2(\x)+\pi^2(\x)} = \pi(\x)\sqrt{1+\sigma^2}\\
%	&\propto \post(\x).
%\end{align}
%As a second example, Let us consider a Bernoulli-type noise where $\widetilde{\pi}(\x) = \epsilon\pi_\text{max}$, and $\epsilon \sim \mbox{Bernoulli}\left(\frac{\pi(\x)}{\pi_\text{max}}\right)$, where $\pi_\text{max} = \max_{\x} \pi(\x)$. Then, $s^2(\x) = \pi(\x)(\pi_\text{max}-\pi(\x))$, and the optimal proposal is
%\begin{align}
%		q_\text{opt}(\x) \propto \pi(\x)\sqrt{1 + (\pi_\text{max}-\pi(\x))^2}.
%\end{align}
%\newline
%%%%%%%%%%%%%%%%%%%%%%%%%%%%%
\subsection{Optimal proposal for standard noisy IS}
%%%%%%%%%%%%%%%%%%%%%%%%%%%%%

%We have already seen that the optimal proposal that minimizes the variance of $\widetilde{Z}$ is $q_\text{opt}(\x) \propto \sqrt{m(\x)^2 + s(\x)^2}$. 
Let us consider now the estimator $\widetilde{{\bf I}}_\text{IS}$. Note that this estimator assumes we can evaluate $\bar{Z}=\int_{\Theta} m(\x)d\x$. 
Since we are considering a vector-valued function, the estimator has $P$ components  $\widetilde{{\bf I}}_\text{IS}=[\widetilde{I}_{\text{IS},1} \dots \widetilde{I}_{\text{IS},P}]^\top$, and $\text{Var}[\widetilde{{\bf I}}_\text{IS}]$ corresponds to a $P \times P$ covariance matrix.
We aim to find the proposal that minimizes the sum of diagonal variances. From the results of the previous section, it is straightforward to show that the variance of the $p$-th component is
\begin{align*}
	\text{Var}[\widetilde{I}_{\text{IS},p}] = \frac{1}{N\bar{Z}^2}\E\left[\frac{f_p(\x)^2(m(\x)^2+s(\x)^2)}{q(\x)^2}\right] - \frac{1}{N\bar{Z}^2}\bar{I}_p^2,
\end{align*}
where $f_p(\x)$ and $\bar{I}_p$ are respectively the $p$-th components of ${\bf f}(\x)$ and $\bar{{\bf I}}$.
Thus,
{\footnotesize$$
	\sum_{p=1}^{P} \text{Var}[\widetilde{I}_{\text{IS},p}] = \frac{1}{N\bar{Z}^2}\E\left[\frac{\sum_{p=1}^{P} f_p(\x)^2(m(\x)^2+s(\x)^2)}{q(\x)^2}\right] - \frac{1}{N\bar{Z}^2}\sum_{p=1}^{P} \bar{I}_p^2.
	$$}
%By Jensen's inequality, we have
%{\footnotesize
%	\begin{align*}
%		\mathbb{E}\left[\frac{\sum_{p=1}^{P} f_p(\x)^2(m(\x)^2+s(\x)^2)}{q(\x)^2}\right] 
%		%	&\geq \left(\mathbb{E}\left[\frac{\sqrt{\sum_{p=1}^\chi f_p(\x)^2(m(\x)^2+s(\x)^2)}}{q(\x)}\right]\right)^2 \\
%		&\geq\left(\mathbb{E}\left[\frac{\sqrt{m(\x)^2+s(\x)^2}\norm{{\bf f}(\x)}_2}{q(\x)}\right]\right)^2,
%	\end{align*}
%}where $\norm{{\bf f}(\x)}_2$ denotes the euclidean norm.
%The equality holds if and only if $\frac{\sqrt{m(\x)^2+s(\x)^2}\norm{{\bf f}(\x)}_2}{q(\x)}$ is constant.
Hence, the optimal proposal is
\begin{align}
	\fbox{$q_\text{opt}(\x)\propto \norm{{\bf f}(\x)}_2\sqrt{m(\x)^2+s(\x)^2}.$}
\end{align}
%\newline
%{\bf Optimal proposal for $\widetilde{{\bf I}}_{SNIS}$.}
%%%%%%%%%%%%%%%%%%%%%%%%%%%%%
\subsection{Optimal proposal for self-normalized noisy IS}
%%%%%%%%%%%%%%%%%%%%%%%%%%%%%
Let us consider the case of the self-normalized estimator $\widetilde{{\bf I}}_\text{SNIS}$.
Recall that $\widetilde{{\bf I}}_\text{SNIS} = \frac{\widetilde{{\bf E}}}{\widetilde{Z}}$,
where $\widetilde{{\bf E}}$ denotes the noisy estimator of ${\bf E} = \int_{\Theta} {\bf f}(\x)m(\x)d\x$, so that we are considering ratios of estimators.
Again, we aim to find the proposal that minimizes the variance of the vector-valued estimator $\widetilde{{\bf I}}_\text{SNIS}$.
When $N$ is large enough, the variance of $p$-th ratio is approximated  as \cite{LLORENTEnoisyIS}
\begin{align*}
	\text{Var}_q[\widetilde{I}_{\text{self},p}] = \text{Var}_q\left[\frac{\widetilde{E}_p}{\widetilde{Z}}\right] 
	\approx \frac{1}{\bar{Z}^2}\text{Var}_q[\widetilde{E}_p] - 2\frac{E_p}{\bar{Z}}\text{Cov}_q[\widetilde{E}_p,\widetilde{Z}] + \frac{E_p^2}{\bar{Z}^4}\text{Var}_q[\widetilde{Z}],
\end{align*}
where $E_p$ is the $p$-th component of ${\bf E}$, and it is possible to show that 
\begin{align*}
	\text{Var}_q[\widetilde{E}_p] &= \frac{1}{N}\E\left[\frac{f_p(\x)^2(m(\x)^2+s(\x)^2)}{q(\x)^2}\right] - \frac{1}{N}E_p^2, \\
	\text{Var}_q[\widetilde{Z}] &= \frac{1}{N}\E_q\left[\frac{m(\x)^2+s(\x)^2}{q(\x)^2}\right] - \frac{1}{N}\bar{Z}^2, \\
	\text{Cov}_q[\widetilde{E}_p,\widetilde{Z}] &= \frac{1}{N}\E_q\left[\frac{f_p(\x)(m(\x)^2+s(\x)^2)}{q(\x)^2}\right] - \frac{1}{N}E_p\bar{Z}.
\end{align*}
%The first two results have been already obtained in the previous sections. The third result is given in Appendix \ref{App1}.
The sum of the variances is thus
\begin{align*}
	\sum_{p=1}^{P}\text{Var}_q[\widetilde{I}_{\text{self},p}] \approx \frac{1}{N\bar{Z}^2}\E_q\left[\frac{(m(\x)^2+s(\x)^2)\sum_{p=1}^{P}(f_p(\x) - \bar{I}_p)^2}{q(\x)^2}\right],
\end{align*}
and, by Jensen's inequality, we obtain that the optimal proposal density is 
\begin{align}
	\fbox{$q_\text{opt}(\x) \propto \norm{{\bf f}(\x) - \bar{{\bf I}}}_2\sqrt{m(\x)^2 + s(\x)^2}.$}	
\end{align}
%%%%%%%%%%%%%%%%%%%%%%%%%%%%%%%%%%%%%%%%%%%%%%
{
\section{{Optimality in energy-based models (EBMs)}}\label{EBMSect}
%%%%%%%%%%%%%%%%%%%%%%%%%%%%%%%%%%%%%%%%%%%%%

%{ result above useful for all the noisy MC schemes that require to generate artificial data (inside...)...... IS-squared Nested IS and others...
%}
%%%%%%%%%%%%%%%%%%%%%
\subsection{{Non-normalized models}}
%%%%%%%%%%%%%%%%%%%%%
Let us introduce the framework of the energy-based models and/or doubly intractable posteriors \cite{dawid2024introduction,Geyer1994Convergence,LlorenteABC_RF}. An energy-based model is represented by a parametrized family of density functions $\psi({\bf y}|{\bm \theta},Z) $, defined for each ${\bm \theta}$ as 
\begin{equation}
\psi({\bf y}|{\bm \theta},Z) =\frac{\phi({\bf y}|{\bm \theta})}{Z({\bm \theta})}, % \quad \mbox{with} \quad \phi({\bf y}| {\bm \theta})=e^{-E({\bf y}|{\bm \theta})}, \quad {\bf y} \in \mathbb{R}^d
\label{px|theta-Z}
\end{equation}
where we can evaluate $ \phi({\bf y}|{\bm \theta})=e^{-E({\bf y}|{\bm \theta})}\geq0$ 
but generally we cannot evaluate the integral:
 %\footnote{All the integrals in this work are definite integrals. However, in the rest of the paper, for simplicity we avoid to write the integration domain.}
\begin{equation}
Z({\bm \theta})=\int_{\mathcal{Y}} \phi({\bf y}|{\bm \theta}) d {\bf y}\label{Z(theta)}.
\end{equation}
Namely, $Z({\bm \theta})$ is unknown because the integral above cannot be solved analytically in closed form, i.e., is intractable.\footnote{We assume that ${\bf y}$ be a continuous vector, although several considerations are also valid  for the discrete case.} Hence, the normalizing constant $Z({\bm \theta})$, often called {\em partition function}, cannot be evaluated point-wise. This represents a challenge for making inference on ${\bm \theta}$.   Let us assume that we have an observed dataset  
${\bf y}_{1:N}=\{{\bf y}_1, \ldots, {\bf y}_N \}$,
 that contains i.i.d. realizations distributed as the  the EBM in Eq. \eqref{px|theta-Z} for a specific unknown vector of parameters ${\bm \theta}_{\texttt{tr}}$ (true vector of parameters), i.e., 
\begin{align}
{\bf y}_1,....,{\bf y}_N \sim \psi({\bf y}|{\bm \theta}_{\texttt{tr}},Z)=\frac{\phi({\bf y}|{\bm \theta}_{\texttt{tr}})}{Z({\bm \theta}_{\texttt{tr}})}.
\end{align}
%We focus on a frequentist approach.  However, all the algorithms in this work can be employed for a pre-Bayesian analysis.
Focusing in a frequentist approach, the most straightforward approach to try to perform inference in EBMs is through maximum likelihood estimation.
  In order to estimate the parameter of the distribution,  the likelihood function of ${\bm \theta}$ given ${\bf y}_{1:N}$  is given by
\begin{equation}\label{likeEq}
L({\bf y}_{1:N}|{\bm \theta},Z)= p({\bf y}_1,{\bf y}_2,\ldots, {\bf y}_N|{\bm \theta}) =\prod_{n=1}^N \psi({\bf y}_n|{\bm \theta},Z) 
=\frac{1}{Z({\bm \theta})^N} \prod_{n=1}^N \phi({\bf y}_n|{\bm \theta}),
\end{equation}
and then corresponding  the log-likelihood is
\begin{align}
\log L({\bf y}_{1:N}|{\bm \theta},Z)&=  \sum_{n=1}^N \log \psi({\bf y}_n|{\bm \theta},Z)  =- \sum_{n=1}^N E({\bf y}_n|{\bm \theta}) -  N\log Z({\bm \theta}),
\label{loglik_theta_logphi_gen}
\end{align}
where we have used $E({\bf y}_n|{\bm \theta})=-\log\phi({\bf y}_n|{\bm \theta})$,
so that $\widehat{{\bm \theta}}_{\texttt{ML}} =\arg  \max L({\bf y}_{1:N}|{\bm \theta},Z)$. 
However,  $Z({\bm \theta})$ is unknown for any ${\bm \theta}$, and  cannot be computed in closed form.  We can  replace $Z({\bm \theta})$ with an IS approximation \cite{Geyer1994Convergence}. For instance, we can draw $M$ auxiliary data, $\{ {\bf x}_1, \ldots, {\bf x}_M\}$,  from an proposal $q({\bf x})$ independent from ${\bm \theta}$, i.e.,
\begin{align}\label{EqZest}
\widehat{Z}(\boldsymbol{\theta})
=\frac{1}{M}\sum_{m=1}^M 
\frac{\phi({\bf x}_m | \boldsymbol{\theta})}{q({\bf x}_m)}, 
\qquad {\bf x}_m \sim q({\bf x}),
\end{align}
which is unbiased and consistent for every $\boldsymbol{\theta}$, even uses the same set  of samples ${\bf x}_1,\ldots,{\bf x}_M$ for each $\boldsymbol{\theta}$. A potentially more efficient estimator is
\begin{align}\label{EqZest_dep}
\widehat{Z}_{\theta}(\boldsymbol{\theta})
=\frac{1}{M}\sum_{m=1}^M 
\frac{\phi({\bf x}_m^{(\boldsymbol{\theta})} | \boldsymbol{\theta})}
     {q({\bf x}_m^{(\boldsymbol{\theta})} | \boldsymbol{\theta})}, 
\qquad {\bf x}_m^{(\boldsymbol{\theta})} \sim q({\bf x} | \boldsymbol{\theta}),
\end{align}
where the proposal depends on $\boldsymbol{\theta}$, and a different set 
$\{{\bf x}_1^{(\boldsymbol{\theta})},\ldots,{\bf x}_M^{(\boldsymbol{\theta})}\}$ 
is generated for each value of $\boldsymbol{\theta}$. 
In this scenario, we know that the optimal proposal is $q({\bf x}| \boldsymbol{\theta})
\propto \phi({\bf x} | \boldsymbol{\theta})$ as shown in Section \ref{Sect_optZ}. 
Although using a proposal distribution that depends on $\boldsymbol{\theta}$ can yield improved performance, it requires a substantially larger number of evaluations of the numerator $\phi({\bf x}|\boldsymbol{\theta})$ compared to the approach based on a $\boldsymbol{\theta}$-independent proposal density $q({\bf x})$ \cite{Geyer1994Convergence}. Therefore, to reduce computational cost, the $\boldsymbol{\theta}$-independent proposal should be preferred in this setting. We discuss below the corresponding optimality.

%%%%%%%%%%%%%%%%%%%%%%%%%%%%%%%%%%%%
\subsection{{Optimal independent proposal density for EBMs}}
%%%%%%%%%%%%%%%%%%%%%%%%%%%%%%%%%%%%
  Our goal is to select an independent proposal density $q_{\text{opt}}({\bf x} )$ that performs adequately for different parameter values (e.g., ${\bm \theta}_1, {\bm \theta}_2,\ldots$), and, if possible, for the entire parameter space ${\bm \theta}$.  For this reason, a possible idea is to consider the following quantity,
\begin{align}
\bar{Z}=\int_{{\bm \theta}} Z({\bm \theta}) d{\bm \theta}=\int_{{\bm \theta}}\left(\int_{\mathcal{Y}} \phi({\bf x} |{\bm \theta}) d {\bf x} \label{Z(theta)}\right) d{\bm \theta}, \quad(\mbox{assuming } \bar{Z}<\infty), 
 \end{align}
 and try to minimize the variance for its estimation. Note that we are assuming that $\bar{Z}$ is a finite value.
Let consider a  finite set ${\bm \theta}_{\texttt{sub}} \subset {\bm \theta}$ with $|{\bm \theta}_{\texttt{sub}}|<\infty$, such that $ \widehat{{\bm \theta}}_{\texttt{ML}} \in {\bm \theta}_{\texttt{sub}}$.  %Hence, we assume that we know a subset of this type, $ \widehat{{\bm \theta}}_{\texttt{ML}} \in {\bm \theta}_{\texttt{sub}}$.
Hence, we can draw uniformly in ${\bm \theta}_{\texttt{sub}}$, i.e., $
{\bm \theta}_i\sim \mathcal{U}({\bm \theta}_{\texttt{sub}}),$ where $i=1,...,Q$. Then, we can write
\begin{align}
\bar{Z}&\approx \frac{1}{Q}\sum_{i=1}^Q \left[\int_{\mathcal{Y}} \phi({\bf x} |{\bm \theta}_i) d {\bf x} \right]= \frac{1}{Q} \sum_{i=1}^Q Z({\bm \theta}_i), 
\end{align}
with $Z({\bm \theta}_i)=\int_{\mathcal{Y}} \phi({\bf x} |{\bm \theta}_i) d {\bf x} $ is a scalar value.
It is possible to show that for minimizing the variance of the finite sum $\sum_{i=1}^Q Z({\bm \theta}_i)$ (estimating each $ Z({\bm \theta}_i)$ by IS), we have to use 
\begin{align}\label{Optimal_ind_propEq}
q_{\text{opt}}({\bf x} ) \propto \sqrt{[\phi({\bf x} |{\bm \theta}_1)]^2+[\phi({\bf x} |{\bm \theta}_2)]^2+...+[\phi({\bf x} |{\bm \theta}_Q)]^2},
\end{align}
as proposal density. For a proof see Section \ref{VariasTargets}.  The optimal density above is relevant from a theoretical point of view, since generally we cannot evaluate point-wise (since we do not know the normalization) it and we cannot draw from it.
However,  these considerations above can drive the construction of an adaptive proposal density. For instance, see \cite{PaperScaffidiMangano} for a practical procedure.

}

%%%%%%%%%%%%%%%%%%%%%%%%%%%%%%%%%%%%%%%%%%%%%%
%{
%\section{{Optimality in tempering strategies}}
%%%%%%%%%%%%%%%%%%%%%%%%%%%%%%%%%%%%%%%%%%%%%
%optimality over $\beta$
%}

%{
%\section{{Optimality in sequential importance sampling}}
%%%%%%%%%%%%%%%%%%%%%%%%%%%%%%%%%%%%%%%%%%%%%
%}

%%%%%%%%%%%%%%%%%%%%%%%%%%%%%%%%%%%%%%%%%%%%%%
\section{Optimality in IS schemes for computing the evidence $Z$}\label{sec_marglike}
%%%%%%%%%%%%%%%%%%%%%%%%%%%%%%%%%%%%%%%%%%%%%
In this section, we focus on the computation of normalizing constants or ratios of normalizing constants. From a practical point of view, these problems appear in the computation of marginal likelihoods,  $Z= \int_\Theta \pi(\x)d\x$, and/or Bayes factors, $Z_1/Z_2$  \cite{gelman1998simulating,llorenteREV_ML,meng2002warp}.
The methods in this section rely on different identities, some of them using multiple proposal pdfs. In some cases, $\post(\x) = \frac{\pi(\x)}{Z}$ is itself employed as a proposal density, from which samples are drawn. 
Clearly, in this scenario, we imply the use of MCMC algorithms or other Monte Carlo schemes from drawing from $\post(\x)$.
\newline
\newline
We recall that  $\widehat{Z}_{\text{IS}}$ in Eq. \eqref{eq_Z_est} is the simplest estimator of $Z= \int_\Theta \pi(\x)d\x$, and its variance is given by
\begin{align}\label{eq_var_Z2}
\text{Var}_q[\widehat{Z}_{\text{IS}}] = \frac{1}{N}\E_q\left[\frac{\pi(\x)^2}{q(\x)^2}\right] - \frac{1}{N}Z^2.
\end{align}
We also recall that optimal proposal pdf in this case is  \fbox{$q_\text{opt}(\x)\propto \pi(\x)$}. Here, we discuss different concepts of optimality of specific IS schemes specifically devoted to the approximation of $Z$, and can improve in some way the performance of $\widehat{Z}$ in Eq. \eqref{eq_Z_est}.

%%%%%%%%%%%%%%%%%%%%%%%%%%%%
\subsection{Reverse Importance Sampling (RIS)}\label{RISsect} 
%%%%%%%%%%%%%%%%%%%%%%%%%%%%%
It is also possible to estimate $Z$ using the so-called reverse importance sampling, also known as {\it reciprocal} IS \cite{gelfand1994bayesian,llorenteREV_ML}.
The RIS scheme can be derived from the identity   
\begin{align}\label{ReverseISidentity}
\frac{1}{Z} =\E_{\post}\left[ \frac{\varphi(\x)}{\pi(\x)} \right] =\int_{\Theta}\frac{\varphi(\x)}{\pi(\x)}\post(\x)d\x= \frac{1}{Z}\int_{\Theta} \varphi(\x) d\x, 
\end{align}
where  we consider an auxiliary normalized density $\varphi(\x)$, i.e., $\int_{\Theta} \varphi(\x) d\x=1$. Then, one could consider the estimator
\begin{align}\label{ReverseIS}
\widehat{Z}_\text{RIS} = \left(\frac{1}{N}\sum_{i=1}^N\frac{\varphi(\x_i)}{\pi(\x_i)}\right)^{-1},
%=\left(\frac{1}{N}\sum_{i=1}^N\frac{\varphi(\x_i)}{\ell(\y|\x_i)g(\x_i)}\right)^{-1}
\quad \x_i \sim \post(\x).
\end{align}
The samples $ \x_i \sim \post(\x)$ can be obtained approximately by an MCMC algorithm, for instance. See also Figure \ref{USfig}.
\newline
\newline
 The expression $\frac{1}{N}\sum_{i=1}^N\frac{\varphi(\x_i)}{\pi(\x_i)}$ is an unbiased estimator of $1/Z$.   The estimator $\widehat{Z}_\text{RIS} $ above is consistent but, however, is a biased estimator of $Z$.
Here,  $\post(\x)$ plays the role of the proposal density from which we need to draw from. Indeed, in this case,  we do not need samples from $\varphi(\x)$, although its choice affects the precision of the approximation \cite{llorenteREV_ML}. 
 Unlike in the standard IS approach, $\varphi(\x)$ must have lighter tails than $\pi(\x)$\cite{llorenteREV_ML}.   See the experiment in Section \ref{IS_vs_RIScomp} for more details.  
Taking into account the inverse estimator $\frac{1}{\widehat{Z}_\text{RIS}}$ that is unbiased with respect to $1/Z$, we can write that the variance of $\frac{1}{\widehat{Z}_\text{RIS}}$ is
\begin{align}\label{eq_var_RIS}
\mbox{Var}\left[\frac{1}{\widehat{Z}_\text{RIS}}\right] = \frac{1}{N}\E_{\post}\left[\frac{\varphi(\x)^2}{\pi(\x)^2}\right] - \frac{1}{NZ^2}.
\end{align}
Then, the optimal choice of the auxiliary density $\varphi(\x)$ is
\begin{align}
\fbox{$\varphi_\text{opt}(\x) = \post(\x).$}
\end{align}
 However,  although $\x_i \sim \post(\x)$, recall that  $\varphi_\text{opt}(\x)$ is not the proposal density but, in this scenario, plays the role of an auxiliary/reference pdf. See also Figure \ref{USfig2}.
\subsection{Ratio Importance Sampling for $Z$ (a.k.a, umbrella sampling)} 
%%%%%%%%%%%%%%%%%%%%%%%%%%%%%%%%%%
%\noindent{{\bf Self-normalized/ratio IS for Z}.} 
%Setting $\widetilde{q}_1(\x)=\pi(\x)$, $q_2(\x)=\widetilde{q}_2(\x) = \varphi(\x)$, $c_2=1$, hence $r=Z$ in Eq. \eqref{eq_US_identity}. Also setting $q_3(\x)=q(\x)$ and $\widetilde{q}_3(\x)=\widetilde{q}(\x)$, the resulting estimator in Eq. \eqref{eq_US_estimator},
Let $\varphi(\x)$ and $q(\x)$ denote two normalized densities, where $\varphi(\x)$ is some normalized auxiliary pdf and $q(\x)$ is the proposal pdf (from which we draw samples from). The following identity expresses $Z$ as the ratio of two expectations, producing the following estimator called ``ratio importance sampling" (and ``umbrella sampling'' in the physics literature) \cite{chen1997monte},
\begin{align}\label{eq_umbrella_para_Z}
Z = \frac{\E_q\left[\frac{\pi(\x)}{q(\x)}\right]}{\E_q\left[\frac{\varphi(\x)}{q(\x)}\right]}\approx\widehat{Z}_\text{ratio} = \frac{\frac{1}{N}\sum_{i=1}^{N}\frac{\pi(\z_i)}{q(\z_i)}}{\frac{1}{N}\sum_{i=1}^{N}\frac{\varphi(\z_i)}{q(\z_i)}},  \qquad {\bf z}_i \sim q(\x).
\end{align}
%where ${\bf z}_i \sim q(\x)$ for $i=1,\dots,N$.

Since $\varphi(\x)$ and $q(\x)$ are normalized, the denominator above is an estimator of the value $1$. Surprisingly, $\widehat{Z}_\text{ratio}$ can be more efficient than $\widehat{Z}_\text{IS}$ in Eq. \eqref{eq_Z_est} \cite{chen1997monte,llorenteREV_ML} (see below).

This general estimator encompasses many known estimators of normalizing constants/marginal likelihoods that use samples from one proposal $q(\x)$ \cite{llorenteREV_ML}. For instance, standard IS and RIS are obtained as special cases by setting $\varphi(\x)=q(\x)$ or $q(\x)=\post(\x)$, respectively. Table \ref{TablaUmbrella} shows different techniques as special case of the estimator $\widehat{Z}_\text{ratio}$.

\begin{table}[!h]	
	 \caption{Famous special cases of the estimator $\widehat{Z}_\text{ratio}$ \cite{llorenteREV_ML}. Recall $g(\x)$ represents a normalized prior density.  }\label{TablaUmbrella}
	 \vspace{-0.2cm}
	\begin{center}
		\begin{tabular}{|c||c|c|} 
		\hline 
			{\bf Methods} & $\varphi(\x)$  & $q(\x)$ \\ 
			 \hline
			 \hline
			 Naive Monte Carlo & $g(\x)$ & $g(\x)$  \\
			 Harmonic Mean  &     $g(\x)$ & $\post(\x)$   \\ 
		% stand IS   & $q(\x)$ & $q(\x)$ (quitar)   \\ 
		 RIS & $\varphi(\x)$ & $\post(\x)$   \\
			\hline
		\end{tabular}
	\end{center}

\end{table}

{\rem The motivation for using the identity \eqref{eq_umbrella_para_Z} is the idea of taking advantage of an intermediate proposal pdf $q(\x)$, that is ``in the middle'' of $\post(\x)$ and $\varphi(\x)$ \cite{chen1997monte,meng1996simulating,gelman1998simulating}. Figure \ref{USfig3} represents the umbrella sampling idea compared to other approaches.
}
\newline
\newline
The optimal choice of the auxiliary density $\varphi(\x)$ is always
\begin{align}\label{optVARphi}
	\fbox{$\varphi_\text{opt}(\x)=\post(\x)$},
\end{align}  
which gives the exact solution $\widehat{Z}_\text{ratio}=Z$, for any choice of $q(\x)$.
Fixing a generic $\varphi(\x)$, the optimal choice of $q(\x)$, that minimizes the asymptotic relative mean-squared error (rel-MSE) of $\widehat{Z}_\text{ratio}$, is \cite{chen1997monte,llorenteREV_ML}
\begin{align}\label{eq:OptProUmbrella}
\fbox{$q_\text{opt}(\x) = \dfrac{|\post(\x) - \varphi(\x)|}{\int_\Theta  |\post(\x') - \varphi(\x')|d\x '} \propto \left|\dfrac{1}{Z}\pi(\x) - \varphi(\x)\right|.$}
\end{align}
With this optimal choice of the (intermediate) proposal pdf $q(\x)$, the relative MSE (rel-MSE) in estimation of $\widehat{Z}_\text{ratio}$  is given by 
\begin{align}
	\mbox{rel-MSE} =\frac{\E\left[(Z - \widehat{Z}_\text{ratio})^2\right]}{Z^2}  &\approx \frac{1}{N} \left[\int_\Theta |\post(\x)-\varphi(\x)|d\x\right]^2, \nonumber \\
	 &\approx \frac{1}{N}L_1^2(\post,\varphi), \quad \mbox{(for $N$  great enough),}
\end{align}
where $L_1(\post,\varphi)$ denotes the $L_1$-distance  between $\post$ and $\varphi$ \cite[Theorem 3.2]{chen1997monte}.  %Note that, in ratio IS, using jointly both $\varphi_\text{opt}(\x)$ in Eq. \eqref{optVARphi} and $q_\text{opt}(\x)$ in Eq. \eqref{eq:OptProUmbrella}, we get
%\begin{equation}
%\fbox{$q_\text{opt}(\x)=\varphi_\text{opt}(\x)= \post(\x),$}
%\end{equation}
%and we obtain a zero variance estimator in this case.

{\rem Since $L_1^2(\cdot,\cdot)\leq D_{\chi^2}(\cdot,\cdot)$ \cite{chen1997monte} and Eq. \eqref{Eqchi0}, the optimal estimator $\widehat{Z}_\text{ratio}$, using $q_\text{opt}(\x)$,  is asymptotically more efficient than a standard IS estimator using $\varphi(\x)$ as proposal, and than a RIS estimator using $\varphi(\x)$ as auxiliary pdf. }
%{ creo que lo he entendido.., se necesitaria una figura...o un ejemplo....}{MIRAR FIG 1 (b)-(c)}  }
\newline
\newline
However, the pdf $q_{\text{opt}}(\x)$ depends on $Z$ (hence we cannot evaluate it), and $q_{\text{opt}}(\x)$  is not easy to draw from.   In order to implement the optimal umbrella estimator in practice, one can employ the following iterative procedure \cite{chen1997monte}:
%First, use a initial approximation $\widehat{r}_0$ to build and sample $q_{3,\text{opt}}(\x)$. Then, use the samples to obtain the final estimator $\widehat{r}_1$.
%The following two-stage procedure is often used in practice: 
\newline
\newline
- Start with an arbitrary density $q^{(1)}(\x) \propto \widetilde{q}^{(1)}(\x)$.
\newline
- For $t=2,...,T:$
\begin{enumerate}
	 \item Draw $N$ samples from $q^{(t-1)}(\x)$ (using an MCMC or other Monte Carlo method), and use them to obtain 
	\begin{align}\label{eq_TwoStageRatioIS_1}
		\widehat{Z}_\text{ratio}^{(t)} = \frac{\sum_{i=1}^{N}\frac{\pi_1(\x_i)}{\widetilde{q}^{(t-1)}(\x_i)}}{\sum_{i=1}^{N}\frac{\varphi(\x_i)}{\widetilde{q}^{(t-1)}(\x_i)}}, \quad \{\x_i \}_{i=1}^{N} \sim q^{(t-1)}(\x),
	\end{align}
	\item Set
	\begin{align}\label{eq_TwoStageRatioIS_2}
		q^{(t)}(\x) \propto \widetilde{q}^{(t)}(\x) = |\pi(\x) - \widehat{Z}_\text{ratio}^{(t)}\varphi(\x)|.
	\end{align}
%	\item  	
%	{\it Stage 2}: Draw $N_2$ samples from $q^{(t)}(\x)$ via, e.g., MCMC and compute
%	\begin{align}\label{eq_TwoStageRatioIS_3}
%		\widehat{Z}_\text{ratio}^{(t+1)} = \frac{\sum_{i=1}^{N_2}\frac{\pi(\x_i)}{\widetilde{q}^{(2)}(\x_i)}}{\sum_{i=1}^{N_2}\frac{\varphi(\x_i)}{\widetilde{q}^{(2)}(\x_i)}}, \quad \{\x_i \}_{i=1}^{N_2} \sim q^{(2)}(\x).
	%\end{align}
\end{enumerate}
%\newline
%\newline
A graphical representation of umbrella sampling is given in Figure \ref{USfig3}. Fixing the reference pdf $\varphi$, the proposal pdf $q$ is a pdf ''in between'' $\varphi$ and $\post$.

%%%%%%%%%%%%%%%%%%%%%%%%%%%%
\subsection{Bridge sampling} \label{BridgeSect}
%%%%%%%%%%%%%%%%%%%%%%%%%%%%%
 In the previous section, devoted to umbrella sampling, we have employed two densities which play the role of the proposal, $q$, and of an auxiliary reference pdf, $\varphi$. We only draw samples from the proposal pdf $q$. From Eq. \eqref{eq:OptProUmbrella}, we can interpreted that the reference pdf and the posterior act as two ``extremes'' (using the analogy of an interval),  and the proposal pdf represents a function  ``in between'' of both extremes.  This is graphically shown in Figure \ref{USfig3}.
\newline
In this section, we describe the bridge sampling technique. Using the same analogy, in this case  the ``extremes'' are the proposal, $q$, and the posterior $\post$. The density  ``in between'' , used as a ``bridge'', is the auxiliary pdf  $\varphi$. Another difference with umbrella sampling is that here we generated from both  ``extremes'' i.e., from $q$, and $\post$. This is depicted in Figure \ref{USfig4}.
% which considers the opposite case: we sample from the ``extremes'' (being one of the proposal pdf and the other one the posterior pdf), and use a ``bridge'' (i.e., an ``intermediate'' density) to improve performance. 
\newline
\newline
Bridge sampling is a technique for computing ratios of constants by using samples drawn from their corresponding densities.  It is based on other identity that can be adapted for computing a single constant, $Z$, as follows \cite{meng1996simulating,llorenteREV_ML}
\begin{align}\label{BridgeSamplingIdentity}
Z = \frac{\E_{q}\left[\frac{\varphi(\x)}{q(\x)}\right]}{\E_{\post}\left[\frac{\varphi(\x)}{\pi(\x)}\right]},
\end{align}
%where $q(\x)$ is the proposal pdf, and $\varphi(\x)$ is an arbitrary pdf  defined on the intersection of the supports of $q(\x)$ and $\post(\x)$. 
and the corresponding estimator uses samples from  $q(\x)$ and $\post(\x)$,
\begin{align}\label{eq_BS_para_Z}
\widehat{Z}_\text{bridge} = \frac{\frac{1}{N_2}\sum_{i=1}^{N_2}\frac{\varphi(\z_i)}{q(\z_i)}}{\frac{1}{N_1}\sum_{j=1}^{N_1}\frac{\varphi(\x_j)}{\pi(\x_j)}}, \quad \x_j\sim\post(\x), \quad \z_i \sim q(\x),
\end{align}
where $j=1,\dots,N_1$ and $i=1,\dots,N_2$. 
The function $\varphi(\x)$ is an arbitrary density, defined on the intersection of the supports of $q(\x)$ and $\post(\x)$.  Note that $\varphi(\x)$ can be evaluated up to a normalizing constant, i.e., it can be an unnormalized pdf.
\newline
\newline
For any $\varphi(\x)$, the optimal proposal is 
\begin{align}
	\fbox{$q_\text{opt}(\x) \propto \pi(\x),$}
\end{align} 
which produces the exact solution $\widehat{Z}_\text{bridge}=Z$ (i.e., a zero variance solution).
Regarding $\varphi(\x)$, keeping fixed a generic $q(\x)$,  the asymptotic relative mean-squared error (rel-MSE) is minimized by the choice 
\begin{align}\label{eq_optBridgeBS}
	\fbox{$\varphi_\text{opt}(\x) = \dfrac{1}{\frac{N_2}{N_1+N_2}\post(\x)^{-1} + \frac{N_1}{N_1+N_2}q(\x)^{-1}} \propto \dfrac{q(\x)\pi(\x)}{N_1\pi(\x) + N_2Zq(\x)},$}
\end{align}
which is a weighted harmonic mean of the ``extreme'' densities $q$ and $\post$. 

{\rem Also in bridge sampling, employing jointly both $q_\text{opt}(\x)$ and $\varphi_\text{opt}(\x)$, we have
\begin{equation}
\fbox{$q_\text{opt}(\x)=\varphi_\text{opt}(\x)\propto \pi(\x),$}
\end{equation}
and we obtain a zero variance estimator.
}
\newline
\newline
Since $q_\text{opt}(\x)$ and $\varphi_\text{opt}(\x)$ depends on the unknown quantity $Z$, in order to use the optimal bridge sampling estimator we need again to use  an iterative procedure \cite{meng1996simulating}.  Starting with an initial estimate $\widehat{Z}^{(0)}$, we iteratively update it as  
\begin{align}
	\label{IterativeBS}
	\widehat{Z}^{(t)} = \frac{\frac{1}{N_2}\sum_{i=1}^{N_2}\dfrac{\pi(\z_i)}{N_1\pi(\z_i) + N_2\widehat{Z}^{(t-1)}q(\z_i)}}{\frac{1}{N_1}\sum_{i=1}^{N_1}\dfrac{q(\x_i)}{N_1\pi(\x_i) + N_2\widehat{Z}^{(t-1)}q(\x_i)}},\quad  \text{for} \quad t=1,...,T,
\end{align}
where $\{ \z_i \}_ {i=1}^{N_2} \sim q(\x)$ and $ \{\x_i\}_{i=1}^{N_1} \sim \post(\x)$. 

{\rem In the iterative procedure above, note that the sampling part and the evaluations of $\pi(\x)$ and $q(\x)$ are performed only once.}
\newline
\newline
The authors in \cite{meng1996simulating} demonstrate that this iterative scheme has a unique limit, and that achieves the same optimal variance of the optimal bridge sampling estimator.

{
\subsection{{Optimal tempering for thermodynamic integration}\label{OptimalTDI_sect}}

\subsubsection{{Thermodynamic integration: background}}

Let us define that the so-called power posterior (i.e., a tempered posterior) at inverse temperature parameter $\beta \in[0,1]$,
\begin{align}
\post_\beta(\x | {\bf y})= \post(\x | {\bf y}, \beta)=\frac{\ell({\bf y} | \x)^\beta g(\x)}{Z(\beta)},
\end{align}
with normalizing constant
\begin{align}
Z(\beta)=\int_\Theta \ell({\bf y} | \x)^\beta g(\x) d \x, \quad \mbox{where $Z(0)=1$ and $Z(1)=Z=p({\bf y})$.}  
\end{align}
 By definition of derivative and integral, we can also write the equality:
\begin{align}\label{ReplaceAQUI_Eq}
\log Z(1)-\log Z(0)=\log p(\mathbf{y})=\int_0^1 \frac{d}{d \beta}[\log Z(\beta)] d \beta .
\end{align}
Moreover, we can also write
\begin{align}
\frac{d}{d \beta} \log \{Z(\beta)\} =\frac{1}{Z(\beta)} \frac{d}{d \beta} Z(\beta)  &=\frac{1}{Z(\beta)} \frac{d}{d \beta} \int_\Theta \ell(\mathbf{y} | \boldsymbol{\theta})^\beta g(\x) d \x  \nonumber\\
& =\frac{1}{Z(\beta)} \int_\Theta \ell(\mathbf{y} | \x)^\beta \log [\ell(\mathbf{y} | \x)] g(\x) d \x \nonumber \\
& =\int_\Theta \log [\ell(\mathbf{y} | \x)] \frac{\ell(\mathbf{y} | \x)^\beta g(\x)}{Z(\beta)} d \x  \nonumber\\
& =\int_{\Theta} \log [\ell(\mathbf{y} | \x)] \post_\beta(\x | {\bf y}) d \x  \nonumber\\
& =\mathbb{E}_{\post_\beta}[\log [\ell(\mathbf{y} | \x)]],
\end{align}
where we have used the derivation rule $\frac{d a^\beta}{d\beta}=a^\beta \log a$, with $a>0$ a positive constant. Replacing in Eq. \eqref{ReplaceAQUI_Eq}, then we get
\begin{align}
\log p({\bf y})=\int_0^1 \mathbb{E}_{\post_\beta}[\log \ell({\bf y} | \x)] d \beta =\int_0^1 \varphi(\beta) d\beta,
\end{align}
where we have set  $ \varphi(\beta)=\mathbb{E}_{\post_\beta}[\log \ell({\bf y} | \x)]$.
This is the thermodynamic integral representation of the log marginal likelihood \cite{Llorente2023_TBI, llorenteREV_ML}. In practice,  the trapezoidal quadrature rule  is often used for approximating the one dimensional integral,
\begin{align}\label{TRAP_song_Rule}
\log p({\bf y})=\int_0^1 \varphi(\beta) d \beta \approx \sum_{j=0}^{N-1}\left(\beta_{j+1}-\beta_j\right) \frac{\varphi\left(\beta_{j+1}\right)+\varphi\left(\beta_j\right)}{2},
\end{align}
where $\varphi(\beta)=\mathbb{E}_{\post_\beta}[\log \ell({\bf y} | \x)]$. Note also that
\begin{align}\label{TRAP_song_Rule2}
\log p(\mathbf{y})&=\int_0^1 \frac{\mathbb{E}_{\post_\beta}[\log \ell(\mathbf{y} \mid \boldsymbol{\theta})]}{q(\beta)} q(\beta) d \beta=\int_0^1 \frac{\varphi(\beta)}{q(\beta)} q(\beta) d \beta =\mathbb{E}_q\left[\frac{\varphi(\beta)}{q(\beta)}\right],
\end{align}
where $q(\beta)$ is the proposal density of tempering parameters (called inverse temperatures).

%%%%%%%%%%%%%%%%%%%%%%%%%%%%%%%%%%%%
\subsubsection{{Optimal density for the inverse temperatures}}
%%%%%%%%%%%%%%%%%%%%%%%%%%%%%%%%%%%%

For large $N$, the expected (local) spacing between two consecutive nodes, i.e., the average of $d_i=\beta_{i+1}-\beta_i$, is approximately \cite{david2003order}
$$
\E[d_i] \approx \frac{1}{N q\left(\xi_i\right)} \quad \text { for some } \xi_i \in\left[\beta_i, \beta_{i+1}\right], \qquad \beta_i \sim q(\beta).
$$
The expression above has an intuitive interpretation since densely sampled regions have small spacing and sparsely sampled regions have large spacing \cite{david2003order}. Moreover, for large $N$, we also have  $d_i \approx \E[d_i]$. Thus, in the $i$-th interval, the error of the trapezoidal rule \cite{davis2007methods} can be written as 
$$
E_i\approx-\frac{d_i^3}{12} |\varphi^{\prime \prime}\left(\xi_i\right)| \approx -\frac{1}{12} \frac{|\varphi^{\prime \prime}\left(\xi_i\right)|}{N q\left(\xi_i\right)^3}, \qquad \beta_i \sim q(\beta),
$$
where we have substituted $d_i \approx \frac{1}{N q\left(\xi_i\right)}$.
The global error is then \cite{davis2007methods}:
$$
E_{\text{trap}}^{(N)}=\sum_{i=0}^{N-1} E_i \approx-\frac{1}{12 N^3} \sum_{i=0}^{N-1} \frac{|\varphi^{\prime \prime}\left(\xi_i\right)|}{q\left(\xi_i\right)^3}, \quad \xi_i \in\left[\beta_i, \beta_{i+1}\right], \quad \beta_i \sim q(\beta).
$$
Furthermore, as  $N\rightarrow \infty$, the global error is expressed by the following integral:
\begin{align} 
E_{\text{trap}}^{(\infty)}&\approx -\frac{1}{12 N^2} E_q\left[\frac{|\varphi^{\prime \prime}(\beta)|}{q(\beta)^3}\right]= -\frac{1}{12 N^2} \int_0^1 \frac{|\varphi^{\prime \prime}(x)|}{q(\beta)^3} q(\beta) d \beta,\\
&\approx-\frac{1}{12 N^2} \int_0^1 \frac{|\varphi^{\prime \prime}(\beta)|}{q(\beta)^2} d \beta.
\end{align}
Now, minimizing the corresponding Lagrangian functional,
$$
\mathcal{L}[q]=\int_0^1 \frac{|\varphi^{\prime \prime}(\beta)|}{q(\beta)^2} d \beta+\lambda\left(\int_0^1 q(\beta) d \beta-1\right),
$$
we obtain, as shown in Appendix \ref{FVsect2},
\begin{align}\label{SuperExperEq}
\fbox{$q_{\mathrm{opt}}(\beta)\propto \left|\varphi^{\prime \prime}(\beta)\right|^{1 / 3}.$}
\end{align}
The idea is that more nodes are placed where $\left|\varphi^{\prime \prime}(\beta)\right|$ is large, where $\varphi(\beta)$ is highly curved. Wheras fewer nodes are placed where $\left|\varphi^{\prime \prime}(\beta)\right|$ is small, where $\varphi(\beta)$ is almost linear. The density Eq. \eqref{SuperExperEq} minimizes the error of the trapezoidal rule  \eqref{TRAP_song_Rule}-\eqref{TRAP_song_Rule2} for a given $N$. The first derivative of $\varphi(\beta)$ is
\begin{align}\label{EqToProve}
\varphi^{\prime}(\beta)=\frac{d}{d \beta} \mathbb{E}_{\post_\beta}[\log p(y | \x)] 
=\operatorname{Var}_{\post_\beta}[\log p(y | \x)],
\end{align}
i.e., the variance under the power posterior (see Appendix \ref{AppC} for the proof).  Hence,  $
\varphi^{\prime\prime}(\beta)=\frac{d}{d \beta} \operatorname{Var}_{\post_\beta}[\log p(y | \x)]$,  and the optimal density of the inverse temperatures is:
\begin{align}
\fbox{$q_{\mathrm{opt}}(\beta)\propto \left|\varphi^{\prime \prime}(\beta)\right|^{1 / 3}= \left|\frac{d}{d \beta} \operatorname{Var}_{\post_\beta}[\log p(y | \x)] \right|^{1 / 3}.$} 
\end{align}
However, estimating $\varphi^{\prime \prime}(\beta)$ by MC samples can be difficult, whereas estimating $\varphi^{\prime}(\beta)$ (i.e., the variance) is easier. Then,
the authors in \cite{CalderheadGirolami2009} propose using the first derivative $\varphi^{\prime}(\beta)=\operatorname{Var}_{\post_\beta}[\log p(y | \x)]$ instead of $\varphi^{\prime \prime}(\beta)$. More specifically, they suggest {\it the heuristic solution}:
\begin{align}
\fbox{$q_{\mathrm{hst}}(\beta) \propto \sqrt{\operatorname{Var}_{\post_\beta}[\log p(y | \x)]}.$}
\end{align}
This can be justified heuristically: where $\varphi^{\prime}(\beta)=\operatorname{Var}_{\post_\beta}[\log p(y | \theta)]$ is large (typically near $\beta=0$, i.e., near the prior), the log-likelihood changes rapidly and many temperature points are needed. However, using just $\varphi^{\prime}(\beta)$ linearly would over-concentrate points in high-variance regions. Hence, perhaps it would be a too aggressive strategy, so that the authors in \cite{CalderheadGirolami2009} included the square root. 
Moreover, without estimating $\varphi^{\prime}(\beta)$,  a simple and practical choice can be the power-law schedule:
$$
\beta_i=\left(\frac{i}{N}\right)^\gamma, \quad \gamma \in[2,5].
$$
A common choice is $\gamma=3$. Clearly, most of the inverse temperatures $\beta_i$'s are located close to $\beta=0$, as suggested by the theory.

%$$
%\underbrace{1 / 3}_{\text {exact cube-root of } \varphi^{\prime \prime}} \Rightarrow \underbrace{1 / 2}_{\text {heuristic using } \operatorname{Var}=\varphi^{\prime}} .
%$$
}

%%%%%%%%%%%%%%%%%%%%%%%%%%%%%%%%%%%%%
\section{Optimality for specific parametric proposal families}\label{SpecPropSect}
%%%%%%%%%%%%%%%%%%%%%%%%%%%%%%%%%%%%%%

In this section, we consider proposals of specific forms and derive their optimal expressions.
Specifically, we first consider the case of a proposal as product of independent univariate densities, one for each  component of $\x$ (as known as  {\it mean-field} approach), and then the case of a proposal which is piece-wise constant (more generally, it could be a linear combination of basis functions). For simplicity, we consider $P=1$ and $M=1$, hence $f(\x)$ and $I$ are scalar-valued throughout all the section. 

%%%%%%%%%%%%%%%%%%%%%%%%%%%%%%%%%%%
\subsection{Optimal separable proposal density }
%%%%%%%%%%%%%%%%%%%%%%%%%%%%%%%%%%

Let consider again the scalar integral $I = \int_\Theta  f(\x)\post(\x)d\x$ for $f(\x)\geq 0$ and denote $\x = [\theta_1,\dots,\theta_{D}]^\top$. We now consider a separable proposal pdf, formed by the product $D$ univariate proposal densities, one for each component of $\x$, %()
$$
q(\x) = \prod_{k=1}^{D}q_k(\theta_k).
%\bar{q}_1(x_1)\bar{q}_2(x_2)\dots \bar{q}_{D}(x_{D}),
$$
%i.e., a proposal with uncorrelated coordinates. 
This kind of proposal is applied in algorithms such as the {\it Vegas algorithm}  and other approaches \cite{lepage1978new,Djuric_a_la_Gibbs}.
This scenario is also related to {\it mean field} approximations \cite{Su2021}. In this section, our goal is to find the optimal proposal of this form $q_\text{opt}(\x) = \prod_{i=k}^{D}q_{k,\text{opt}}(\theta_k)$, such that each partial proposal $q_{k,\text{opt}}$ individually minimizes the variance of the corresponding IS estimator, i.e., we look for
 \begin{align*}
	q_{i,\text{opt}}(\theta_i) &= \arg \min_{q_i} \mbox{Var}[\widehat{I}],\quad \text{for all} \quad i=1,...,D.
\end{align*}
\newline
{\bf Knowing $Z$.}
We consider the standard IS estimator $\widehat{I}$ in Eq. \eqref{eq_std_IS_est}. Hence, for each $i = 1,\dots,D$
\begin{align}
	q_{i,\text{opt}} = \arg \min_{q_i} \mbox{Var}[\widehat{I}_\text{IS}]&= \arg \min_{q_i} \mbox{var}_{q}\left[\frac{f(\x)\post(\x)}{\prod_{k=1}^{D}q_k(\x)}\right],  \nonumber\\
	&= \arg \min_{q_i} \E_{q}\left[\frac{f(\x)^2\post(\x)^2}{\prod_{k=1}^{D}q_k(\theta_k)^2}\right].
\end{align}
Note that 
\begin{align}
	\E_{q}\left[\frac{f(\x)^2\post(\x)^2}{\prod_{k=1}^{D}q_k(\theta_k)^2}\right] 
	&= 
	\bigintsss \frac{f(\x)^2\post(\x)^2}{\prod_{k=1}^{D}q_k(\theta_k)^2} \left(\prod_{k\neq i}^{D}q_k(\theta_k)d\theta_k\right)q_i(\theta_i)d\theta_i,  \nonumber\\
	&= \E_{q_i}\left[
	\frac{\bigintss{\frac{f(\x)^2(\x)\post(\x)^2}{\prod_{k\neq i}^{D}q_k(\theta_k)} \prod_{k\neq i}^{D}d\theta_k}}{q_i(\theta_i)^2}
	\right].
\end{align}
Hence, by Jensen's inequality, we have that the variance is minimized when
\begin{align}\label{OP_IS_vegas}
	\fbox{$q_{i,\text{opt}}(\theta_i) \propto
	\left[\bigintsss{\dfrac{f(\x)^2\post(\x)^2}{\prod_{k\neq i} q_{k,\text{opt}}(\theta_k)} \prod_{k\neq i}d\theta_k}\right]^\frac{1}{2}, \qquad i=1,...,D.$}
	\end{align}
{\rem Note that each $q_{i,\text{opt}}$ depends on the other $q_{k,\text{opt}}$. Then, generally an iterative procedure should be applied to obtain these optimal densities. }
\newline
\newline
The popular Vegas algorithm employs these types of optimal partial proposal densities \cite{lepage1978new}. 
More specifically, the Vegas algorithm starts with constant, uniform partial proposals (i.e. $q_i(\theta_i) = \mathcal{U}_{[a_i,b_i]}(\theta_k)$ for some $a_i,b_i\in \mathbb{R}$, $i=1,\dots,D$), and then iteratively refines piece-wise constant approximations to each $q_{i,\text{opt}}$. 

{\rem If in Eq. \eqref{OP_IS_vegas} we consider $f(\x)=1$, we obtain  the optimal mean-field partial proposals for estimating $Z$. }
{%\newline
%{\bf Optimal mean-field proposal for estimating $Z$.}
%We have to find $q_{i,\text{opt}}$ that minimizes the variance of $\widehat{Z}$ in Eq. \eqref{eq_Z_est}, i.e., that minimizes the term
%\begin{align}
%\E_{q}\left[\frac{\pi(\x)^2}{\prod_{k=1}^{D}q_k(\theta_k)^2}\right] 
%&= 
%\bigintsss\frac{\pi(\x)^2}{\prod_{k=1}^{D}q^2_k(\theta_k)} \left(\prod_{k\neq i}q_k(\theta_k)d\theta_k\right)q_i(\theta_i)d\theta_i \\
%&= \E_{q_i}\left[
%\dfrac{\bigintss{\dfrac{\pi(\x)^2}{\prod_{k\neq i}q_k(\theta_k)} \prod_{k\neq i}d\theta_k}}{q_i(\theta_i)^2}
%\right].
%\end{align}
%The optimal $i$-th proposal is thus
%\begin{align}
%q_{i,\text{opt}}(\theta_i) \propto
%\left[\bigintsss{\dfrac{\pi(\x)^2}{\prod_{k\neq i}q_k(\theta_k)} \prod_{k\neq i}d\theta_k}\right]^\frac{1}{2}.
%\end{align}
\newline
\newline
{\bf Unknown $Z$.} 
In this scenario, we consider the SNIS estimator in Eq. \eqref{eq_SNIS_est}.
Recall that 
\begin{align}
	\mbox{Var}\left[\widehat{I}_\text{SNIS}\right] \approx \frac{1}{N}\E_q\left[\frac{\post(\x)^2(f(\x)-I)^2}{q(\x)^2}\right].
\end{align}
We can replace  $q(\x)^2 = \prod_{k=1}^{D}q_k(\theta_k)^2$ into expression above, 
\begin{align}
\E_{q}\left[\frac{\post(\x)^2(f(\x)-I)^2}{\prod_{k=1}^{D}q_k(\theta_k)^2}\right] 
&= 
\bigintsss\frac{\post(\x)^2(f(\x)-I)^2}{\prod_{k=1}^{D}q^2_k(\theta_k)} \left(\prod_{k\neq i}^Dq_k(\theta_k)d\theta_k\right)q_i(\theta_i)d\theta_i \nonumber \\
&= \E_{q_i}\left[
\frac{\bigintss{\frac{\post(\x)^2(f(\x)-I)^2}{\prod_{k\neq i}q_k(\theta_k)} \prod_{k\neq i}^D d\theta_k}}{q_i(\theta_i)^2}
\right].
\end{align}
By Jensen's inequality, we have that the variance is minimized when
\begin{align}
\fbox{$q_{i,\text{opt}}(\theta_i) \propto
\left[\bigints{\dfrac{\post(\x)^2(f(\x)-I)^2}{\prod_{k\neq i}q_{k,\text{opt}}(\theta_k)} \prod_{k\neq i}d\theta_k}\right]^\frac{1}{2}, \qquad i=1,...,D,$}
\end{align}
that is $i$-th optimal proposal.

%%%%%%%%%%%%%%%%%%%%%%%%%%%%%%%%%%%%%%%
\subsection{Optimal coefficients for a piecewise constant proposal pdf}\label{SectOC_PCProp}
%%%%%%%%%%%%%%%%%%%%%%%%%%%%%%%%%%%%%%%

The use of proposal densities formed by  piecewise constant pieces has been proposed in different works in the literature \cite{Sticky13,FUSS,MartinoA2RMS,llorente2021deep,pantaleoni2017notes}. 
This family of proposals has the great advantage that can be easily employed in adaptive Monte Carlo schemes {\it but} considering a non-parametric construction, allowing a complete convergence  of the sequence of proposal functions to the target density (thanks to the flexibility of a non-parametric function) \cite{Sticky13,MartinoA2RMS,llorente2021deep}.
A piecewise constant proposal pdf can be expressed as a mixture of {\it disjoint} uniform densities, i.e.,
\begin{align}\label{StickyEq1}
	q(\x) = \sum_{i=1}^M \alpha_i b_i(\x),
\end{align}
with $\sum_i \alpha_i = 1$, and $\mathcal{C}_i \cap \mathcal{C}_j = \emptyset$ for $i\neq j$, such that  $\Theta = \bigcup_{i=1}^M \mathcal{C}_i$, i.e., the sets $\mathcal{C}_i$ form together a partition of $\Theta$. Moreover, the functions $b_i(\x)$ are defined as
\begin{align}\label{StickyEq2}
	b_i(\x) = \frac{1}{|\mathcal{C}_i|}\mathbb{I}_{\mathcal{C}_i}(\x), 
\end{align}
where 
$$
%\mbox{$\mathbb{I}_{C_i}(\x)=1$ if $\x\in C_i$ or $\mathbb{I}_{C_i}(\x)=0$ if $\x\notin C_i$,}
\mathbb{I}_{\mathcal{C}_i}(\x)=
\left\{
\begin{array}{ccc}
1 & \text{if} & \x\in \mathcal{C}_i, \\
0 & \text{if} & \x\notin \mathcal{C}_i. \\
\end{array}
\right.
$$
Namely, $b_i(\x)$ are constant non-linear functions (that we will call {\it bases}) with non-overlapping supports.
%The optimal proposal for estimating  $Z = \int_\Theta _\Theta\pi(\x)d\x$ is $q(\x) \propto \pi(\x)$. 
We aim to find the set of optimal coefficients that minimize the variance of the estimators $\widehat{Z}$, $\widehat{I}_\text{IS}$ and $\widehat{I}_\text{SNIS}$. That is, considering a pre-established (fixed) partition $\{\mathcal{C}_i\}_{i=1}^M$,  we seek the optimal coefficients $\alpha_i$, which completely defines $q(\x)$ (jointly with the partition).  This type of {\it optimality} is related to optimal sampling/allocation and weighting in mixture proposals \cite{pantaleoni2017notes,he2014optimal,ElviraMIS15}. Another approach would be to consider also the optimization of the partition $\{\mathcal{C}_i\}_{i=1}^M$. 
\newline
\newline
{\bf Optimal $\alpha_i$'s for $\widehat{Z}$.}
Here, our goal is to obtain the optimal coefficients $\alpha_i$ (employed in $q(\x)$) in order to obtain the most efficient estimator $\widehat{Z}$. Hence, we need to minimize the term
\begin{align}
	\E\left[\frac{\pi(\x)^2}{q(\x)^2}\right] = \int_\Theta \frac{\pi(\x)^2}{\sum_{i=1}^M\alpha_ib_i(\x)}d\x = \sum_{i=1}^M\frac{1}{\alpha_i}\int_{\mathcal{C}_i}\frac{\pi(\x)^2}{b_i(\x)}d\x,
\end{align}
with respect to $\bm{\alpha}=[\alpha_1,\dots,\alpha_M]$ subject to $\sum_{i=1}^M\alpha_i=1$. 
The solution to this constrained optimization problem are  (see \cite{pantaleoni2017notes}) 
\begin{align}
	\alpha_{i,\text{opt}} = \dfrac{\left[\bigintsss_{\mathcal{C}_i}\frac{\pi(\x)^2}{b_i(\x)}d\x\right]^\frac{1}{2}}{\sum_{j=1}^M \left[\bigintsss_{\mathcal{C}_j}\frac{\pi(\x)^2}{b_j(\x)}d\x\right]^\frac{1}{2}}, \quad i=1,\dots, M, \nonumber 
%	= \dfrac{|\mathcal{C}_i|^{\frac{1}{2}}\left[\bigintsss_{\mathcal{C}_i}\pi(\x)^2d\x\right]^{\frac{1}{2}}}{\sum_{j=1}^M|\mathcal{C}_j|^{\frac{1}{2}}\left[\bigintsss_{\mathcal{C}_j}\pi(\x)^2d\x\right]^{\frac{1}{2}}}, \nonumber 
	\end{align}
	and, using the equality $b_k(\x)= \frac{1}{|\mathcal{C}_k|}\mathbb{I}_{\mathcal{C}_k}(\x)$, we arrive to the following final formula, 
\begin{align}\label{AlfaOptAqui}	
	\fbox{$\alpha_{i,\text{opt}} = \dfrac{\left[|\mathcal{C}_i|\bigintsss_{\mathcal{C}_i}\pi(\x)^2d\x\right]^{\frac{1}{2}}}{\sum_{j=1}^M\left[|\mathcal{C}_j|\bigintsss_{\mathcal{C}_j}\pi(\x)^2d\x\right]^{\frac{1}{2}}}, \quad i=1,\dots,M$.}
\end{align} 
%\begin{align}
%	m_i^{(2)} =  = |C_i|\int_{C_i}\pi(\x)^2d\x.
%\end{align}
These  coefficients of the mixture correspond to the normalized $\ell_2$-norm projection of $\pi(\x)$ onto the space of basis functions \cite{pantaleoni2017notes}. 
%{the paper also considers the extension to general basis functions with probably overlapping supports...}. 
\newline
\newline
{\bf Optimal $\alpha_i$'s for $\widehat{I}_\text{IS}$.}
Now, our goal is to obtain the optimal coefficients $\alpha_i$ (employed in $q(\x)$) in order to obtain the most efficient estimator $\widehat{I}_\text{IS}$. Thus, we need to minimize the term
\begin{align}
\E\left[\frac{f(\x)^2\post(\x)^2}{q(\x)^2}\right] = \int_\Theta\frac{f(\x)^2\post(\x)^2}{\sum_{i=1}^M\alpha_ib_i(\x)}d\x = \sum_{i=1}^M\frac{1}{\alpha_i}\int_{\mathcal{C}_i}\frac{f(\x)^2\post(\x)^2}{b_i(\x)}d\x,
\end{align}
with respect to $\bm{\alpha}=[\alpha_1,\dots,\alpha_M]$ subject to $\sum_{i=1}^M\alpha_i=1$.
Following the previous result, the optimal coefficients of the mixture are 
%\begin{align}
%\alpha_{i,\text{opt}} = \dfrac{\left[\bigintsss_{C_i}\frac{f(\x)^2\post(\x)^2}{b_i(\x)}d\x\right]^\frac{1}{2}}{\sum_{j=1}^M \left[\bigintsss_{C_j}\frac{f(\x)^2\post(\x)^2}{b_j(\x)}d\x\right]^\frac{1}{2}} = \dfrac{|C_i|^{\frac{1}{2}}\left[\bigintsss_{C_i}f(\x)^2\post(\x)^2d\x\right]^{\frac{1}{2}}}{\sum_{j=1}^M|C_j|^{\frac{1}{2}}\left[\bigintsss_{C_j}f(\x)^2\post(\x)^2d\x\right]^{\frac{1}{2}}}.
%\end{align}
\begin{align}	
	\fbox{$\alpha_{i,\text{opt}} = \dfrac{\left[|\mathcal{C}_i|\bigintsss_{\mathcal{C}_i}f(\x)^2\pi(\x)^2d\x\right]^{\frac{1}{2}}}{\sum_{j=1}^M\left[|\mathcal{C}_j|\bigintsss_{\mathcal{C}_j}f(\x)^2\pi(\x)^2d\x\right]^{\frac{1}{2}}},$}
\end{align}
where we can see that, comparing with Eq. \eqref{AlfaOptAqui}, the term $f(\x)^2$ appears. 
\newline
\newline
{\bf Optimal coefficients for SNIS.} Lastly, we desire to obtain the optimal coefficients $\alpha_i$ (employed in $q(\x)$) in order to obtain the most efficient estimator $\widehat{I}_\text{SNIS}$. Recalling that $\mbox{Var}\left[\widehat{I}_\text{SNIS}\right] \approx \frac{1}{N}\E_q\left[\frac{\post(\x)^2(f(\x)-I)^2}{q(\x)^2}\right]$, we aim to minimize
\begin{align*}
	\E_q\left[\frac{\post(\x)^2(f(\x)-I)^2}{q(\x)^2}\right] = \int_{\Theta}\frac{(f(\x)-I)^2\post(\x)^2}{\sum_{i=1}^M\alpha_ib_i(\x)}d\x = \sum_{i=1}^M\frac{1}{\alpha_i}\int_{\mathcal{C}_i}\frac{(f(\x)-I)^2\post(\x)^2}{b_i(\x)}d\x.
\end{align*}
Thus, it is possible to show that the optimal coefficients are %{(see \cite{optCoeffaqui})}
\begin{align}	
	\fbox{$\alpha_{i,\text{opt}} = \dfrac{\left[|\mathcal{C}_i|\bigintsss_{\mathcal{C}_i}(f(\x)-I)^2\pi(\x)^2d\x\right]^{\frac{1}{2}}}{\sum_{j=1}^M\left[|\mathcal{C}_j|\bigintsss_{\mathcal{C}_j}(f(\x)-I)^2\pi(\x)^2d\x\right]^{\frac{1}{2}}}.$}
\end{align}

\subsection{Optimal $\alpha_i$'s for the auxiliary function $\varphi$ in RIS}
%{\bf }

Considering again a density formed by piecewise constant pieces, i.e.,
\begin{align*}
	\varphi(\x) = \sum_{i=1}^M \alpha_i b_i(\x), \quad \mbox{ where }  \quad  b_i(\x) = |C_i|^{-1}\mathbb{I}_{C_i}(\x), 
\end{align*}
given also in Eqs \eqref{StickyEq1}-\eqref{StickyEq2}. Now, we derive the optimal coefficients $\alpha_i$'s in the RIS estimator for estimating $Z$ (described in Section \ref{RISsect}), when the auxiliary function in Eq. \eqref{ReverseIS} has this form, i.e., $\varphi(\x) = \sum_{i=1}^M\alpha_ib_i(\x)$.
In order to obtain the optimal coefficients, we consider the variance of $\frac{1}{\widehat{Z}_\text{RIS}}$, i.e.,
\begin{align}
	\text{Var}\left[\frac{1}{\widehat{Z}_\text{RIS}}\right] = \frac{1}{N}\E_{\post}\left[\frac{\varphi(\x)^2}{\pi(\x)^2}\right] - \frac{1}{NZ^2}.
\end{align}
We need to minimize the term
\begin{align}
	\E_{\post}\left[\frac{\varphi(\x)^2}{\pi(\x)^2}\right] = 
	\frac{1}{Z}\int_\Theta\frac{(\sum_{i=1}^M\alpha_ib_i(\x))^2}{\pi(\x)}d\x = \frac{1}{Z}\sum_{i=1}^M\alpha_i^2\int_{C_i}\frac{b_i(\x)^2}{\pi(\x)}d\x,
\end{align}
with respect to $\bm{\alpha}=[\alpha_1,\dots,\alpha_M]$ subject to $\sum_{i=1}^M\alpha_i=1$. 
%Define $$
In order to find $\bm{\alpha}_\text{opt}$, we build the Lagrangian function, derive it and solve the following equation
\begin{align}
	\nabla_{\bm{\alpha},\lambda}\left[\frac{1}{Z}\sum_{i=1}^M\alpha_i^2\int_{C_i}\frac{b_i(\x)^2}{\pi(\x)}d\x - \lambda\left(\sum_{i=1}^M\alpha_i - 1\right)\right] = 0.
\end{align}
Let us define $\gamma_i = \int_{C_i}\frac{b_i(\x)^2}{\pi(\x)}d\x$. The solution can be expressed as
\begin{align}
\fbox{$\alpha_{i,\text{opt}} = \frac{\frac{1}{\gamma_i}}{\sum_{j=1}^M \frac{1}{\gamma_j^2}}$,} \quad \mbox{ where }  \quad  \gamma_i = \int_{C_i}\frac{b_i(\x)^2}{\pi(\x)}d\x.%{= \frac{1}{|C_i|^{2}}\int_{C_i}\frac{1}{\pi(\x)}d\x.}
\end{align}
The method in \cite{wang2018new} considers this type of proposal and estimator for computing the marginal likelihood, and is known as {\it partition weighted kernel estimator}.

{
%%%%%%%%%%%%%%%%%%%%%%%%%%%%%%%%%%%%%%%
\section{{Optimal proposal densities via variational inference}}\label{VarInference_Sect}
%%%%%%%%%%%%%%%%%%%%%%%%%%%%%%%%%%%%%%%

\subsection{{Unconstrained functional minimization}}

 Given the integral $I=\int_\Theta f(\x) \bar{\pi}(\x) d \x$, the variance of the estimator $\widehat{I}_\text{IS}=\frac{1}{N}\sum_{i=1}^N \frac{f(\x_i)\post(\x_i)}{q(\x_i)}$ is given in Eq. \eqref{eq_var_std_IS}, that is,  
\begin{align}
	\mbox{Var}_q[\widehat{I}_\text{IS}]&= \frac{1}{N}\left(\E_{q}\left[\left(\frac{f(\x)\post(\x)}{q(\x)}\right)^2\right] - I^2 \right), \nonumber \\
	&= \frac{1}{N}\left(\int_\Theta \frac{h(\x)^2}{q(\x)^2} q(\x) d\x - I^2 \right), \nonumber \\	
&= \frac{1}{N}\left(\int_\Theta \frac{h(\x)^2}{q(\x)} d\x - I^2 \right), \quad h(\x)=|f(\x)| \post(\x).	
\end{align}
Ignoring constants, minimizing variance is exactly equivalent to minimize the term $
\int_\Theta \frac{h(\x)^2}{q(\x)} d \x$, hence the optimal proposal of minimum variance is 
\begin{align}
q_{\text{opt}}(\x)=\arg\min_q  \mbox{Var}_q[\widehat{I}_\text{IS}] \propto h(\x),
\end{align}
as shown in Section \ref{Op_in_StandIS} (recall that $ h(\x)=|f(\x)| \post(\x)\geq 0$).

{\rem  Note that the quantity
\begin{align}
\int_\Theta \frac{h(\x)^2}{q(\x)}  d\x=D_{\chi^2}\left(h \| q\right)+1, 
\end{align}
is related to the  $\chi^2$ (Pearson) divergence between $h$ and $q$, i.e.,   $ D_{\chi^2}\left(h  \| \ q\right)=\int_\Theta \frac{h(\x)^2}{q(\x)^2} q(\x)d\x$.  
Hence, from a theoretical standpoint (without any constraints or restrictions), minimizing the $\chi^2$ divergence,
 is exactly equivalent to minimize the variance of $\widehat{I}_\text{IS}$,
\begin{align}
q_{\chi^2}^*(\x)=\arg\min_q D_{\chi^2}\left(h \| q\right)=\arg\min_q  \mbox{Var}_q[\widehat{I}_\text{IS}], 
\end{align} 
i.e., $q_{\chi^2}^*(\x)=q_{\text{opt}}(\x) \propto h(\x)$.
 }
\newline
\newline
{\bf Other divergences.} Without any constraints/restrictions, i.e., in terms of functional optimization, other 
divergences could be used finding the same minimum, that is $q_{\text{opt}}(\x) \propto h(\x)$.  For instance, let us consider the {\it reverse} and/or {\it forward Kullback-Leibler} (KL) divergences \cite{akyildiz2024global,Akyildiz2021},
\begin{align*}
 D_{KL}\left(q \| h\right)&=\int_{\Theta} q(\x) \log \frac{q(\x)}{h(\x)} d \x =-H(q)-\int_{\Theta} q(\x) \log h(\x) d \x  \quad  \mbox{(reverse),} \\
 D_{KL}\left(h \| q\right)&=\int_{\Theta} h(\x) \log \frac{h(\x)}{q(\x)} d \x= -H(h)-\int_{\Theta} h(\x) \log q(\x) d \x  \quad \mbox{(forward),} 
\end{align*} 
where we have denoted as $H(b)=-\int_{\Theta} b(\x) \log b(\x) d \x$ the entropy of the generic density $b(\x)$. In unconstrained infinite-dimensional space, all three divergences (including Pearson) have the same global minimum:
$$
q_{\chi^2}^*(\x)=q_{rKL}^*(\x)=q_{fKL}^*(\x) \propto h(\x),
$$
that also coincides with $q_{\text{opt}}(\x)=\arg\min_q  \mbox{Var}_q[\widehat{I}_\text{IS}]$. Clearly, we have denoted with $q_{rKL}^*(\x)=\arg\min_q D_{KL}\left(q \| h\right)$ and  $q_{fKL}^*(\x)=\arg\min_q D_{KL}\left(h \| q\right)$.

{\rem Although several divergences may admit the same functional minimum, in constrained settings such as adaptive schemes based on parametric proposal densities, the resulting performance can differ significantly. See next section for more details.} 
 
%%%%%%%%%%%%%%%%%%%%%%%%%%%%%%%%%%%%%
\subsection{{Constrained scenarios: adapting a parametric proposal pdf}}
\label{SuperIMP_Sect}
%%%%%%%%%%%%%%%%%%%%%%%%%%%%%%%%%%%%%

Let us consider the parametric family of densities 
$\left\{q_{\bm \xi}(\x): {\bm \xi} \in \mathbb{R}^{d_\xi} \right\}$
where ${\bm \xi}$ is a vector of parameters to adapt. Given a generic divergence $D(h, q_{{\bm \xi}})$  the goal is to obtain the optimal vector:
$$
\widehat{{\bm \xi}}= \arg\min_{{\bm \xi}} D(h, q_{{\bm \xi}}),
$$
where also symmetric divergences can be considered. In this setting, the performance of  $ D_{\chi^2}\left(h  \| \ q\right)$, $D_{KL}\left(q \| h\right)$ or $D_{KL}\left(h \| q\right)$ to name a few, could be completely different:
\begin{itemize}
\item {\bf Reverse KL.} It penalizes in particular regions where $q_{\bm \xi}(\x)>0$ but $h(\x)$ is small, due to the presence of the term $\log h(\x)$. It can also allow for missing regions, in the sense that $q_{\bm \xi}(\x) \approx 0$  may occur even when  $h(\x)>0$. On the other hand, the reverse KL divergence encourages mode-seeking behavior. Generally, it  leads to proposals tightly concentrated around regions where $h(\x)$ is large.
\item {\bf Forward KL.} In this case, the only factor that affects the minimization is the cross-entropy, i.e., $-\int_{\Theta} h(\x) \log q_{\bm \xi}(\x) d \x$. Due to the presence of the term $\log q_{\boldsymbol{\xi}}(\x)$, the forward KL divergence 
strongly penalizes regions where $h(\x) > 0$ but $q_{\boldsymbol{\xi}}(\x) \approx 0$. 
As a result, it promotes mass-covering behavior and discourages to miss regions with  non-negligible probability $h(\x)>0$. Hence,  it generally produces broader proposal pdfs that cover the full support of $h(\x)$. Although it may exhibit higher variance than its reverse counterpart in certain scenarios, 
it is generally more stable and robust in high-dimensional or multimodal settings. 
Finally, its results are closely related to moment-matching approaches.
\end{itemize} 
Clearly, alternative divergences can be considered to combine the respective strengths of the KL divergences. Indeed, the {\it symmetrized} KL (a.k.a., Jeffreys divergence) or {\it Jensen-Shannon} divergence (JSD)  could be used, i.e.,
\begin{align*}
D_S(h,q_{{\bm \xi}})&= D_{KL}(h \| q_{{\bm \xi}})+D_{KL}(q_{{\bm \xi}} \| h)  \quad  \mbox{(symmetrized)}, \\
 D_{JS}(h,q_{{\bm \xi}})&= \frac{1}{2} D_{KL}(h \| s)+\frac{1}{2} D_{KL}(q_{{\bm \xi}} \| s)  \quad \mbox{(Jensen-Shannon),} 
\end{align*} 
where $s(\x)=\frac{1}{2}h(\x)+\frac{1}{2}q_{{\bm \xi}}(\x)$. Furthermore, modern adaptive IS methods often interpolate between the two KL divergences using the families called $\alpha$-divergence (a.k.a., R{\'e}nyi divergence) of the $f$-divergence \cite{Akyildiz2021,akyildiz2024global,pmlr-v238-guilmeau24a,pmlr-v161-jerfel21a,perello2023adaptively}.

{\rem The choice of a specific divergence (or family of divergences) depends also on its differentiability, convexity and the resulting robustness of the adaptive procedure. }

}

%%%%%%%%%%%%%%%%%%%%%%%%%%%%%%%
\section{Some numerical and theoretical comparisons}\label{NumSect}
%%%%%%%%%%%%%%%%%%%%%%%%%%%%%%%

This section is devoted to provide some numerical and theoretical comparisons, in order to highlight the importance of the notion of optimality in IS. We can avoid catastrophic situations (when the variance of the estimators explodes to infinity), and improve the baseline, ideal Monte Carlo scenario. Theoretical and numerical results are provided and checked. In section \ref{FirstSectNum}, we consider different functions $f(\x)$ and different proposal densities $q(\x)$, including the optimal ones. 
{ In Section~\ref{Divergence_Num_Sect}, we illustrate the landscapes of three different divergences, highlighting their distinct behaviors during a parametric proposal adaptation.}
In section \ref{IS_vs_RIScomp}, we focus on the comparison between IS and RIS for the estimation of the marginal likelihood $Z$. { All these sections provide examples and remarks that are useful for a complete understanding of the concepts presented above in the rest of the work.}

%%%%%%%%%%%%%%%%%%%%%%%%%%%%%%%%%%%%%
\subsection{The reason why IS is a variance reduction method}\label{FirstSectNum}
%%%%%%%%%%%%%%%%%%%%%%%%%%%%%%%%%%%%%
This section is divided in two parts. In the first part,  we show the shape of optimal densities considering different functions $f$  (different integrals), and having the same target density $\bar{\pi}$. In the second part, we show that the IS estimators can have better performance (in terms of smaller MSE) than the ideal Monte Carlo, using the optimal proposal density (or a proposal close to the optimal one).  %In the last part, we show that the IS estimators can beat the ideal Monte Carlo even with a proposal different from the optimal one, but close to it (in some sense). 
\newline
\newline
{\bf First part.} For the sake of simplicity, let us consider a one-dimensional Gaussian target  distribution, i.e.,
$$
\bar{\pi}(\theta)=\frac{1}{\sqrt{2\pi}} \exp\left( -\frac{(\theta+1)^2}{2}\right),
$$
i.e., with mean $\mu=-1$ and variance $\sigma^2=1$. We assume the following integrals of interest,
\begin{align*}
I_1 = \int_\Theta  \theta\post(\theta)d\theta, \quad  I_2 = \int_\Theta  \sqrt{|\theta|}\post(\theta)d\theta, \quad I_3 = \int_\Theta  \theta^2\post(\theta)d\theta,
\end{align*}
i.e., $f_1(\theta)=\theta$, $f_2(\theta)=\sqrt{|\theta|}$ and $f_3(\theta)=\theta^2$, respectively. The optimal proposals for the standard IS and the SNIS schemes are, respectively,
\begin{align*}
q_\text{opt}(\theta) \propto |f_k(\theta)|\post(\theta), \quad \mbox{ and } \quad q_\text{opt}(\theta) \propto |f_k(\theta)-I_k|\post(\theta), \quad k=1,2,3.
\end{align*}
The corresponding optimal proposal densities are depicted in Figures \ref{LucaEx1}, \ref{LucaEx2} and \ref{LucaEx3}. If compared with the ideal MC  (where the proposal coincides with the target $\post(\theta)$), their shapes are quite surprising, since some of them present regions of low probabilities around the mode of $\post(\theta)$. More generally, they differ substantially to the shape of the target density $\post(\theta)$: for instance, all of them are at least bimodal (in Figures \ref{Tremodes1}-\ref{Tremodes2}, there are three modes), instead of just unimodal as $\post(\theta)$.
\newline
\newline
{\bf Second part.} Assuming now $f(\theta)=\theta$, we compute the theoretical effective sample size (ESS) \cite{ESSmartino,ESSelvira} defined as 
$$
\mbox{ESS}=N \cdot \frac{\mbox{MSE of ideal MC}}{\mbox{MSE of $\widehat{I}_{\text{SNIS}}$}}.
$$
We have always that $\mbox{ESS}>0$ and, with a bad or regular choice of the proposal $q$, we generally have $\mbox{ESS} < N$, i.e., the  ideal MC performs better than an IS scheme. However, it is possible to obtain $\mbox{ESS} \geq N$. Indeed, we show that with a good choice of the proposal density $q$, the IS estimators can have better performance that the baseline MC estimators, so that we obtain $\mbox{ESS}\geq N$. This is the reason why IS is often included within the class of variance reduction techniques \cite{Arouna2004,Lapeyre2011,owen2013monte}. Firstly, we employ the optimal proposal $q_{\text{opt}}(\theta)$ for SNIS, assuming $f(\theta)=\theta$. Then, we also consider
 \begin{align*}
q(\theta) = \mathcal{N}(\theta|-1,h^2) = \frac{1}{\sqrt{2\pi h^2}}\exp\left(- \frac{1}{2h^2}(\theta+1)^2\right),
\end{align*}
 as proposal density in SNIS. Note that $q(\theta)$ has the same mean of 
  $\post(\theta)$ and variance $h^2$. We test the values $h=1.5$ and $h=5$. Finally, recall that in the baseline MC we employ $q(\theta)=\post(\theta)$. We set different values of $N \in \{10,50,100, 500,1000, 5000\}$. 
  The results averaged  over $1000$ independent simulations, are given in Table \ref{TablaResultsESS}. The MSE of the SNIS estimators with $q_{\text{opt}}(\theta)$ and with $q(\theta), h=1.5$,  is always lower than the MSE of the ideal MC scheme and, as a consequence, the ESS is always bigger than 1, in these cases. Whereas the MSE of the SNIS estimator with $q(\theta)$ and $h=5$ is bigger  than the MSE of the ideal MC scheme. The reason of this change in the performance is that $q(\theta)$ with $h=1.5$ covers the two modes of the optimal proposal in Figure \ref{CitabLuca}. Hence, with $h=1.5$, the proposal $q(\theta)$ is more similar to $q_{\text{opt}}(\theta)$, than $q(\theta)$ with $h=5$ and also than $q(\theta)$ with $h=1$ (that coincides with $\post(\theta)$).
  
 \begin{figure}[!h]
		\centering
		\centerline{
		\subfigure[Standard IS with $f(\theta)=\theta$.]{\includegraphics[width=6cm]{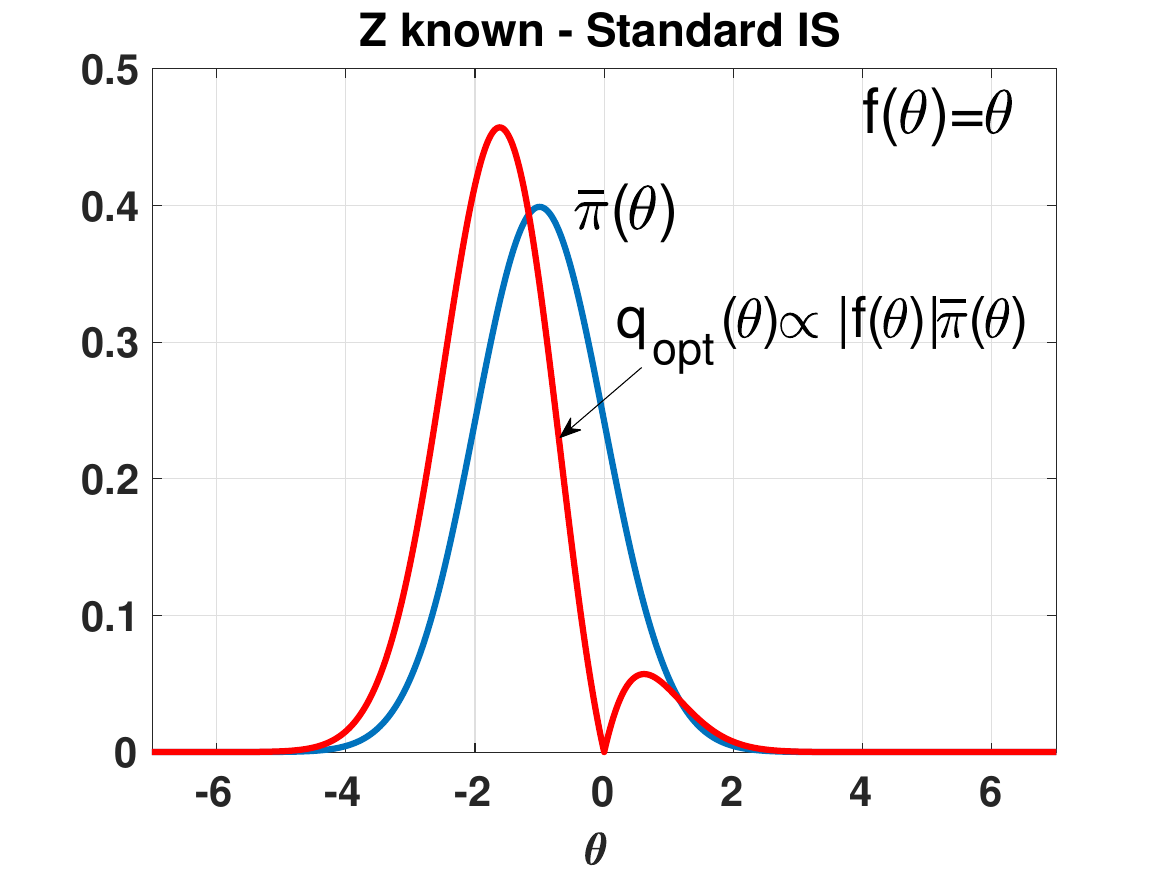}}
		\subfigure[\label{CitabLuca}Self-Normalized IS with $f(\theta)=\theta$.]{\includegraphics[width=6cm]{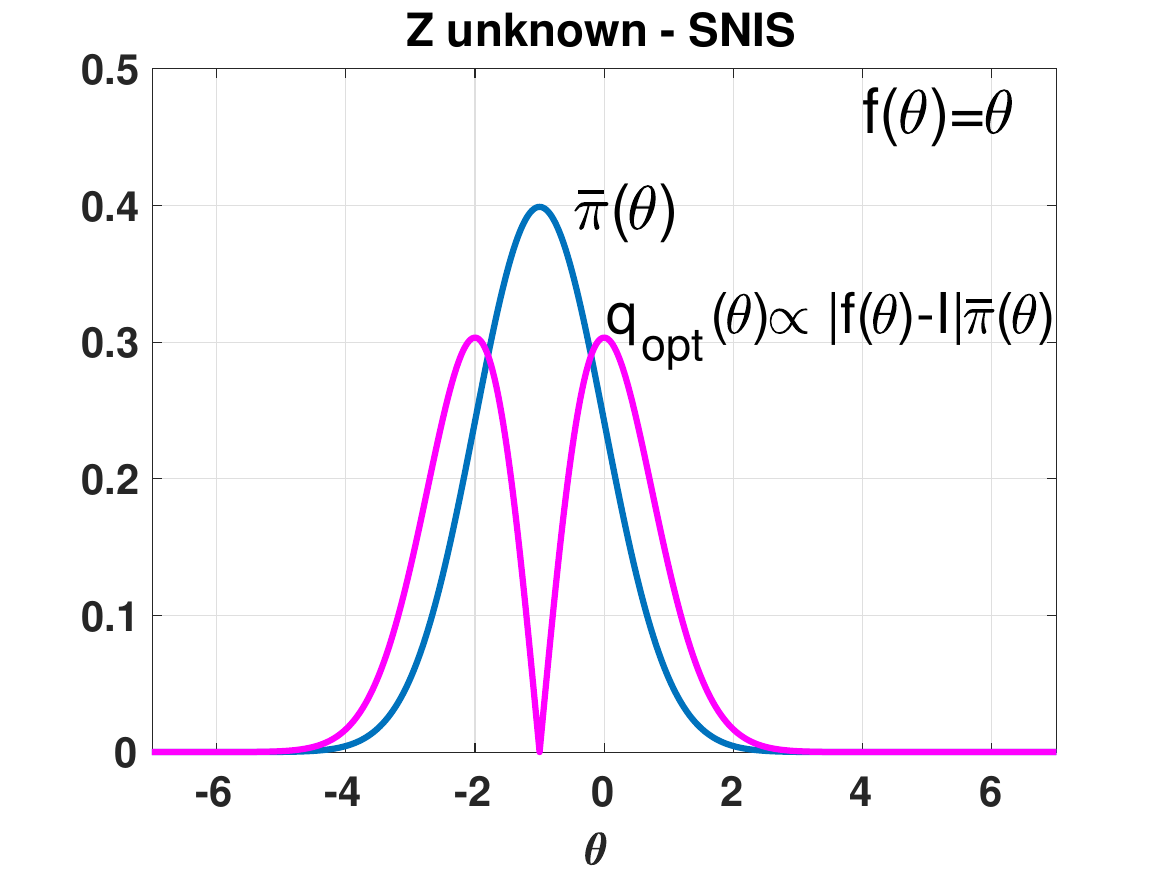}}
		}	
		%\vspace{-0.4cm}
		\caption{\footnotesize Target density $\post(\theta)$ (blue line) and optimal proposal densities $q_{\text{opt}}(\theta)$ for the standard IS (red line) and SNIS (magenta line) schemes, when $f(\theta)=\theta$.  }
		\label{LucaEx1}
	\end{figure}

	\begin{figure}[!h]
		\centering
		\centerline{
		\subfigure[Standard IS with $f(\theta)=\sqrt{|\theta|}$.]{\includegraphics[width=6cm]{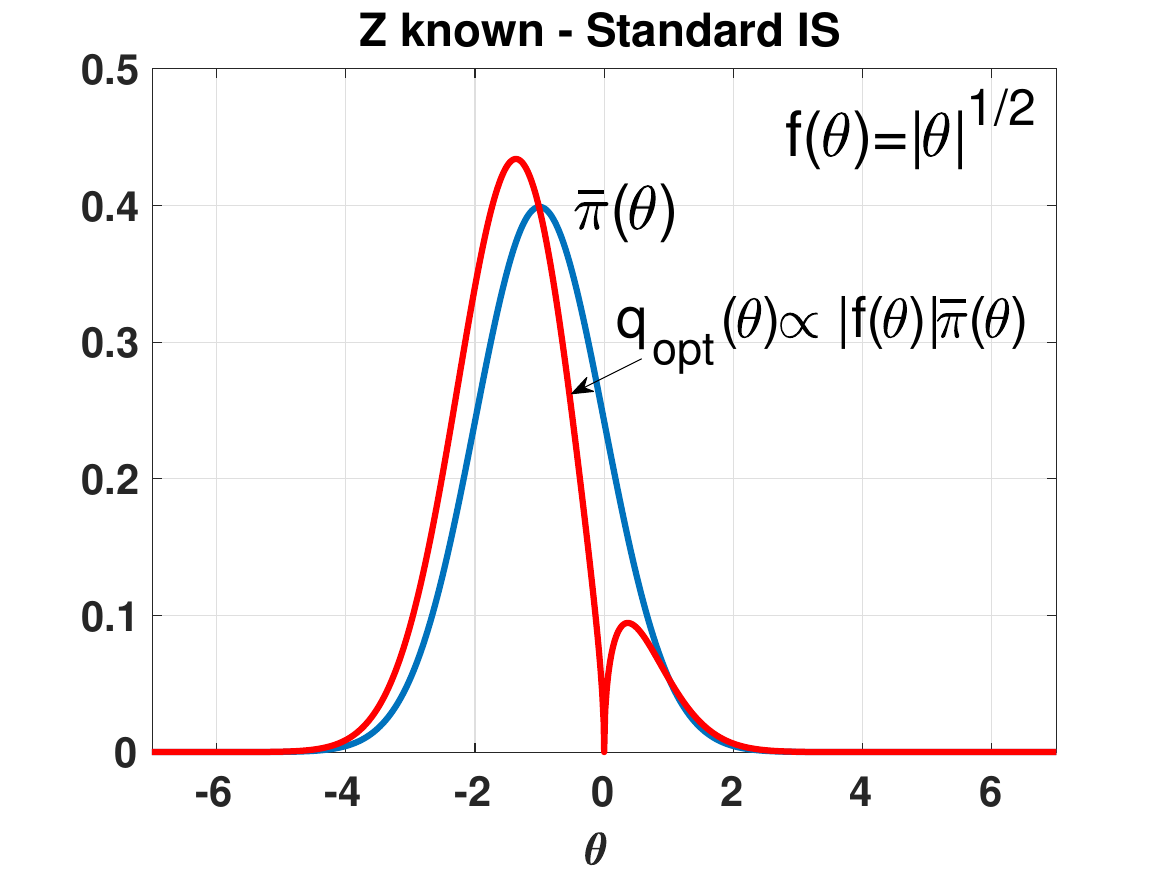}}
		\subfigure[\label{Tremodes1}Self-Normalized IS with $f(\theta)=\sqrt{|\theta|}$]{\includegraphics[width=6cm]{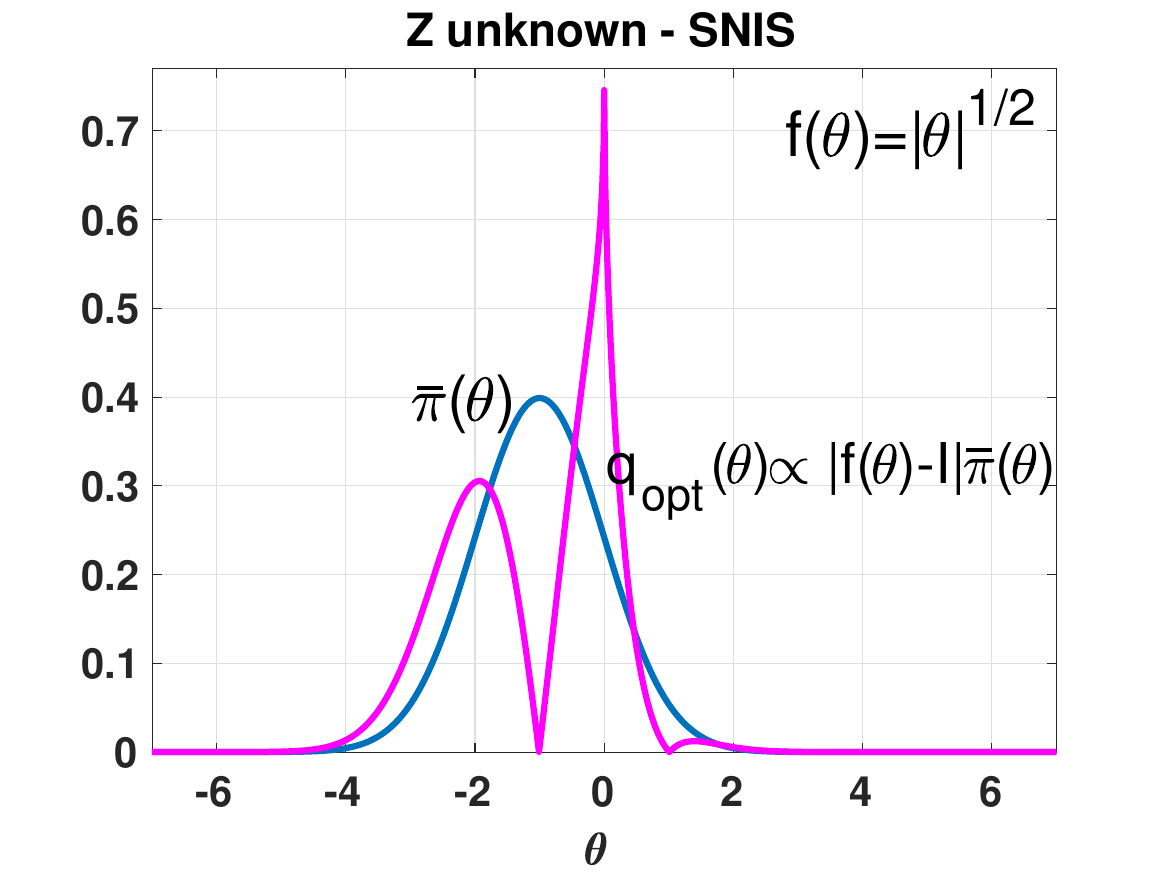}}
		}	
		%\vspace{-0.4cm}
		\caption{\footnotesize Target density $\post(\theta)$ (blue line) and optimal proposal densities $q_{\text{opt}}(\theta)$ for the standard IS (red line) and SNIS (magenta line) schemes, when $f(\theta)=\sqrt{|\theta|}$. }
		\label{LucaEx2}
	\end{figure}
	
	\begin{figure}[!h]
		\centering
		\centerline{
		\subfigure[Standard IS with $f(\theta)=\theta^2$.]{\includegraphics[width=6cm]{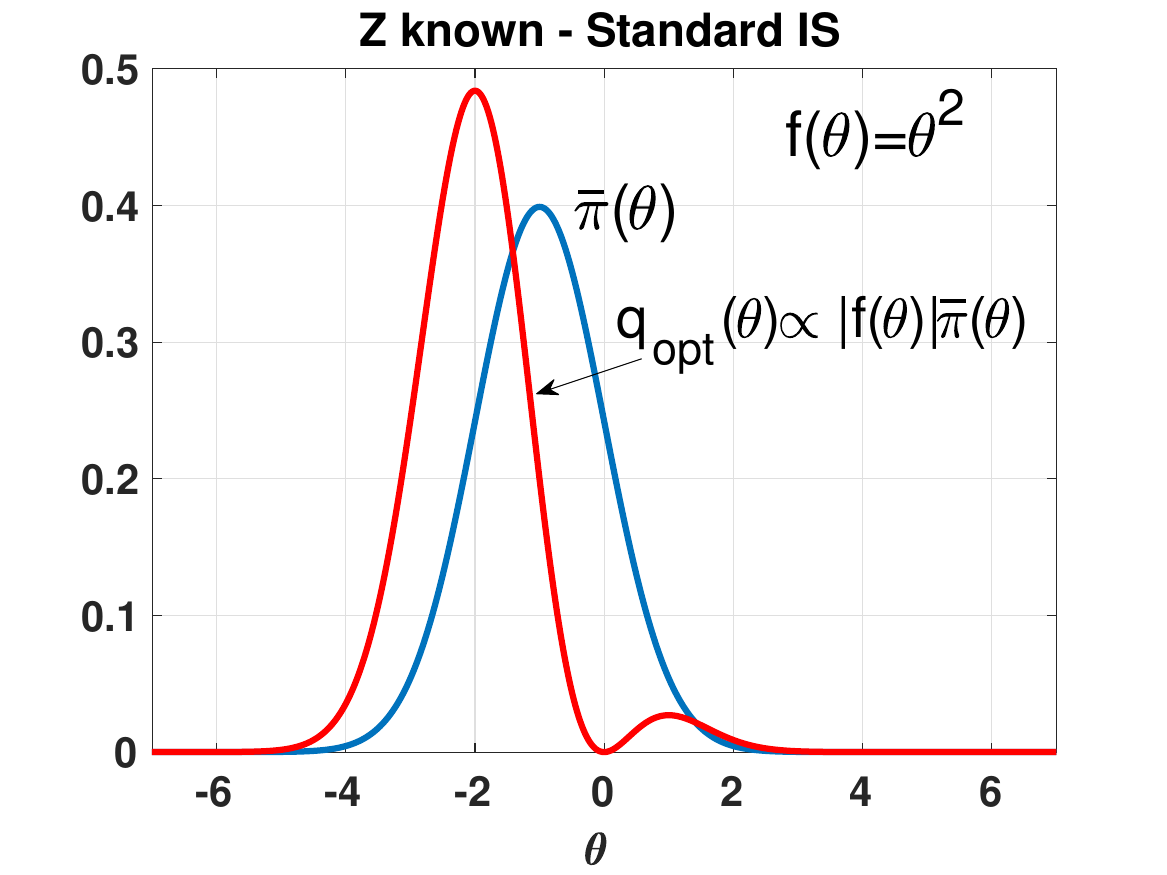}}
		\subfigure[\label{Tremodes2}Self-Normalized IS with $f(\theta)=\theta^2$]{\includegraphics[width=6cm]{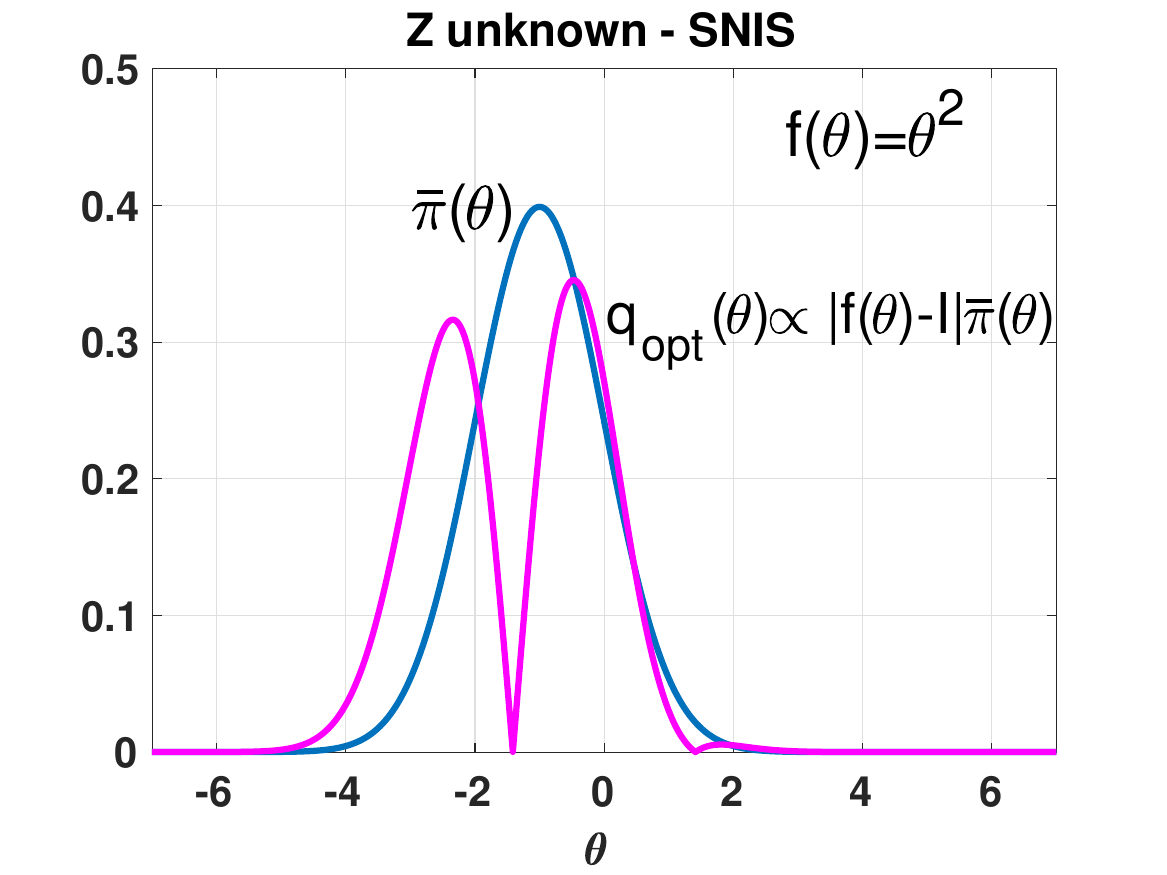}}
		}
		%\vspace{-0.4cm}
		\caption{\footnotesize Target density $\post(\theta)$ (blue line) and optimal proposal densities $q_{\text{opt}}(\theta)$ for the standard IS (red line) and SNIS (magenta line) schemes, when $f(\theta)=\theta^2$. }
		\label{LucaEx3}
	\end{figure}

%308.65   62.20   32.30    6.33    3.21    0.66    0.32
%276.73   46.79   23.22    4.39    2.11    0.44    0.21

 \begin{table}[!h]	
	 \caption{\footnotesize MSE and ESS comparing the ideal MC and the SNIS estimators, as function of the number of samples $N$ and for  different proposal densities. We can see that the MSE of the SNIS estimator with the optimal proposal (and with a $q$ close to the the optimal proposal, i.e., with $h=1.5$)  is always lower and, as a consequence, the ratio $\frac{\mbox{ESS}}{N}$ is always bigger than 1.}\label{TablaResultsESS}
	 \vspace{-0.2cm}
	\begin{center}
	\footnotesize
		\begin{tabular}{|c||c|c|c|c|c|c|} 
		\hline 
		Number of samples, $N$ 	 & $10$  & $50$ & $100$ &  $500$ & $1000$ & $5000$   \\ 
			 \hline
			 \hline
			 Ideal Monte Carlo, i.e., $q(\x)=\post(\x)$ & 
			 0.0986  &   0.0201 &   0.0101 &    0.0020  &  0.0010    & 0.0002		       \\
\hline 
\hline 			 
			 SNIS with $q_{\text{opt}}(\x)$  &    0.0834    & 0.0146 &   0.0067  &  0.0013   & 0.0006 &    0.0001    			    \\ 
              %      \hline 
                  %    \hline 
		 ESS$/N$ with $q_{\text{opt}}(\x)$ & 1.1823
		   &  1.3809  &  1.5103 &   1.5912  &  1.5475  &  1.5262   \\
			\hline 
 \hline 			 
			 SNIS, $q(\x)$ with $h=1.5$  & 0.0910 & 0.0158 &    0.0078  &  0.0016   & 0.0008  &   0.0002 \\    			      
		 ESS$/N$, $q(\x)$ with $h=1.5$ & 1.0834  &   1.2722 &   1.2949    &1.2500  &  1.2500 &    1.0000         \\
			\hline	
			\hline 			 
			 SNIS, $q(\x)$ with $h=5$  & 0.3916   & 0.0400&    0.0189    &0.0037   & 0.0019  &   0.0004        			    \\  
		 ESS$/N$, $q(\x)$ with $h=5$ &  0.2518  &  0.5031  &  0.5342    &0.5456   & 0.5404  &  0.5479         \\
			\hline		
		\end{tabular}
	\end{center}

\end{table}

{
 %%%%%%%%%%%%%%%%%%%%%%%%%%%%%%%%
 \subsection{{Behaviors of different divergences for the proposal adaptation}} \label{Divergence_Num_Sect} 
%%%%%%%%%%%%%%%%%%%%%%%%%%%%%%%%

The goal of this example is to illustrate in a simple, controlled and replicable scenario the behaviors of  the $\chi^2$- Pearson divergence and the reverse-forward KL divergences, described in Section \ref{SuperIMP_Sect}. Let us consider a bimodal target density,
\begin{align}
\post(\theta)=\frac{1}{2} \mathcal{N}(\theta|-3,0.5)+\frac{1}{2} \mathcal{N}(\theta|3,0.5).
\end{align}
We desire to adapt a parametric Gaussian proposal density,
\begin{align}
q_{{\bm \xi}}(\theta)=\mathcal{N}(\theta|\xi_1,\xi_2^2),
\end{align}
finding some ``optimal'' mean $\xi_1$ and standard deviation  $\xi_2$ (hence ${\bm \xi}=[\xi_1,\xi_2]$), in order to compute the normalizing constant $Z$. In this case, we know from the theory that $q_{\text{opt}}(\theta)=\post(\theta)$.  Note that here $h(\theta)\propto \post(\theta)$. The idea is to minimize the reverse and forward KL divergences, obtaining the three estimators:
\begin{align*}
\widehat{{\bm \xi}}_{\chi^2}&= \arg\min_{{\bm \xi}} D_{\chi^2}(q_{{\bm \xi}} || h), \\ 
\widehat{{\bm \xi}}_{\text{rev}}&= \arg\min_{{\bm \xi}} D_{KL}(q_{{\bm \xi}} || h),   \\
\widehat{{\bm \xi}}_{\text{for}}&= \arg\min_{{\bm \xi}} D_{KL}(h || q_{{\bm \xi}}). 
\end{align*}
We have numerically evaluated the three divergences $D_{\chi^2}(q_{{\bm \xi}} || h)$, $D_{KL}(q_{{\bm \xi}} || h)$ and $D_{KL}(h ||Êq_{{\bm \xi}} )$, as functions of the two scalar parameters $\xi_1$,  $\xi_2$. The last two divergences are depicted in Figure \ref{FigDiv}.  In Figures \ref{ForKLfig} and \ref{CHIfig}, the red point denotes moment matching result (which is very close to the true minimum). In Figure \ref{RevKLfig},  the two red points denote the two possible solutions/minima.
\newline
The forward KL divergence provides a unique minimum at $\xi_1\approx 0$ and $\xi_2 \approx \sqrt{3^2+0.5^2}=3.04$, corresponds also to the moment matching result. Hence, the forward KL divergence provides a stable/robust solution with a smooth landscape. The reserve KL divergence provides two possible solutions each one corresponding to one mode. This confirms clearly its mode-seeking behavior and the possibility of missing some region with a high probability mass, as discussed in Section \ref{SuperIMP_Sect}. 
\newline
The $\chi^2$-Pearson  divergence exactly minimizes the variance of the IS weights. However, within parametric families it can behave in a highly aggressive manner. In particular, its optimization landscape typically exhibits very sharp decay and large derivative magnitudes at some specific standard deviations $\xi_2$ of the proposal. See the decays in Figure \ref{CHIfig}. Moreover, in a neighborhood of the moment-matching solution, the $\chi^2$-Pearson divergence becomes extremely flat, leading to weak curvature and poor local identifiability.
This is because it gives a strong penalties when the proposal $q$  misses any region where $h$ is large. Hence, the $\chi^2$-Pearson divergence encourages {\it over-dispersion} of the proposal $q$, i.e., higher values of $\xi_2$  to control the ratio $h(\theta)^2/q(\theta)$ and, as a consequence the variance of the IS weights. Indeed, the solution in this case is $\xi^*_1=0$ for the mean, and $\x_2^*\geq \sqrt{3^2+0.5^2}=3.04$ for the standard deviation of the proposal.}

\begin{figure}[!h]
		\centering
		\centerline{
		\subfigure[\label{ForKLfig}Forward KL divergence.]{\includegraphics[width=7cm]{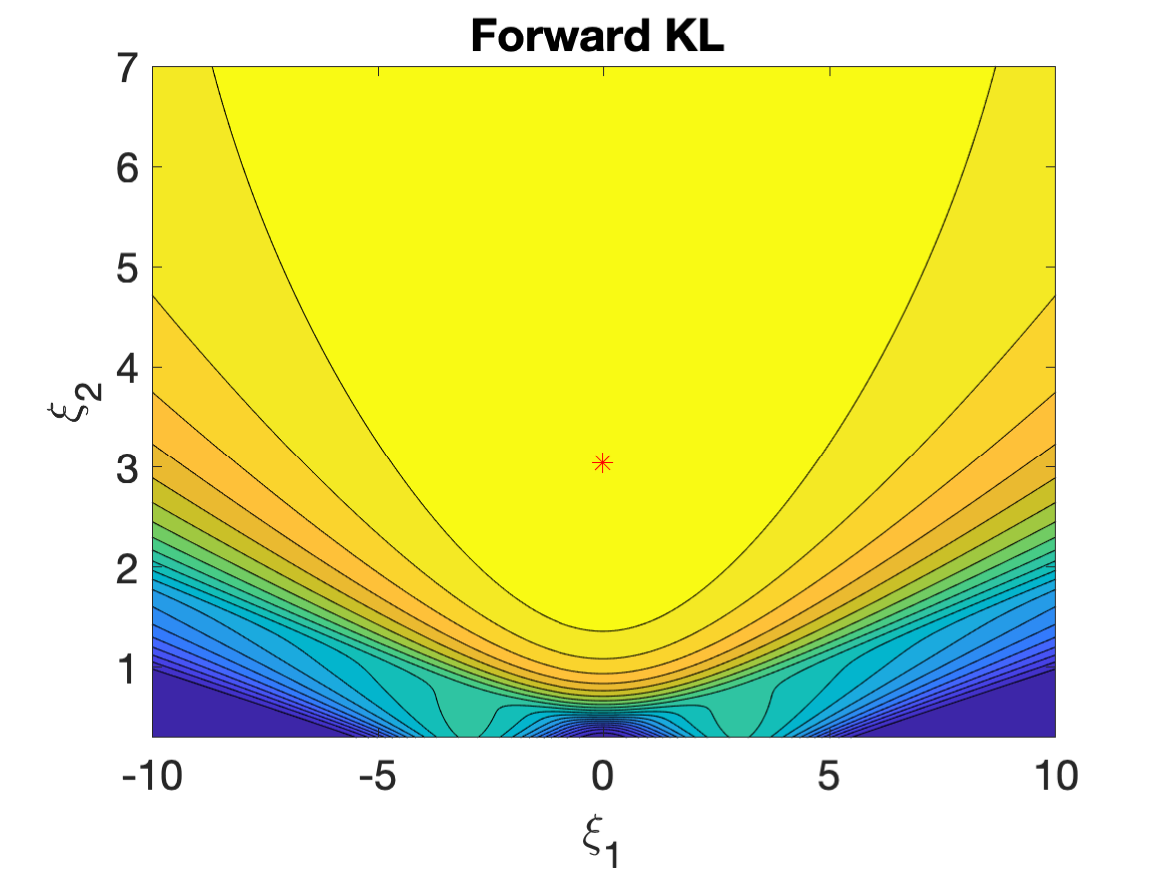}}
		\subfigure[\label{RevKLfig}Reverse KL divergence.]{\includegraphics[width=7cm]{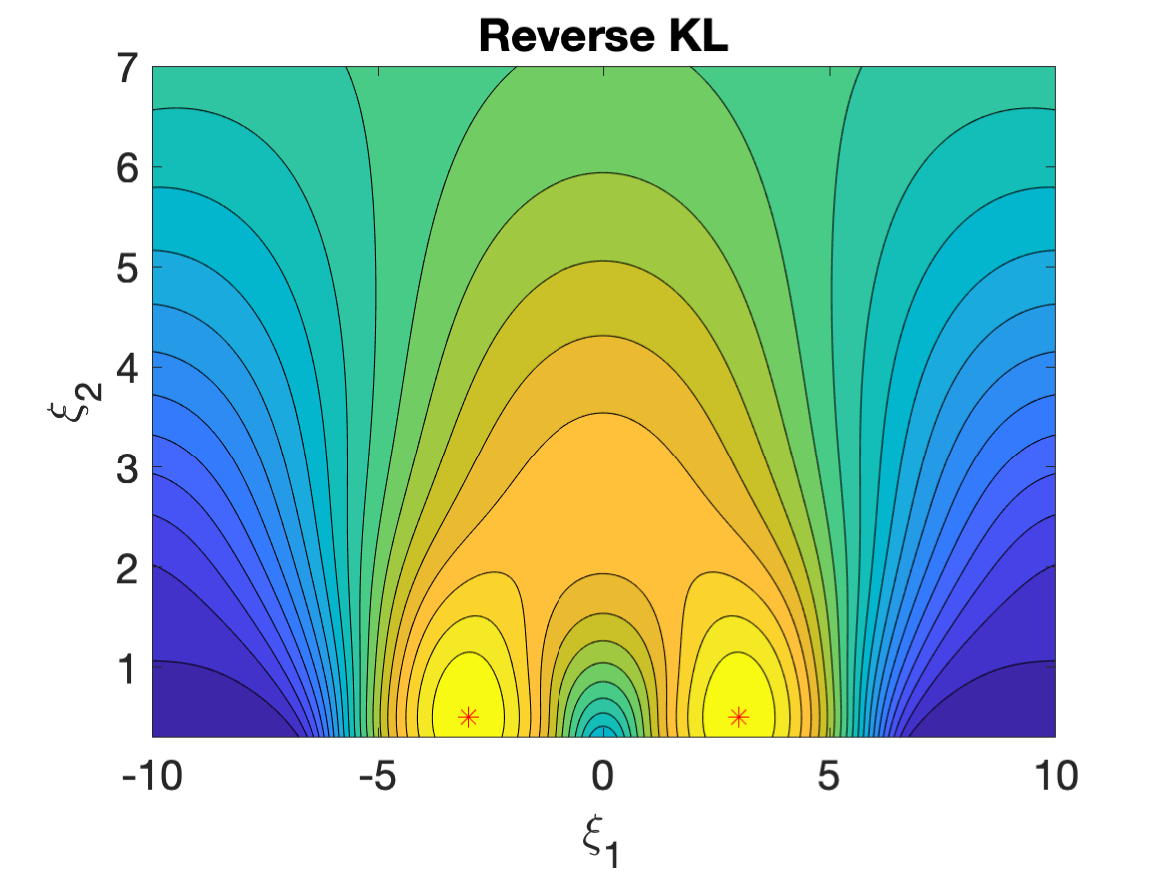}}	
		}
			\centerline{
		\subfigure[\label{CHIfig}$\chi^2$-Pearson divergence.]{\includegraphics[width=7cm]{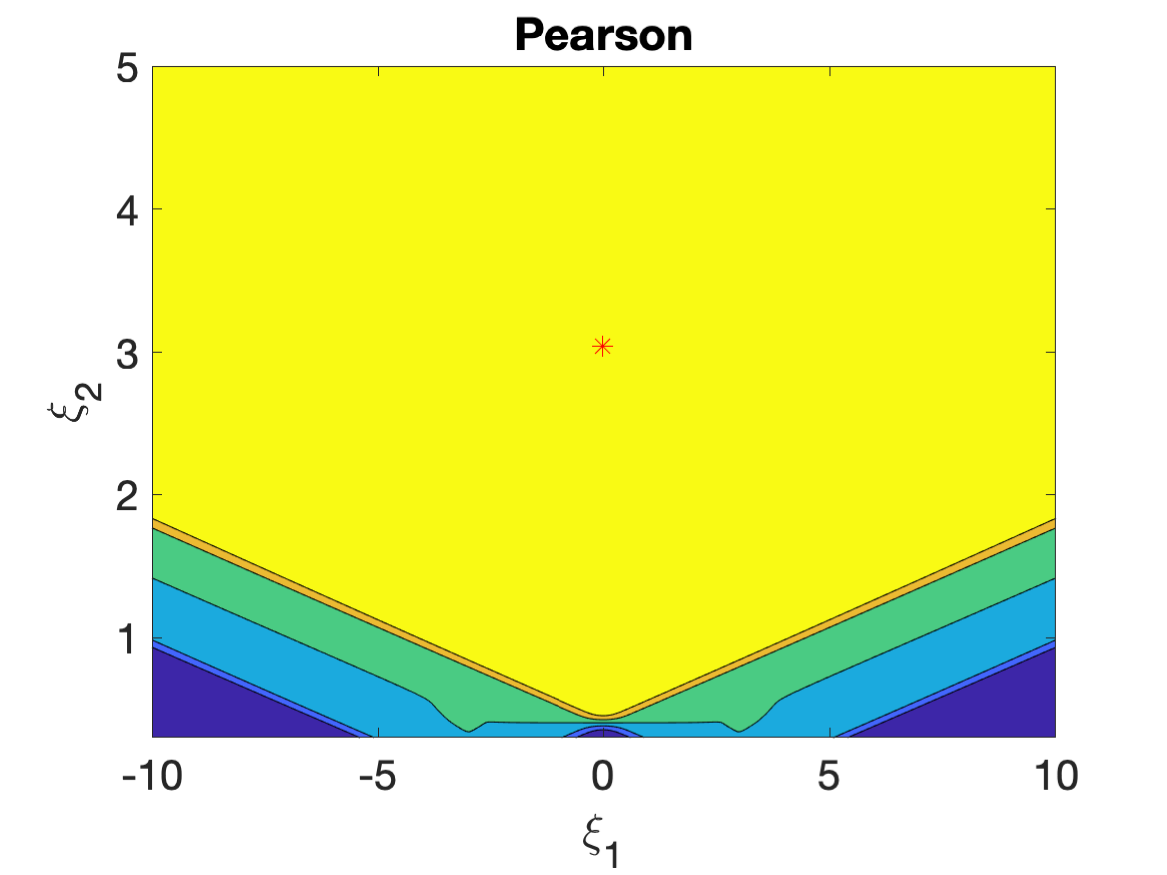}}	
		}
		%\vspace{-0.4cm}
		\caption{\footnotesize {Visualization of the three divergences $D_{\chi^2}(q_{{\bm \xi}} || h)$, $D_{KL}(q_{{\bm \xi}} || h)$ and $D_{KL}(h ||Êq_{{\bm \xi}} )$, as functions of the two scalar parameters: the mean $\xi_1$, and the standard deviation  $\xi_2$ of the proposal density. The red point in (c) is actually the moment matching result. The solution for the Pearson divergence is $\xi^*_1=0$ for the mean, and $\x_2^*\geq \sqrt{3^2+0.5^2}=3.04$ for the standard deviation of the proposal.}    }
		\label{FigDiv}
	\end{figure}

}

%\subsection{Second experiment}\label{Comparison IS vs RIS}

%%%%%%%%%%%%%%%%%%%%%%%%%%%%%%%%%%%%
\subsection{Theoretical and numerical comparisons between IS and RIS} \label{IS_vs_RIScomp} 
%%%%%%%%%%%%%%%%%%%%%%%%%%%%%%%%%%%%

{  The purpose of this example is not to show performance on a complex model, but rather to illustrate the variance behavior of IS and RIS under controlled conditions. In particular, the example is designed to validate the theoretical variance results in the simplest possible setting, thereby making the presentation clearer and more pedagogically effective.
\newline
More specifically, we choose  the estimation of the marginal likelihood $Z$ in a simple one-dimensional Gaussian setting, which serves as an instructive test case.
Focusing on the estimation of the normalizing constant $Z$ allows the variance properties of IS and RIS to be isolated and compared directly, without the confounding effects of an arbitrary integrand $f(\theta)$. The Gaussian target- Gaussian proposal configuration also ensures full analytical tractability: both 
$Z$ and the variances of the corresponding estimators can be computed in closed form, enabling a precise match between theoretical predictions and simulation results. }

%%%%%%%%%%%%%%%%%%%%%%%%%%%%%%%%%%%%%%%%%%%%
\subsubsection{Theoretical comparison}
%%%%%%%%%%%%%%%%%%%%%%%%%%%%%%%%%%%%%%%%%%%%

\begin{figure}[!h]
	\centering
	\centerline{
	\subfigure[\label{Fig4a}Variance of $\widehat{Z}_\text{IS}$.]{\includegraphics[width=6cm]{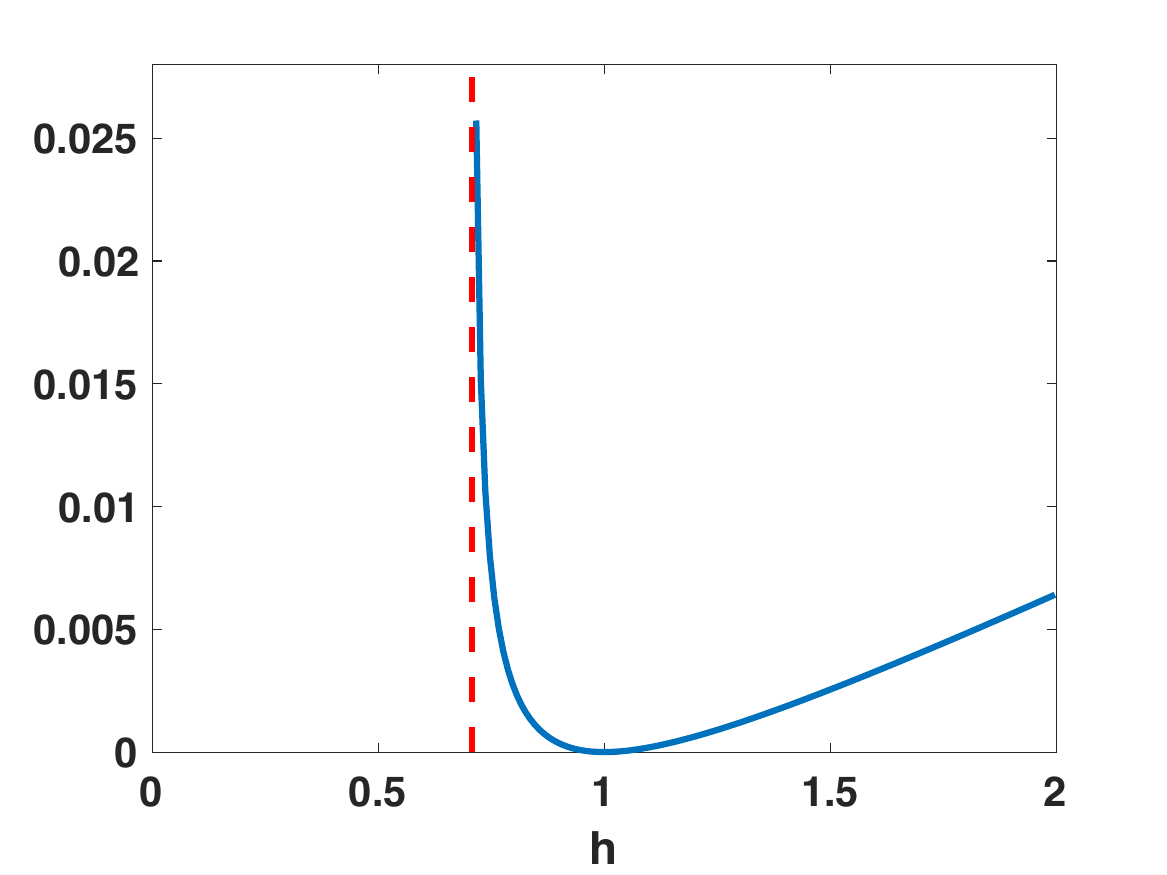}}
	%\hspace{0.5cm}
	\subfigure[\label{Fig4b}Variance of $\frac{1}{\widehat{Z}_\text{RIS}}$.]{\includegraphics[width=6cm]{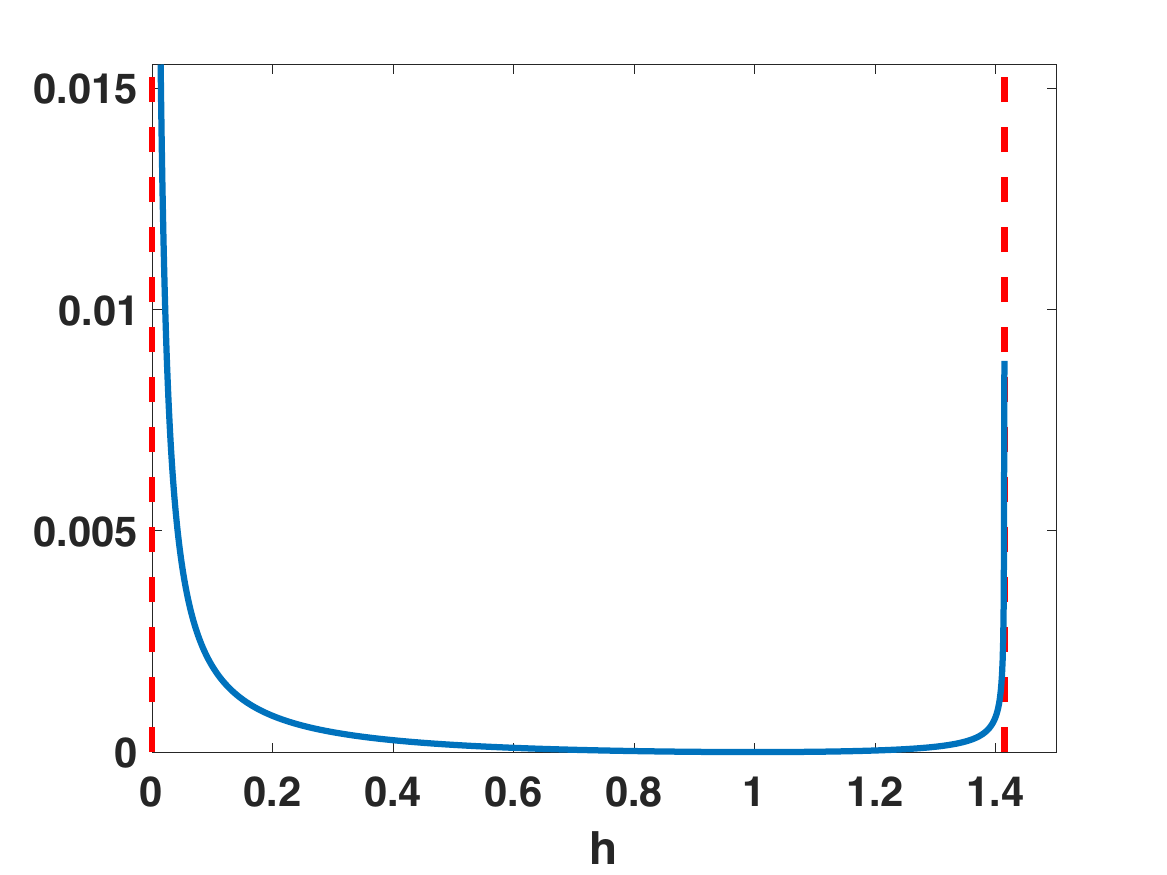}}
	}
	%\vspace{-0.4cm}
	\caption{The variances  $\text{Var}_{\post}[\widehat{Z}_\text{IS}]$ and $\text{Var}_{\post}[1/\widehat{Z}_\text{RIS}]$ in Eqs. \eqref{VAR_IS_teo} and \eqref{VAR_RIS_inv}, respectively ($N=500$).
	}
	\label{varZteo_ISandRIS}
\end{figure}

%We consider the scenario in the example in Section \ref{Comparison IS vs RIS}.
%{decimos que $\pi(\theta)$ no tiene dependencia en $\y$ en este toy example?}
In this section, the goal is to compare theoretically the standard IS and RIS schemes for estimating the normalizing constant of a target density $Z$. For simplicity, we consider again Gaussian target $\pi(\theta) = \exp( -\frac{1}{2}\theta^2)$, since we know the ground-truth  $Z =\int_{-\infty}^{\infty}\pi(\theta)d\theta = \sqrt{2\pi}$, so that $\post(\theta) = \frac{\pi(\theta)}{Z} = \mathcal{N}(\theta|0,1)$. 
%{Since this is a data-independent example, $\pi(\theta)$ and $\post(\theta)$ have no dependence on $\y$.}
The standard IS estimator of $Z$ with proposal $q(\theta)$ and the RIS estimator with auxiliary density $\varphi(\theta)$ are the following:
\begin{align*}
\widehat{Z}_\text{IS} = \frac{1}{N}\sum_{i=1}^{N} \frac{\pi(\theta_i)}{q(\theta_i)}, \quad  \theta_i \sim q(\theta), \quad \widehat{Z}_\text{RIS} = \frac{1}{\frac{1}{N}\sum_{k=1}^{N}\frac{\varphi(\theta_k)}{\pi(\theta_k)}}, \quad \theta_k \sim \post(\theta).
\end{align*}
For a fair theoretical and empirical comparison, we consider
\begin{align*}
\varphi(\theta)=q(\theta) = \mathcal{N}(\theta|0,h^2) = \frac{1}{\sqrt{2\pi h^2}}\exp\left(- \frac{1}{2h^2}\theta^2\right),
\end{align*}
where $h>0$ is the standard deviation. Thus,  both estimators depend on $q(\theta)$, although the density $q(\theta)$ plays a different role inside each estimator. We desire to study the performance of the two estimators as $h$ varies.
\newline
Now, we study the variances of the estimators $\widehat{Z}_\text{IS}$ and $\widehat{Z}_\text{RIS}$ as function of $h$, starting from $\widehat{Z}_\text{IS}$. Note that by the i.i.d. assumption, we can write
\begin{align}\label{VarIS_ex}
\text{Var}_{q}[\widehat{Z}_\text{IS}]&= \frac{1}{N}\text{Var}_{q}\left[\frac{\pi(\theta)}{q(\theta)}\right] = \frac{1}{N}\left\{ \mathbb{E}_{q}\left[\frac{\pi(\theta)}{q(\theta)}\right]^2 - Z^2 \right\},
\end{align}
Substituting $\pi(\theta) = \exp(-\frac{1}{2}\theta^2)$ and $q(\theta)=\frac{1}{\sqrt{2\pi h ^2}}\exp(-\frac{1}{2h^2}\theta^2)$, then we obtain
\begin{align*}
\mathbb{E}_{q}\left[\frac{\pi(\theta)}{q(\theta)}\right]^2 = \int_{-\infty}^{\infty}\frac{\pi(\theta)^2}{q(\theta)}d\theta &= \sqrt{2\pi h^2}\int_{-\infty}^{\infty}\exp\left\{-\left(1-\frac{1}{2h^2}\right)\theta^2 \right\}d\theta, \\
&=  2\pi \frac{h}{\sqrt{2-\frac{1}{h^2}}}. 
\end{align*}
Replacing the last expression above in Eq. \eqref{VarIS_ex}, we obtain that the variance of $\widehat{Z}_\text{IS}$ is given by 
\begin{align}\label{VAR_IS_teo}
\text{Var}_{q}\left[\widehat{Z}_\text{IS}\right] = \frac{2\pi}{N}\left\{ \frac{h}{ \sqrt{2 - \frac{1}{h^2}}} - 1,
\right\}.
\end{align}
that is depicted in Figure \ref{Fig4a}.
This variance reaches its minimum, $\text{Var}_q[\widehat{Z}_\text{IS}]=0$, at $h=1$, i.e., when the proposal is optimal, coinciding exactly the posterior $q(\theta)=\mathcal{N}(\theta|0,1)=\post(\theta)$ as expected (recall that we are estimating $Z$). For $h < 1$, $\text{Var}_q[\widehat{Z_\text{IS}}]$ grows exponentially until reaching $h = \frac{1}{\sqrt{2}}$ where is infinite. For $0 < h < \frac{1}{\sqrt{2}}$, $\text{Var}_q[\widehat{Z_\text{IS}}]$ is not defined. Finally,  $\text{Var}_q[\widehat{Z_\text{IS}}]$ grows linearly from $h=1$ onwards, i.e., diverges to infinity as $h\rightarrow \infty$. Figure \ref{Fig4a} shows that behavior when $N=500$.  Clearly, this is  perfectly in line the well-known theoretical requirement that the proposal pdf must have fatter tails than the posterior density in a IS  scheme. Moreover, this confirms that the use of proposals with variance bigger than that of the target is generally not catastrophic. The opposite could yield  catastrophic results.  Recall also that $\mathbb{E}_{q}[\widehat{Z}_\text{IS}]=Z$, i.e., the bias of $\widehat{Z}_\text{IS}$ is zero. 
\newline
\newline
Regarding RIS, it is easier to compute analytically the variance of $\widehat{r} =\frac{1}{\widehat{Z}_\text{RIS}}$, rather than $\widehat{Z}_\text{RIS}$ itself. Namely,  we consider the estimator $\widehat{r} =\frac{1}{N} \sum_{i=1}^N \frac{q(\theta_k)}{\pi(\theta_k)}$, with $\theta_k \sim \post(\theta)$, which is an unbiased estimator of $\frac{1}{Z}$. 
%Note that
%$$
%\widehat{Z}_\text{RIS} = \frac{1}{\widehat{r}} \ .
%$$
Since $\theta_k$'s are i.i.d. from $\post(\theta)$, then we have 
%{Mediante el metodo delta, podemos decir que el error de RIS es $\mathbb{E}[(\widehat{Z}_\text{RIS}-Z)^2]=\mbox{var}\left[\widehat{r}\right] + \mathcal{O}(\frac{1}{N^2})$ y asi nos ahorramos lo de la inversa....}
\begin{align*}
\text{Var}_{\post}\left[\widehat{r}\right] = \text{Var}_{\post}\left[\frac{1}{\widehat{Z}_\text{RIS}}\right] &= \frac{1}{N}\text{Var}_{\post}\left[ \frac{f(\theta)}{\pi(\theta)} \right], \\
&= \frac{1}{N}\left\{ \mathbb{E}_{\post}\left[ \frac{f(\theta)}{\pi(\theta)} \right]^2 - \frac{1}{Z^2} \right\},
\end{align*}
Substituting $\pi(\theta) = \exp\left(-\frac{1}{2}\theta^2\right)$ and $f(\theta)=\frac{1}{\sqrt{2\pi h ^2}}\exp\left(-\frac{1}{2h^2}\theta^2\right)$, we obtain
\begin{align*}
\mathbb{E}_{\post}\left[ \frac{f(\theta)}{\pi(\theta)} \right]^2 &= \int_{-\infty}^{\infty} \left(\frac{f(\theta)}{\pi(\theta)} \right)^2 \frac{\pi(\theta)}{Z}d\theta, \\
&= \frac{1}{Z}\int_{-\infty}^{\infty} \frac{f(\theta)^2}{\pi(\theta)} d\theta, \\
&= \frac{1}{2\pi h^2 \sqrt{2\pi}}\int_{-\infty}^{\infty} \exp \left\{ -\left( \frac{1}{h^2} - \frac{1}{2}\right)\theta^2 \right\}d\theta, \\
&= \frac{1}{2\pi}\frac{1}{h^2\sqrt{\frac{2}{h^2}-1}}.
\end{align*}
Hence the variance of $\widehat{r}$ is given by 
\begin{align}\label{VAR_RIS_inv}
\text{Var}_{\post}[\widehat{r}] = \text{Var}_{\post}\left[\frac{1}{\widehat{Z}_\text{RIS}}\right] = \frac{1}{2\pi N}\left\{ \frac{1}{h^2\sqrt{\frac{2}{h^2}-1}} - 1 \right\},
\end{align}
which reaches its minimum, $\text{Var}_{\post}[\widehat{r}]=0$, again at $h=1$ as expected. Recall that in RIS, $q(\theta)$ is playing the role of an auxiliary density, and it is not a proposal pdf. 
 Note that $\text{Var}[\widehat{r}]$ is defined when $0<h<\sqrt{2}$ (there are two vertical asymptotes). Moreover, $\text{Var}_{\post}[\widehat{r}]$ grows more quickly in $1<h<\sqrt{2}$ than in $0<h<1$. In Figure \ref{Fig4b}, we show $\text{Var}_{\post}[1/\widehat{Z}_\text{RIS}]$ for $N=500$. Observe that  $\widehat{r} =\frac{1}{\widehat{Z}_\text{RIS}}$ has the same behavior as the IS estimator when the variance of the denominator (in this case $\post(\theta)$) is smaller than the numerator (in this case $q(\theta)$), but the asymptote is reached only at $h=0$ (not before). Therefore, the case $h<1$ is less catastrophic than in the standard IS scheme. However, RIS presents an additional catastrophic scenario for $h>1$, at $h=1.4$, where there is another vertical asymptote. However, studying numerically $\widehat{Z}_\text{RIS}$ instead of $1/\widehat{Z}_\text{RIS}$, we can see that second  vertical asymptote disappears (see below).   The variance around the optimal value $h=1$ is flatter than in the standard IS. 
\newline
Therefore, choosing properly $h$, RIS can provide better performance than standard IS. However, it seems that the only safe region for avoiding catastrophic scenarios of infinite variance (for estimation of $Z$) is given by the use of a standard IS scheme with a variance of the proposal density greater than the  variance of the target density.
  
%Then, we see that $q(\theta)$ should have non-zero variance and less variance than $\post(\theta)$, in order to avoid infinite variance of the resulting estimator $\widehat{r} =\frac{1}{\widehat{Z}_\text{RIS}}$.  We can observe two vertical asymptotes when we analyze $\text{Var}_{\post}[1/\widehat{Z}_\text{RIS}]$.  However, note that  $\text{Var}_{\post}[\widehat{Z}_\text{RIS}]$ has just one vertical asymptote at $h=0$ as shown below.

%{ 
%In Figure \ref{varZteo_ISandRIS}, we show var$[\widehat{Z}_\text{IS}]$ for $M=500$. In fact, the effect of the sample size in this case is simply scaling the whole curve.} 

%%%%%%%%%%%%%%%%%%%%%%%%%%%%%%%
\subsubsection{Numerical comparison} 
%%%%%%%%%%%%%%%%%%%%%%%%%%%%%%%

\begin{figure}[h!] 
	\centering
	\centerline{
	\subfigure[\label{FigRIS1}Bias and variance of $\widehat{Z}_\text{RIS}$.]{\includegraphics[width=6cm]{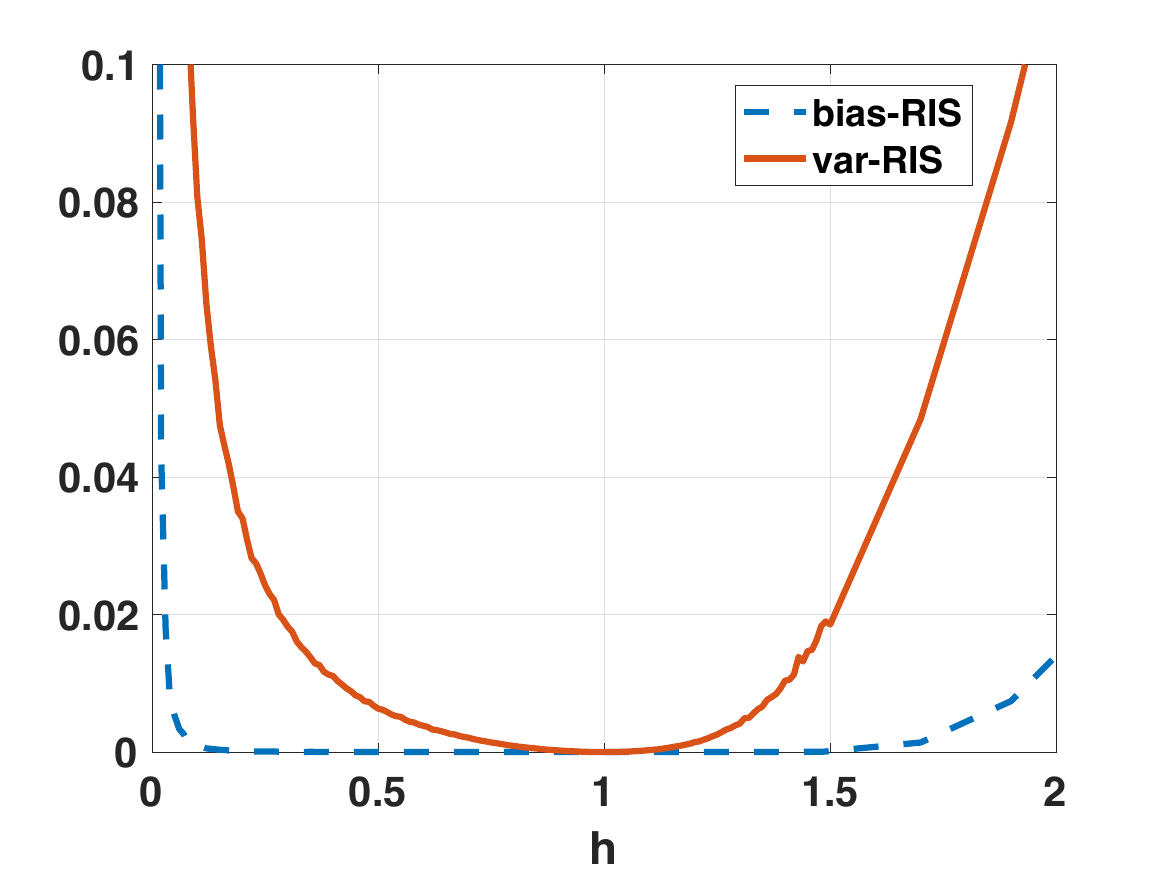}}
	\subfigure[MSE of $\widehat{Z}_\text{IS}$ and $\widehat{Z}_\text{RIS}$.]{\label{FigRIS2}\includegraphics[width=6cm]{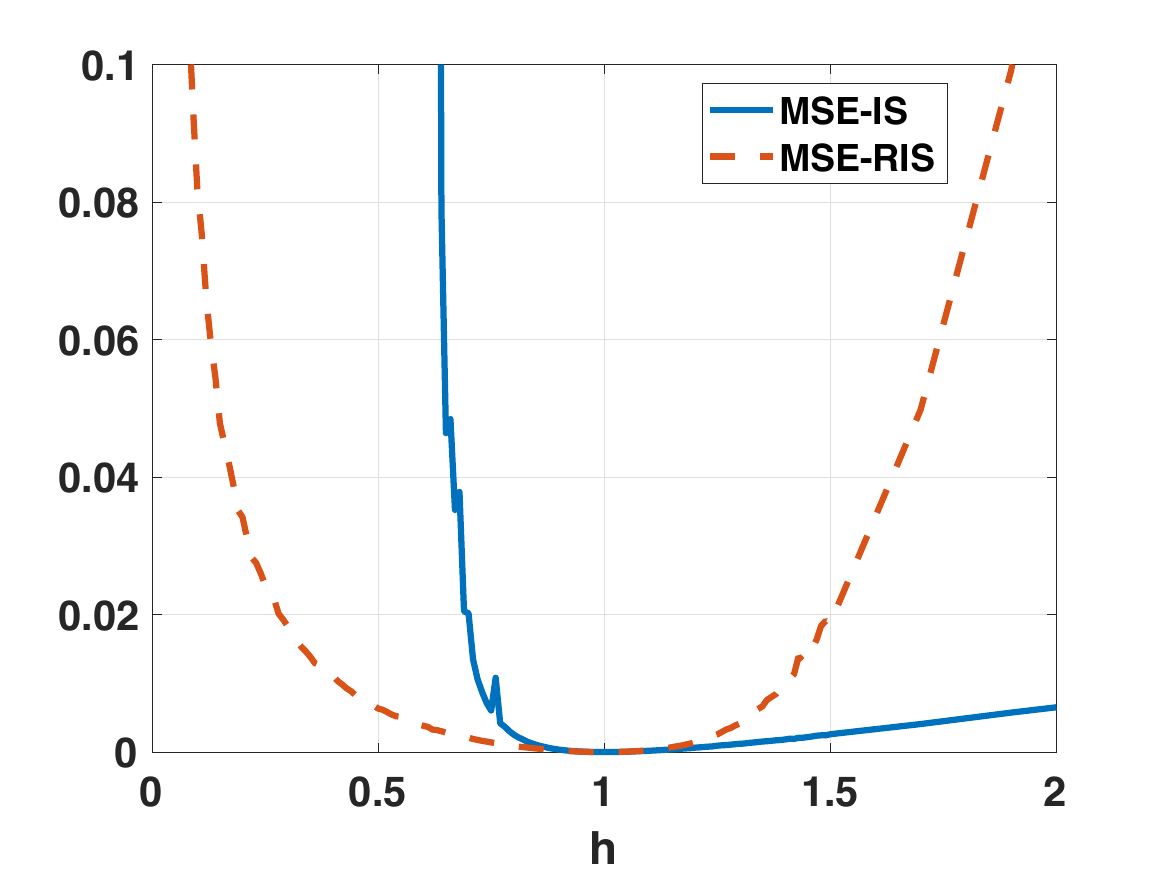}}
	}
	\caption{(a) Bias (dashed line) and variance (solid line) of $\widehat{Z}_\text{RIS}$ as a function of $h$ ($N=500$). (b) MSE of $\widehat{Z}_\text{IS}$ (solid line) and $\widehat{Z}_\text{RIS}$ (dashed line) as a function of $h$ ($N=500$).}
	%\label{MSE, var and bias numerically of IS and RIS}
	\label{cacalabel}
\end{figure}

In the previous part of this section, we have compared $\text{Var}_{q}\left[\widehat{Z}_\text{IS}\right]$ with $\text{Var}_{\post}\left[\frac{1}{\widehat{Z}_\text{RIS}}\right]$. Recall that 
 $\mathbb{E}_{q}[\widehat{Z}_\text{IS}]=Z$ and $\mathbb{E}_{\post}\left[\frac{1}{\widehat{Z}_\text{RIS}}\right]=Z$ but $\mathbb{E}_{\post}\left[\widehat{Z}_\text{RIS}\right]\neq Z$,  i.e., the bias is non-zero in this last case. 
 \newline
Thus, setting $N=500$, we compute numerically the mean square error (MSE) of both $\widehat{Z}_\text{IS}$ and $\widehat{Z}_\text{RIS}$, and the variance and bias of both $\widehat{Z}_\text{RIS}$, averaging the results over 5000 independent runs. We show the results in Fig. \ref{cacalabel}.  In Figure \ref{FigRIS1}, we provide bias and variance of the estimator $\widehat{Z}_\text{RIS}$. Note that its variance the bias have only one asymptote at $0$ instead of two asymptotes, unlike the variance of $\widehat{r}=1/\widehat{Z}_\text{RIS}$. Indeed, the variance and bias  of  $\widehat{Z}_\text{RIS}$ diverge  also as $h\rightarrow \infty$ (bur without an additional vertical asymptote). Observe also that the bias is negligible for $0.1<h<1.6$,  with respect to the value of the variance.  
\newline
 In Fig. \ref{FigRIS2}, we can see that the MSE of $\widehat{Z}_\text{IS}$ corresponds to its theoretical variance shown in Fig. \ref{varZteo_ISandRIS}, as we expect since $\widehat{Z}_\text{IS}$ has zero bias, hence $\text{MSE}_q(\widehat{Z}_\text{IS}) = \text{Var}_q(\widehat{Z}_\text{IS})$. Although $\widehat{Z}_\text{RIS}$ is not unbiased, we see that its MSE, also shown in Fig. \ref{FigRIS2}, is virtually identical to its variance shown in  Fig. \ref{FigRIS1}, where the bias seems to be negligible for the majority of values of $h$. {In fact, applying the Delta method to $\widehat{Z}_\text{RIS}$ shows that $\mathbb{E}[(\widehat{Z}_\text{RIS}-Z)^2]=\mbox{var}\left[\widehat{r}\right] + \mathcal{O}(\frac{1}{N^2})$, i.e., the MSE of $\widehat{Z}_\text{RIS}$ coincides with $\mbox{var}\left[\widehat{r}\right]$ up to $\mathcal{O}(\frac{1}{N^2})$ terms.}

% See also Figure for a numerical comparison of the MSE of $\widehat{Z}_\text{IS}$ and $\widehat{Z}_\text{RIS}$ using $M=N=500$ for a range of $h$ from $h=0.1$ to $h=5$. The results were averaged over 1000 independent simulations. 

\section{Conclusions}\label{FinalDiscSect}

The choice of the proposal density is crucial  for the performance of Monte Carlo sampling methods and, specifically, in IS schemes. Hence, knowing the optimal proposal density in the specific scenario of interest is essential in order to design suitable adaptive procedures within modern IS schemes. In this review, we have provided an exhaustive and accessible introduction to different results about the optimality in IS schemes, that were spread in the literature during the last decades. We have also included novel variants and several settings, including the noisy target scenario, the marginal likelihood estimation, { optimal tempering schedule, models with intractable likelihoods, to name a few}. The relationships among the different frameworks and schemes have been widely described in the text, by means of several summary tables and figures. Theoretical and empirical comparisons have been also provided.  { The range of applications is particularly broad, encompassing Bayesian statistics, machine learning, computer graphics, and many other scientific and engineering domains.}
\newline
This work also should be of particular interest for practitioners and researchers involved in the development of new methods that seek to address the growing list of challenges modern day statistical science is being called upon to address.
{  The presentation is  intentionally self-contained, making the material accessible to both newcomers and experienced researchers.}
 As an example of future work and research challenge, we suggest the analysis of the relevant connection between importance sampling and {\it contrastive learning} \cite{gutmann12a}, where the concept of {\it optimal reference density} has been recently started to study \cite{chehab2023}.

\bibliographystyle{plain}
\bibliography{bibliografia}

{
\appendix
%%%%%%%%%%%%%%%%%%%%%%%%%%%%%%%%%%%%%%%
\section{{ First variation for the optimal proposal}}\label{FVsect}
%%%%%%%%%%%%%%%%%%%%%%%%%%%%%%%%%%%%%%
Let us recall the following Lagrangian functional: 
$$
\mathcal{L}[q]=\int_{\Theta} \frac{\|h(\x)\|_2^2}{q(\x)} d \x+\lambda\left(\int_{\Theta}  q(\x) d \x-1\right).
$$
The first variation is analogous to a derivative in finite dimensions.
Given a functional $\mathcal{L}[q]$, the first variation $\delta \mathcal{L}[q]$ in the direction of a small perturbation $\delta q(\x)$ is defined as
$$
\delta \mathcal{L}[q ; \delta q]= \left.\frac{d}{d \varepsilon} \mathcal{L}[q+\varepsilon \delta q]\right|_{\varepsilon=0}=\lim _{\epsilon \rightarrow 0} \frac{\mathcal{L}[q+\epsilon \delta q]-\mathcal{L}[q]}{\epsilon} .
$$
The first variation only keeps the terms that are linear in the perturbation $\delta q(\x)$, so that 
\begin{itemize}
\item anything independent of $\varepsilon$ vanishes when differentiated;
\item only terms proportional to $\varepsilon$ survive.
\end{itemize}
Moreover, we desire that $\delta \mathcal{L}[q ; \delta q]=0$ for all perturbations $\delta q$. Thus, we can consider a small perturbation $q(\x) \longrightarrow  q(\x)+\epsilon \delta q(\x)$ so that, for instance, we can replace
$$
\frac{\|h(\x)\|_2^2}{q(\x)} \longrightarrow  \frac{\|h(\x)\|_2^2}{q(\x)+\epsilon \delta q(\x)},
$$
in the functional above. Furthermore, using the first-order Taylor expansion, we get
$$
\frac{1}{q+\epsilon \delta q} \approx \frac{1}{q}-\frac{\epsilon \delta q}{q^2},
$$
so that the first integral of the Lagrangian above becomes
$$
\int_\Theta \frac{\|h(\x)\|_2^2}{q(\x)+\epsilon \delta q(\x)} d \x \approx \int_\Theta \frac{\|h(\x)\|_2^2}{q(\x)} d \x-\epsilon \int_\Theta \frac{\|h(\x)\|_2^2}{q(\x)^2} \delta q(\x) d \x,
$$
whereas the Lagrange multiplier term becomes
$$
\lambda \int_\Theta \left[q(\x)+\epsilon \delta q(\x)\right] d\x=\lambda \int_\Theta q(\x) d\x+\epsilon \lambda \int_\Theta \delta q(\x) d\x.
$$
Hence, collecting the linear term in $\epsilon$ and equalling to zero, we have:
$$
\delta \mathcal{L}[q ; \delta q]= \int_\Theta\left(-\frac{\|h(\x)\|_2^2}{q(\x)^2}+\lambda\right) \delta q(\x) d \x=0.
$$
Since this must hold for all perturbations $\delta q(\x)$, we need that:
$$
-\frac{\|h(\x)\|_2^2}{q(\x)^2}+\lambda=0 \qquad \Longrightarrow \quad q(\x)^2=\frac{\|h(\x)\|_2^2}{\lambda}.
$$
Taking the positive root (since $q(\x) \geq 0$ ), we finally arrive to $q_{\mathrm{opt}}(\x)=\frac{\|h(\x)\|_2}{\sqrt{\lambda}}$.

%%%%%%%%%%%%%%%%%%%%%%%%%%%%%%%%%%%%%%%
\section{{ First variation for the optimal tempering}}\label{FVsect2}
%%%%%%%%%%%%%%%%%%%%%%%%%%%%%%%%%%%%%%
We desire to minimize the following Lagrangian,
$$
\mathcal{L}[q]=\int_0^1 \frac{|\varphi^{\prime \prime}(\beta)|}{q(\beta)^2} d \beta+\lambda\left(\int_0^1 q(\beta) d \beta-1\right).
$$
Taking the functional derivative w.r.t. $q(\beta)$ and setting it to zero,
$$
\frac{\delta \mathcal{L}}{\delta q}=-2 \frac{\left|\varphi^{\prime \prime}(\beta)\right|}{q(\beta)^3}+\lambda=0,
$$
we can  solve for $q(\beta)$ and obtain
\begin{align}\label{Eq161}
q(\beta)^3=\frac{2\left|\varphi^{\prime \prime}(\beta)\right|}{\lambda} \quad \Longrightarrow \quad q(\beta)=\left(\frac{2\left|\varphi^{\prime \prime}(\beta)\right|}{\lambda}\right)^{1 / 3}.
\end{align}
Moreover, using the constraint $\int_0^1 q(\beta) d \beta=1$, we obtain
\begin{align*}
\int_0^1\left(\frac{2\left|\varphi^{\prime \prime}(\beta)\right|}{\lambda}\right)^{1 / 3} d \beta=1, \quad  \Longrightarrow \quad \lambda=2\left(\int_0^1\left|\varphi^{\prime \prime}(\beta)\right|^{1 / 3} d \beta\right)^3.
\end{align*}
Hence, replacing $\lambda$ above in Eq. \eqref{Eq161}, we finally get
\begin{align}
q_{\mathrm{opt}}(x)=\frac{\left|\varphi^{\prime \prime}(\beta)\right|^{1 / 3}}{\int_0^1\left|\varphi^{\prime \prime}(t)\right|^{1 / 3} d t} \propto \left|\varphi^{\prime \prime}(\beta)\right|^{1 / 3}.
\end{align}

%%%%%%%%%%%%%%%%%%%
\section{{Proof of Eq. \eqref{EqToProve}}}\label{AppC}
%%%%%%%%%%%%%%%%%%%
Let us recall the expression in Eq. \eqref{EqToProve},
\begin{align}
\varphi^{\prime}(\beta)=\frac{d}{d \beta} \mathbb{E}_{\post_\beta}[\log p({\bf y} | \x)] 
=\operatorname{Var}_{\post_\beta}[\log p({\bf y} | \x)],
\end{align}
and simplify the notation $ \ell(\x)=\log \ell({\bf y}|\x)$. Then, we can write
$$
\varphi(\beta)=\frac{\int_\Theta \ell(\x) e^{\beta \ell(\x)} g(\x) d \theta}{Z(\beta)}=\frac{\int_\Theta \ell(\x) e^{\beta \ell(\x)} g(\x) d \theta}{\int_\Theta e^{\beta \ell(\x)} g(\x) d \theta}=\frac{N(\beta)}{Z(\beta)},
$$
where $N(\beta)=\int_\Theta \ell(\x) e^{\beta \ell(\x)} g(\x) d \theta$ and $Z(\beta)=\int_\Theta e^{\beta \ell(\x)} g(\x) d \theta$. For the derivation rule of the ratio of functions, we have 
$$
\varphi^{\prime}(\beta)=\frac{d\varphi(\beta)}{d\beta}=\frac{N^{\prime}(\beta) Z(\beta)-N(\beta) Z^{\prime}(\beta)}{Z(\beta)^2}.
$$
Deriving $Z(\beta)$ and $N(\beta)$, we have
\begin{align*}
&Z^{\prime}(\beta)=\int_\Theta  \ell(\x) e^{\beta \ell(\x)} g(\x) d \theta
=Z(\beta)\int_\Theta  \ell(\x) \post_\beta(\x|{\bf y}) d \theta =Z(\beta) \mathbb{E}_{\post_\beta}[\ell(\x)],\\
&N^{\prime}(\beta)=\int_\Theta  \ell(\x)^2 e^{\beta \ell(\x)} g(\x) d \theta=Z(\beta)\int_\Theta  \ell(\x)^2 \post_\beta(\x|{\bf y}) d \theta=Z(\beta) \mathbb{E}_{\post_\beta}\left[\ell(\x)^2\right],
\end{align*}
that can replace in the formula of $\varphi^{\prime}(\beta)$ above, i.e.,
\begin{align*}
\varphi^{\prime}(\beta)=\frac{Z(\beta)^2 \mathbb{E}_{\post_\beta}\left[\ell(\x)^2\right]-N(\beta)Z(\beta) \mathbb{E}_{\post_\beta}[\ell(\x)] }{Z(\beta)^2}.
\end{align*}
Since by definition $N(\beta)=Z(\beta) \mathbb{E}_{\post_\beta}[\ell(\x)]$, we can finally write:
\begin{align*}
\varphi^{\prime}(\beta)=\frac{Z(\beta)^2 \left(\mathbb{E}_{\post_\beta}\left[\ell(\x)^2\right]- \mathbb{E}_{\post_\beta}[\ell(\x)]^2 \right) }{Z(\beta)^2}=\operatorname{Var}_{\post_\beta}[\log p(y | \x)].
\end{align*}
}

\newpage
\newpage

	\begin{figure}[!h]
		\centering
		%\subfigure[]{\includegraphics[width=8cm]{vsBS_MIS_MselfIS.png}}	
		\subfigure[\label{USfig1}]{\includegraphics[width=12cm]{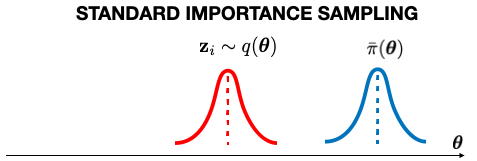}}
		\subfigure[\label{USfig2}]{\includegraphics[width=12cm]{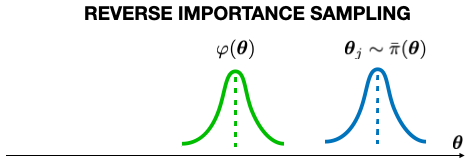}}
	\subfigure[\label{USfig3}]{\includegraphics[width=12cm]{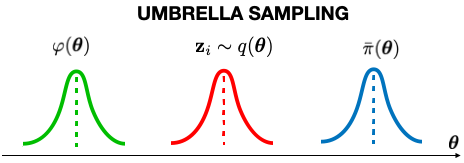}}
	\subfigure[\label{USfig4}]{\includegraphics[width=12cm]{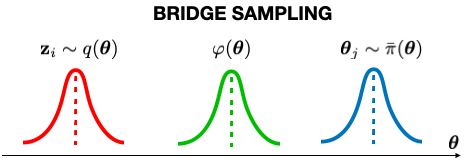}}		
		\caption{Graphical representation and comparison of {\bf (a)} standard IS, {\bf (b)} RIS, {\bf (c)} the umbrella sampling and  {\bf (d)} bridge sampling, for estimating $Z$.}	
		\label{USfig}
	\end{figure}

\end{document}